\documentclass[twocolumn,tighten,astrosymb,trackchanges]{aastex631}
\newcommand{\LCO}{\affiliation{Las Cumbres Observatory, 6740 Cortona Drive, Suite 102, Goleta, CA 93117-5575, USA}}
\newcommand{\UCSB}{\affiliation{Department of Physics, University of California, Santa Barbara, CA 93106-9530, USA}}

\newcommand{\UCD}{\affiliation{Department of Physics and Astronomy, University of California, Davis, 1 Shields Avenue, Davis, CA 95616-5270, USA}}
\newcommand{\WIS}{\affiliation{Department of Particle Physics and Astrophysics, Weizmann Institute of Science, 76100 Rehovot, Israel}}
\newcommand{\OKC}{\affiliation{Department of Physics, Oskar Klein Centre, Stockholm University, SE-106 91, Stockholm, Sweden}}
\newcommand{\OKCAstro}{\affiliation{Department of Astronomy, Oskar Klein Center, Stockholm University, SE-106 91 Stockholm, Sweden}}

\newcommand{\CaltechPhys}{\affiliation{Division of Physics, Mathematics and Astronomy, California Institute of Technology, Pasadena, CA 91125, USA}}

\newcommand{\UCB}{\affiliation{Department of Astronomy, University of California, Berkeley, CA 94720-3411, USA}}

\newcommand{\STScI}{\affiliation{Space Telescope Science Institute, 3700 San Martin Drive, Baltimore, MD 21218-2410, USA}}
\newcommand{\UT}{\affiliation{Department of Astronomy, The University of Texas at Austin, 2515 Speedway, Stop C1400, Austin, TX 78712, USA}}

\newcommand{\Tsinghua}{\affiliation{Physics Department and Tsinghua Center for Astrophysics, Tsinghua University, Beijing, 100084, People's Republic of China}}

\newcommand{\CfA}{\affiliation{Center for Astrophysics \textbar{} Harvard \& Smithsonian, 60 Garden Street, Cambridge, MA 02138-1516, USA}}
\newcommand{\IAIFI}{\affiliation{The NSF AI Institute for Artificial Intelligence and Fundamental Interactions, USA}}
\newcommand{\UA}{\affiliation{Steward Observatory, University of Arizona, 933 North Cherry Avenue, Tucson, AZ 85721-0065, USA}}

\newcommand{\Carnegie}{\affiliation{Observatories of the Carnegie Institute for Science, 813 Santa Barbara Street, Pasadena, CA 91101-1232, USA}}

\newcommand{\UCSC}{\affiliation{Department of Astronomy and Astrophysics, University of California, Santa Cruz, CA 95064-1077, USA}}
\newcommand{\Purdue}{\affiliation{Department of Physics and Astronomy, Purdue University, 525 Northwestern Avenue, West Lafayette, IN 47907-2036, USA}}

\newcommand{\JHU}{\affiliation{Department of Physics and Astronomy, The Johns Hopkins University, 3400 North Charles Street, Baltimore, MD 21218, USA}}

\newcommand{\CIERA}{\affiliation{Center for Interdisciplinary Exploration and Research in Astrophysics (CIERA), 1800 Sherman Ave., Evanston, IL 60201, USA}}
\newcommand{\Northwestern}{\affiliation{Department of Physics and Astronomy, Northwestern University, 2145 Sheridan Rd, Evanston, IL 60208, USA}}
\newcommand{\SkAI}{\affiliation{NSF-Simons AI Institute for the Sky (SkAI), 172 E. Chestnut St., Chicago, IL 60611, USA}}

\newcommand{\IIA}{\affiliation{Indian Institute of Astrophysics, Koramangala 2nd Block, Bangalore 560034, India}}

\newcommand{\GeminiNorth}{\affiliation{Gemini Observatory/NSF's NOIRLab, 670 North A`ohoku Place, Hilo, HI 96720-2700, USA}}

\newcommand{\Rutgers}{\affiliation{Department of Physics and Astronomy, Rutgers, the State University of New Jersey,\\136 Frelinghuysen Road, Piscataway, NJ 08854-8019, USA}}

\newcommand{\Melbourne}{\affiliation{School of Physics, The University of Melbourne, Parkville, VIC 3010, Australia}}

\newcommand{\ICE}{\affiliation{Institute of Space Sciences (ICE, CSIC), Campus UAB, Carrer de Can Magrans, s/n, E-08193 Barcelona, Spain}}
\newcommand{\IEEC}{\affiliation{Institut d’Estudis Espacials de Catalunya (IEEC), E-08034 Barcelona, Spain}}

\newcommand{\USzeged}{\affiliation{Department of Experimental Physics, Institute of Physics, University of Szeged, D\'om t\'er 9, 6720 Szeged, Hungary}}
\newcommand{\BAOUSzeged}{\affiliation{Baja Astronomical Observatory of the University of Szeged, Szegedi {\'u}t, Kt. 766, 6500 Baja, Hungary}}

\newcommand{\MTAELTE}{\affiliation{MTA-ELTE Lend\"ulet "Momentum" Milky Way Research Group, Hungary}}
\newcommand{\Konkoly}{\affiliation{Konkoly Observatory, HUN-REN Research Centre for Astronomy and Earth Sciences (CSFK), Konkoly-Thege Mikl\'os \'ut 15-17, 1121 Budapest, Hungary}}

\newcommand{\AMNH}{\affiliation{Department of Astrophysics, American Museum of Natural History, Central Park West and 79th Street, New York, NY 10024-5192, USA}}
\newcommand{\UDublin}{\affiliation{School of Physics, Trinity College Dublin, The University of Dublin, Dublin 2, Ireland}}

\newcommand{\ICG}{\affiliation{Institute of Cosmology and Gravitation, University of Portsmouth, Dennis Sciama Building, Burnaby Road, Portsmouth PO1 3FX, UK}}

\newcommand{\GSI}{\affiliation{GSI Helmholtzzentrum f\"ur Schwerionenforschung, Planckstra\ss{}e 1, 64291 Darmstadt, Germany}}
\newcommand{\KyotoU}{\affiliation{Department of Astronomy, Kyoto University, Kitashirakawa-Oiwake-cho, Sakyo-ku, Kyoto, 606-8502. Japan}}

\newcommand{\Thailand}{\affiliation{National Astronomical Research Institute of Thailand, 260 Moo 4, Donkaew, Maerim, Chiang Mai 50180, Thailand}}

\newcommand{\OABR}{\affiliation{Istituto Nazionale di Astrofisica, Osservatorio Astronomico di Brera, via E. Bianchi 46, 23807 Merate (LC), Italy}}
\newcommand{\PUC}{\affiliation{Instituto de Astrof\'{i}sica, Facultad de F\'{i}sica, Pontificia Universidad Cat\'{o}lica de Chile, Av. Vicu\~{n}a Mackenna 4860, Santiago, Chile}}

\newcommand{\ELTE}{\affiliation{ELTE E\"otv\"os Lor\'and University, Institute of Physics and Astronomy, P\'azm\'any P\'eter s\'et\'any 1/A, Budapest, 1117 Hungary}}

\newcommand{\UVa}{\affiliation{Department of Astronomy, University of Virginia, 530 McCormick Rd, Charlottesville, VA 22904, USA}}
\newcommand{\CaltechOO}{\affiliation{Caltech Optical Observatories, California Institute of Technology, Pasadena, CA 91125, USA}}
\newcommand{\NOT}{\affiliation{Nordic Optical Telescope, Rambla José Ana Fernández Pérez 7, ES-38711 Breña Baja, Spain}}
\newcommand{\ARIES}{\affiliation{Aryabhatta Research Institute of Observational Sciences (ARIES), Manora Peak, Nainital - 263001, India}}
\newcommand{\ANU}{\affiliation{Research School of Astronomy and Astrophysics, Australian National University, Canberra, ACT 2611, Australia}}
\newcommand{\ANUCGA}{\affiliation{Centre for Gravitational Astrophysics, College of Science, Australian National University, ACT 2601, Australia}}
\newcommand{\UCO}{\affiliation{University of California Observatories, 550 Red Hill Rd, Santa Cruz, CA 95064, USA}}
\newcommand{\Lick}{\affiliation{UCO/Lick Observatory, PO Box 85, Mount Hamilton, CA 95140, USA}}
\newcommand{\ZhejiangU}{\affiliation{Institute for Advanced Study in Physics, Zhejiang University, Hangzhou 310027, China}}

\newcommand{\UIUC}{\affiliation{Department of Astronomy, University of Illinois, Urbana, IL 61801, USA}}
\newcommand{\NCSA}{\affiliation{Center for Astrophysical Surveys, National Center for Supercomputing Applications, Urbana, IL 61801, USA}}
\newcommand{\ICASU}{\affiliation{Illinois Center for Advanced Studies of the Universe; Urbana, IL 61801, USA}}
\newcommand{\UTurku}{\affiliation{Tuorla Observatory, Department of Physics and Astronomy, FI-20014 University of Turku, Finland}}
\newcommand{\FINCA}{\affiliation{Finnish Centre for Astronomy with ESO (FINCA), FI-20014 University of Turku, Finland}}
\newcommand{\CITEVA}{\affiliation{Centro de Astronomía (CITEVA), Universidad de Antofagasta, Av. Angamos 601, Antofagasta, Chile}}
\newcommand{\Oxford}{\affiliation{Astrophysics sub-Department, Department of Physics, University of Oxford, Keble Road, Oxford, OX1 3RH, UK}}
\newcommand{\UChicago}{\affiliation{Department of Astronomy and Astrophysics, University of Chicago, William Eckhart Research Center, 5640 South Ellis Avenue, Chicago, IL 60637, USA}}
\usepackage{appendix}
\usepackage{xcolor}
\usepackage{longtable}
\usepackage{booktabs}
\usepackage{ltablex}
\usepackage{gensymb}
\usepackage{apjfonts}
\usepackage{hyperref}
\keepXColumns

\shorttitle{An intermediate luminosity Type Iax SN~2024pxl}
\shortauthors{Singh et al.}

\begin{document}

\title{Photometry and Spectroscopy of SN 2024pxl: A Luminosity Link Among Type Iax Supernovae} 

\correspondingauthor{Mridweeka Singh}
\email{mridweeka.singh@iiap.res.in, yashasvi04@gmail.com}

\author[0000-0001-6706-2749]{Mridweeka Singh}
\IIA

\author[0000-0003-3108-1328]{Lindsey A.\ Kwok}
\thanks{CIERA Fellow}
\CIERA

\author[0000-0001-8738-6011]{Saurabh W.\ Jha}
\Rutgers


\author[0000-0001-6191-7160]{R.~Dastidar}
\OABR

\author[0000-0003-2037-4619]{Conor~Larison}
\Rutgers

\author[0000-0003-3460-0103]{Alexei V.\ Filippenko}
\UCB

\author[0000-0003-0123-0062]{Jennifer E.\ Andrews}
\GeminiNorth

\author[0000-0002-1895-6639]{Moira~Andrews}
\LCO
\UCSB

\author[0000-0003-3533-7183]{G. C. Anupama}
\IIA

\author[0000-0002-6688-3307]{Prasiddha Arunachalam}
\UCSC

\author[0000-0002-4449-9152]{Katie Auchettl}
\Melbourne
\UCSC

\author[0000-0001-9275-0287]{Dominik B\'anhidi}
\USzeged
\BAOUSzeged

\author[0000-0003-4769-4794]{Barnabas Barna}
\USzeged

\author[0000-0002-4924-444X]{K.\ Azalee Bostroem}
\thanks{LSST-DA Catalyst Fellow}
\UA

\author[0000-0001-5955-2502]{Thomas G.\ Brink}
\UCB

\author[0000-0003-4553-4033]{R\'egis Cartier}
\CITEVA

\author[0000-0003-0853-6427]{Ping Chen}
\ZhejiangU
\WIS

\author[0000-0003-0528-202X]{Collin~T.~Christy}
\UA

\author[0000-0003-4263-2228]{David~A.~Coulter}
\STScI

\author[0000-0003-1858-561X]{Sofia~Covarrubias}
\CaltechPhys

\author[0000-0002-5680-4660]{Kyle W.\ Davis}
\UCSC

\author[0000-0001-9749-4200]{Connor~B.~Dickinson}
\UCSC

\author[0000-0002-7937-6371]{Yize Dong}
\CfA

\author[0000-0003-4914-5625]{Joseph Farah}
\LCO
\UCSB

\author[0000-0003-2024-2819]{Andreas Fl\"ors}
\GSI

\author[0000-0002-2445-5275]{Ryan J.\ Foley}
\UCSC

\author[0000-0003-4537-3575]{Noah Franz}
\UA

\author[0000-0002-4223-103X]{Christoffer~Fremling}
\CaltechPhys
\CaltechOO

\author[0000-0002-1296-6887]{Llu\'is Galbany}
\ICE
\IEEC

\author[0000-0002-3884-5637]{Anjasha Gangopadhyay}
\OKCAstro

\author[0009-0002-4441-3192]{Aarna~Garg}
\UCSC


\author[0000-0002-3739-0423]{Elinor~L.~Gates}
\Lick

\author[0000-0002-4391-6137]{Or Graur}
\ICG
\AMNH

\author[0000-0002-5025-4645]{Alexa~C.~Gordon}
\Northwestern
\CIERA

\author[0000-0002-1125-9187]{Daichi~Hiramatsu}
\CfA
\IAIFI

\author[0000-0003-2744-4755]{Emily~Hoang}
\UCD


\author[0000-0003-4253-656X]{D.~Andrew~Howell}
\LCO
\UCSB

\author[0000-0002-9454-1742]{Brian Hsu}
\UA

\author[0000-0001-5975-290X]{Joel Johansson}
\OKC

\author[0000-0001-9275-0287]{Arti Joshi}
\PUC

\author[0009-0007-5296-4046]{Lordrick~A.~Kahinga}
\UCSC

\author[0009-0005-1871-7856]{Ravjit Kaur}
\UCSC

\author[0000-0001-8367-7591]{Sahana~Kumar}
\UVa

\author[0009-0004-7572-5679]{Piramon~Kumnurdmanee}
\UCSC

\author[0000-0002-1132-1366]{Hanindyo Kuncarayakti}
\UTurku
\FINCA

\author[0000-0002-2249-0595]{Natalie~LeBaron}
\UCB

\author[0000-0003-1731-0497]{C.~Lidman}
\ANU
\ANUCGA

\author[0000-0002-7866-4531]{Chang~Liu}
\Northwestern
\CIERA

\author[0000-0003-2611-7269]{Keiichi Maeda}
\KyotoU

\author[0000-0002-9770-3508]{Kate Maguire}
\UDublin

\author[0009-0006-4963-3206]{Bailey Martin}
\ANU

\author[0000-0001-5807-7893]{Curtis McCully}
\LCO
\UCSB

\author[0009-0008-9693-4348]{Darshana Mehta}
\UCD

\author[0000-0001-7771-4624]{Luca~M.~Menotti}
\UCSC

\author[0009-0007-8154-6863]{Anne~J.~Metevier}
\UCO

\author[0000-0001-9515-478X]{A.~A.~Miller}
\Northwestern
\CIERA
\SkAI

\author[0000-0003-1637-267X]{Kuntal Misra}
\ARIES

\author[0009-0006-5214-0736]{C.~Tanner~Murphey}
\UIUC
\NCSA
\ICASU

\author[0000-0001-9570-0584]{Megan Newsome}
\LCO
\UT

\author[0000-0003-0209-9246]{Estefania Padilla~Gonzalez}
\JHU

\author[0000-0002-1092-6806]{Kishore~C.~Patra}
\UCSC

\author[0000-0002-0744-0047]{Jeniveve Pearson}
\UA

\author[0000-0001-6806-0673]{Anthony~L.~Piro}
\Carnegie

\author[0000-0002-1633-6495]{Abigail~Polin}
\Purdue

\author[0000-0002-7352-7845]{Aravind~P.~Ravi}
\UCD

\author[0000-0002-4410-5387]{Armin Rest}
\STScI
\JHU

\author[0000-0002-5683-2389]{Nabeel~Rehemtulla}
\Northwestern
\CIERA
\SkAI

\author[0000-0002-7015-3446]{Nicolas~Meza~Retamal}
\UCD

\author[0009-0006-3342-6181]{O.~M.~Robinson}
\UCSC

\author[0000-0002-7559-315X]{C\'{e}sar Rojas-Bravo}
\UCSC

\author[0000-0002-6688-0800]{Devendra~K.~Sahu}
\IIA

\author[0000-0003-4102-380X]{David~J.~Sand}
\UA

\author[0000-0002-8538-9195]{Brian~P.~Schmidt}
\ANU

\author[0000-0001-6797-1889]{Steve Schulze}
\CIERA

\author[0009-0002-5096-1689]{Michaela Schwab}
\Rutgers

\author[0000-0002-4022-1874]{Manisha Shrestha}
\UA

\author[0000-0003-2445-3891]{Matthew~R.~Siebert}
\STScI

\author[0000-0003-3801-1496]{Sunil Simha}
\CIERA
\UChicago

\author[0000-0001-5510-2424]{Nathan Smith}
\UA

\author[0000-0003-1546-6615]{Jesper Sollerman}
\OKCAstro

\author[0000-0003-4524-6883]{Shubham~Srivastav}
\Oxford

\author[0000-0001-8073-8731]{Bhagya~M.~Subrayan}
\UA

\author[0000-0003-4610-1117]{Tam\'as Szalai}
\USzeged
\MTAELTE

\author[0000-0002-5748-4558]{Kirsty Taggart}
\UCSC

\author[0000-0002-0525-0872]{Rishabh Singh Teja}
\IIA


\author[0000-0001-9834-3439]{Jacco~H.~Terwel}
\UDublin
\NOT

\author[0000-0002-1481-4676]{Samaporn Tinyanont}
\Thailand

\author[0000-0001-8818-0795]{Stefano Valenti}
\UCD


\author[0000-0001-8764-7832]{J\'{o}zsef Vink\'{o}}
\Konkoly
\USzeged
\ELTE
\UT

\author[0009-0003-8229-0127]{Aya L. Westerling}
\UCSC

\author[0000-0003-1349-6538]{J.\ Craig Wheeler}
\UT

\author[0000-0002-6535-8500]{Yi Yang}
\Tsinghua
\UCB

\author[0000-0002-2636-6508]{WeiKang Zheng}
\UCB



\begin{abstract}

We present extensive ultraviolet to optical photometric and optical to near-infrared (NIR) spectroscopic follow-up observations of the nearby intermediate-luminosity ($M_V = -16.81\pm0.19$~mag) Type Iax supernova (SN) 2024pxl in NGC 6384. SN~2024pxl exhibits a faster light curve than the high-luminosity members of this class, and slower than low-luminosity events. The observationally well-constrained rise time of $\sim$11 days and an estimated synthesized $^{56}$Ni mass of 0.03\, M$_\odot$, based on analytical modeling of the integrated spectral energy distribution light curve, are consistent with models of the weak deflagration of a carbon-oxygen white dwarf. Our optical spectral sequence of SN~2024pxl shows weak \ion{Si}{2} lines and spectral evolution similar to other high-luminosity Type Iax SNe, but also a prominent early-time \ion{C}{2} line, like lower-luminosity Type Iax SNe. The late-time optical spectrum of SN~2024pxl closely matches that of SN~2014dt, and its NIR spectral evolution aligns with that of other well-studied, high-luminosity Type Iax SNe. The spectral-line expansion velocities of SN~2024pxl are at the lower end of the Type Iax SN velocity distribution, and the velocity distribution of iron-group elements compared to intermediate-mass elements suggests that the ejecta are mixed on large scales, as expected in pure deflagration models. SN~2024pxl exhibits characteristics intermediate between those of high-luminosity and low-luminosity Type~Iax SNe, further establishing a link across this diverse class.  

\end{abstract}



\vspace{0.5cm}
\section{Introduction}
\label{sec:intro}

Type Iax supernovae (SNe) are the fainter, peculiar cousins of Type Ia SNe, both arising from the explosive thermonuclear fusion of carbon-oxygen (CO) white dwarfs \citep[WDs; for a review, see][]{2017hsn..book..375J}. Owing to their homogeneity, ``normal'' Type Ia SNe serve as standardizable candles in cosmology \citep{1993ApJ...413L.105P,1999AJ....118.1766P}. Type Iax SNe, however, display comparatively lower luminosities, explosion energies, and ejecta velocities \citep{2003PASP..115..453L,Filippenko2003,2013ApJ...767...57F,2017hsn..book..375J}. Their physical properties are also more heterogeneous, with luminosities ranging from $M_{r} = -12.7$~mag \citep{Karambelkar_2021} to $M_{r} = -18.6$~mag \citep{2015A&A...573A...2S}, and line absorption velocities at maximum light between 2000 and 8000 km s$^{-1}$ \citep{2009AJ....138..376F,2014A&A...561A.146S}. These distinct characteristics suggest that they arise from different progenitor systems and also have explosion mechanisms that differ in detail from those of normal Type Ia SNe.

The light curves of Type Iax SNe differ markedly from those of normal Type Ia SNe, rising faster and declining more rapidly in bluer bands \citep{2016A&A...589A..89M,2017A&A...601A..62M,2017hsn..book..375J,2018MNRAS.478.4575L}. The light-curve-shape variations and the observed diversity in brightness \citep{2010ApJ...720..704M,2011ApJ...731L..11N,2013ApJ...767...57F,2016A&A...589A..89M,2023ApJ...953...93S} suggest that the synthesized $^{56}$Ni, powering the light curve, spans a wide range from 8$^{+4}_{-5} \times$10$^{-4}$ M${_\odot}$ \citep{Karambelkar_2021} to 0.3 M${_\odot}$ \citep{2015A&A...573A...2S}.

Spectroscopically, Type Iax SNe resemble 91T-like Type Ia SNe \citep{2007PASP..119..360P} at early times, characterized by strong lines due to \ion{Fe}{3} and \ion{Fe}{2} and typically weak \ion{Si}{2}. As they evolve, their late-time spectra diverge significantly from those of Type Ia SNe, displaying both permitted and forbidden lines of Fe and Ca \citep{2014ApJ...786..134M,2015A&A...573A...2S}. Notably, no fully nebular spectrum of a Type Iax SN has been observed, owing to the persistence of lines with a pronounced P-Cygni profile for hundreds of days \citep[e.g.,][]{2023ApJ...951...67C}. Their spectral evolution slows considerably after 200$-$400 days past maximum light \citep{2016MNRAS.461..433F}. \cite{2022ApJ...941...15M} discussed the existence of an Fe-rich innermost region and its association with a bound WD (remnant) using the day 500 spectrum of SN~2019muj.

High-luminosity and low-luminosity Type Iax SNe differ spectroscopically. For example, high-luminosity objects do not show a strong \ion{C}{2} feature at 6580~\AA, whereas low-luminosity ($M_{V} \gtrsim -15$ mag) Type Iax SNe do. Low-luminosity objects such as SNe~2008ha \citep{2009AJ....138..376F}, 2010ae \citep{2014A&A...561A.146S}, 2019gsc \citep{2020ApJ...892L..24S,2020MNRAS.496.1132T}, 2020kyg \citep{2022MNRAS.511.2708S,2023ApJ...953...93S}, and 2021fcg \citep{Karambelkar_2021} also exhibit rapid spectroscopic evolution, lower expansion velocities, and a faster transition to the partial nebular phase. 

Weak deflagrations of CO WDs are promising explosion models for Type Iax SNe, roughly reproducing many observed properties of brighter objects, such as their $^{56}$Ni masses, peak luminosities, rise times, and early-time spectra \citep{2014MNRAS.438.1762F, 2013MNRAS.429.2287K, 2004PASP..116..903B, 2012ApJ...761L..23J, Lach2022a}. However, these models struggle to match low-luminosity Type Iax SNe that may instead be explained by weak deflagrations in hybrid carbon-oxygen-neon (CONe) WDs \citep{2014ApJ...789L..45M,2015MNRAS.450.3045K}.

SN~2012Z offered a valuable opportunity to investigate the progenitor systems of Type Iax SNe. Using deep {\it Hubble Space Telescope (HST)} pre-explosion images of its host galaxy NGC~1309, \cite{2014Natur.512...54M} detected a luminous blue source at the location of the SN and proposed a progenitor system consisting of a CO WD with a helium-star companion. Follow-up observations of SN~2012Z confirmed that the SN is still brighter than pre-explosion, suggesting that the helium-star companion survived the explosion and that a bound remnant was left behind \citep{McCully2022,schwab2025remarkablelatetimefluxexcess}. A similar progenitor scenario was suggested for SN~2014dt \citep{2015ApJ...798L..37F}.  Moreover, a faint red source was identified in {\it HST} images taken 4\,yr after the explosion of SN~2008ha that may be a companion star or remnant \citep{2014ApJ...792...29F}. In the case of  SN~2020udy, another bright Type Iax, constraints on interaction with a companion star from light curves obtained soon after the explosion also favor a helium-star companion \citep{2023MNRAS.525.1210M}.

SN~2024pxl ($\alpha = 17^{\mathrm{h}} 32^{\mathrm{m}} 27^{\mathrm{s}}_{^{\centerdot}}350$, $\delta = +07^\circ03' 44\farcs68$, J2000) was first discovered and reported to the Transient Name Server\footnote{\url{https://www.wis-tns.org}} by the \texttt{BTSbot} machine-learning model \citep{Rehemtulla+2024} on 23 July 2024 at 09:41:18 (UTC dates are used throughout this paper) using data from the Zwicky Transient Facility \citep[ZTF;][]{Bellm+2019a, Bellm+2019b, Dekany+2020, Masci+2019, Graham+2019}. As part of the \texttt{BTSbot-nearby} program \citep{Rehemtulla+2025}, \texttt{BTSbot} triggered a target-of-opportunity photometric and spectroscopic request to the spectral energy distribution machine spectrograph \citep[SEDM;][]{Blagorodnova+2018, Kim+2022}. SN\,2024pxl is located 45\farcs5 E and 7\farcs8 N from the nucleus of its host galaxy, NGC~6384. Other designations of this SN are ATLAS24lpk and ZTF24aawrofs. \cite{2024TNSCR2588....1S} classified SN~2024pxl as a Type Iax SN, noting a strong resemblance to the bright Type Iax SN~2005hk one week before maximum light.

This study presents an extensive follow-up campaign of SN~2024pxl, integrating observations from both ground- and space-based telescopes. SN~2024pxl is perhaps the best-sampled Type Iax SN to date in terms of its photometric and spectroscopic data. \cite{2025ApJ...989L..33K} analyze optical and near-infrared (NIR) spectra from this study together with mid-infrared (MIR) data from the {\it James Webb Space Telescope (JWST)}, providing important clues about the explosion mechanism and progenitor system, complementing the results presented here. \autoref{observations_data_reduction} details the observations and data reduction, and \autoref{distance_extinction_explosion_epoch} describes the distance determination, line-of-sight extinction, and explosion epoch estimation. \autoref{light_curve_properties} presents the light curve, color evolution, and analytical modeling of the integrated spectral energy distribution (SED) light curve. In \autoref{spectral_evolution}, we examine the spectral features and line-velocity evolution of SN~2024pxl and compare it with other well-studied Type Iax SNe. \autoref{summary} summarizes the key findings of this study.

\section{Observations and Data Reduction}
\label{observations_data_reduction}

\subsection{Photometry}
\label{photometry}

A high-cadence dataset for SN~2024pxl was acquired with both ground- and space-based telescopes as part of an extensive photometric observing campaign initiated on the day of discovery. Photometry in $BgVri$ bands was obtained through the Global Supernova Project (GSP) collaboration using the Las Cumbres Observatory (LCO; \citealt{2013PASP..125.1031B}) 0.4\,m and 1\,m telescopes. Preprocessing, including bias correction and flat fielding, was handled by the BANZAI pipeline \citep{McCully2018}. Further data reduction was carried out using \texttt{lcogtsnpipe} \citep{Valenti2016}, a photometric reduction pipeline that uses point-spread-function (PSF) photometry \citep{Stetson1987} to calculate zero points, color terms, and extracted magnitudes. Photometry in $BV$ bands is reported in Vega magnitudes \citep{1992AJ....104..340L}, while $gri$ band data are presented in AB magnitudes \citep{1983ApJ...266..713O}, calibrated against Sloan Digital Sky Survey (SDSS) sources \citep{Smith2002}. Given the proximity of SN~2024pxl to its host galaxy, host contamination was mitigated by subtracting template images acquired on March 7, 2018 — before the explosion. These templates, originally taken for SN~2017drh (which exploded in the same galaxy), were subtracted using the PyZOGY algorithm \citep{Zackay2016, Guevel2017}, integrated within the \texttt{lcogtsnpipe} pipeline. 

Additional imaging of SN~2024pxl was obtained in $BVri$ bands with the 1~m Nickel telescope at Lick Observatory. The images were calibrated using bias and sky flat-field frames following standard procedures. PSF photometry was performed and calibrated relative to Pan-STARRS1 photometric standards \citep{Flewelling16}. 

Early-time photometric observations of SN~2024pxl were also obtained through the Distance Less Than 40 Mpc (DLT40) survey \citep{Tartaglia2018}, using the PROMPT-MO 0.4\,m telescope operated via the Skynet Robotic Telescope Network \citep{Reichart2005}. These observations were carried out in a broad, unfiltered “Open” mode, and the resulting data were transformed to the SDSS $r$ band following the calibration and reduction procedures described by \cite{Tartaglia2018}.

Multiband photometry was further acquired using the 0.76\,m Katzman Automatic Imaging Telescope \citep[KAIT;][]{Filippenko2001} at Lick Observatory in the {\em BVRI} bands. All images were reduced with a custom pipeline\footnote{\url{https://github.com/benstahl92/LOSSPhotPypeline}} described by \cite{Ganeshalingam_2010} and \cite{Stahl_2019}. Host-galaxy contamination was removed via image subtraction, using a pre-explosion template obtained on 2018 June 22. PSF photometry was performed with DAOPHOT \citep{Stetson_1987} from the IDL Astronomy User’s Library\footnote{\url{http://idlastro.gsfc.nasa.gov/}}. Photometric calibration was achieved using local standard stars from the Pan-STARRS1 Surveys \citep{Schlafly_2012}, with magnitudes transformed into the Landolt standard system \citep{Landolt_1992} using the relations from \cite{Tonry_2012}.

In addition, observations of SN~2024pxl were conducted in the $uBVgri$ bands using the Direct 4k $\times$ 4k camera on the 1\,m Henrietta Swope telescope at Las Campanas Observatory, Chile. A full description of the data-reduction procedure is provided by \citet{Kilpatrick_2018}.

SN~2024pxl was observed with the Dark Energy Camera (DECam) in the $griz$ bands as part of the Young Supernova Experiment (YSE DECam; NOIRLab propID 2023A-237157). Data reduction, image differencing, and forced photometry were performed using photpipe \citep{Rest2014}. Templates included public DECam Local Volume Exploration Survey (DELVE; \citealt{Drlica-Wagner2021}) pre-explosion images from 2023 September 9 ($g$, $r$) and 2023 April 7 ($z$), as well as a public BLINK (Billion Lines INdexing in a clicK; \citealt{Kamennoff2012}) image from 2018 April 13 ($i$). 

SN~2024pxl was also observed in the {\it z} band with the RC80 and BRC80 robotic telescopes at Piszkesteto station of Konkoly Observatory and at Baja Observatory of University of Szeged, Hungary (a complete {\it BgVriz} light curve will be presented by Banhidi et al., in prep.). The magnitudes were computed via aperture photometry and tied to Pan-STARRS photometry of local stars, including a color term in the transformation. 

The multiband photometry from the different instruments is consistent within the quoted uncertainties, with no significant systematic offsets observed. Optical photometry of SN~2024pxl is presented in Table \ref{tab:photometric_observational_log_2024pxl}. The {\it z}-band photometry of SN~2024pxl is provided in \autoref{tab:z_band_photometry_24pxl}. 

The {\it Neil Gehrels Swift Observatory} \citep[{\it Swift}][]{2004ApJ...611.1005G} observed SN~2024pxl with its Ultra-Violet/Optical Telescope (UVOT; \citealt{2005SSRv..120...95R}) in both ultraviolet (UV) and optical filters.  We reduced the UVOT images using the High-Energy Astrophysics Software (HEASoft\footnote{https://heasarc.gsfc.nasa.gov/docs/software/heasoft/}). A circular source region centered at the position of the SN with a radius of 3$\arcsec$ was used for aperture photometry. We measured the background contribution from a circular region (aperture radius of 5$\arcsec$) that is not contaminated by any other sources. Zeropoints for photometry were chosen from \cite{2010MNRAS.406.1687B} with time-dependent sensitivity corrections updated in 2020. \autoref{tab:swift_photometry_24pxl} details the photometric observations of SN~2024pxl collected by {\it Swift}.

\subsection{Spectroscopy}
\label{spectroscopy}

\subsubsection{Optical Spectra}

We have compiled an extensive spectral dataset for SN~2024pxl using multiple ground-based telescopes in the optical and near-infrared (NIR) domains. Spectra of SN~2024pxl obtained with the Double Beam Spectrograph (DBSP; \citealt{1982PASP...94..586O}) mounted on the 5\,m Hale Telescope at Palomar Observatory were reduced using \texttt{DBSP-DRP} \citep{Mandigo-Stoba2021,Mandigo-Stoba2022,dbsp_drp_zenodo} and \texttt{PypeIt} \citep{Prochaska2020}. 
Spectra acquired with the Wide Field Spectrograph (WiFeS; \citealt{2007Ap&SS.310..255D, 2010Ap&SS.327..245D}) on Australian National University (ANU) 2.3\,m telescope at Siding Spring Observatory were reduced with the data-reduction pipeline \texttt{PyWiFeS} \citep{2014Ap&SS.349..617C,2024PASA...41...68C}. Spectroscopic observations were also triggered with the FLOYDS spectrograph on the 2\,m Faulkes Telescope North and South (FTN and FTS; \citealt{2013PASP..125.1031B}) through the GSP collaboration. The spectra were reduced with the \texttt{floydsspec}\footnote{https://www.authorea.com/users/598/articles/6566} pipeline, using standard reduction techniques. 

Gemini IRAF\footnote{IRAF is distributed by the National Optical Astronomy Observatories, operated by the Association of Universities for Research in Astronomy, Inc., under a cooperative agreement with the National Science Foundation.} packages were used to reduce the spectra of SN~2024pxl from the Gemini Multi-Object Spectrograph mounted at Gemini North (GMOS-N; \citealt{2004PASP..116..425H}) and Gemini South (GMOS-S) Observatories. Two spectra of SN~2024pxl were acquired using the SPectrograph for the Rapid Acquisition of Transients (SPRAT; \citealt{Piascik2014}), mounted on the Liverpool Telescope (LT; \citealt{Steele2004}). Reduction of the SPRAT spectra was carried out using the SPRAT pipeline.

SN~2024pxl was observed with the Robert Stobie Spectrograph (RSS) attached to the 9.2\,m Southern African Large Telescope (SALT). All RSS spectra were reduced using a custom pipeline based on standard Pyraf \citep{Pyraf} routines and the \texttt{PySALT} package \citep{2010SPIE.7737E..25C}.

A few spectra of SN~2024pxl were taken with the Low Resolution Spectrograph 2 (LRS2) \citep{chonis_lrs2:_2014} mounted on the Hobby-Eberly Telescope (HET) \citep{1998SPIE.3352...34R} located at McDonald Observatory. LRS2 is a two-beam spectrograph covering the blue (LRS2-B) and the red (LRS2-R) part of the optical spectrum with an average resolution of $\lambda / \Delta \lambda \sim 1500$. The data were reduced by the Panacea pipeline\footnote{\url{ https://github.com/grzeimann/Panacea}}.

As part of our observing campaign, we have utilized the Hanle Faint Object Spectrograph and Camera (HFOSC) mounted on the Himalayan Chandra Telescope (HCT; \citealt{2010ASInC...1..193P}). The HCT spectra were reduced using standard IRAF routines. 

SN~2024pxl was observed with Binospec on the MMT telescope \citep{bino}; these spectra were reduced using the Binospec IDL pipeline \citep{binopipe}. A spectrum was taken with the Low Resolution Imaging Spectrometer \citep[LRIS;][]{Oke95} on the Keck~I 10\,m telescope, which was reduced in a standard way using the LPipe pipeline \citep{Perley19}. Many spectra of SN~2024pxl were obtained with the Kast double spectrograph \citep{KAST} on the Shane 3\,m telescope at Lick Observatory, either at low airmass or with the slit aligned along the parallactic angle to minimize the effects of atmospheric dispersion \citep{1982PASP...94..715F}.  The reduction of all Kast spectra was performed following standard procedures\footnote{\url{https://github.com/msiebert1/UCSC\_spectral\_pipeline}}\textsuperscript{,}\footnote{\url{https://github.com/ishivvers/TheKastShiv}} as outlined by \cite{Siebert2019} and \cite{Silverman2012}. 

In our observing campaign, we acquired data from the Gran Telescopio Canarias (GTC) with OSIRIS and EMIR, and reduced the spectra using a dedicated pipeline based on PypeIt \citep{2020zndo...3743493P,pypeit:zenodo,Prochaska2021,2025arXiv250119108G}. PypeIt was also used to reduce the spectra of SN~2024pxl gathered with the 2.56\,m Nordic Optical Telescope (NOT) deploying the Alhambra Faint Object Spectrograph and Camera (ALFOSC). We collected one spectrum of SN~2024pxl with the XShooter spectrograph mounted on the Very Large Telescope \citep{2011A&A...536A.105V}; the EsoReflLex pipeline \citep{2013A&A...559A..96F}, including the XShooter module, was used to reduce it.

The reduced spectra were scaled to the photometry of the corresponding epoch using a linear fit. All spectra were dereddened and corrected for host-galaxy redshift. A log of the spectroscopic observations of SN~2024pxl is given in \autoref{tab:spectroscopic_observations_24pxl}.

\subsubsection{NIR Spectra}

The first NIR spectrum in our observing campaign of SN~2024pxl was obtained from Near-InfraRed Echellette Spectrometer (NIRES) on the Keck~II 10\,m telescope through the Keck Infrared Transient Survey (KITS) program and reduced using procedures outlined by \cite{2024PASP..136a4201T}. Two spectra were obtained by the Folded-port InfraRed Echellette Spectrograph \citep[FIRE;][]{Simcoe2013} mounted on the 6.5\,m Magellan-Baade Telescope at Las Campanas Observatory in Chile, and reduced using the IDL pipeline {\tt firehose} \citep{Simcoe2013}. NIR spectra of SN~2024pxl were acquired from the NASA InfraRed Facility Telescope (IRTF) using the SpeX spectrograph \citep{2003PASP..115..362R}. Data reduction of SpeX data was performed with Spextool and included tasks such as flat fielding, wavelength calibration, background subtraction, and spectra extraction \citep{2004PASP..116..362C}. 

We obtained two NIR spectra of SN~2024pxl with the Espectrógrafo Multiobjeto Infra-Rojo spectrograph (EMIR; \citealt{2022A&A...667A.107G}) mounted on the GTC, which were reduced using a dedicated pipeline based on PyEMIR \citep{2010ASPC..434..353P,2019ASPC..523..317C}. The Southern Astrophysical Research (SOAR) telescope, equipped with TripleSpec \citep{Schlawin2014}, was also deployed to obtain NIR spectra. Spectra acquired by SOAR were reduced using the IDL-based Spextool package \citep{2004PASP..116..362C}. All NIR observations for SN~2024pxl are reported in \autoref{tab:nir_spectroscopic_observations_24pxl}.

\section{Distance, extinction, and explosion epoch}
\label{distance_extinction_explosion_epoch}

SN~2024pxl is hosted by NGC~6384 at a spectroscopic redshift of $z = 0.0056$ \citep{2013ApJ...771...88L}. Fortunately, a normal (though highly dust-extinguished) SN~Ia with a well-sampled light curve from LCO and Lick Observatory, SN~2017drh, was also hosted by NGC~6384 \citep{Stahl2019}. We estimate the distance to SN~2017drh using BayeSN, a hierarchical Bayesian SN Ia light-curve model \citep{mandel_hierarchical_2022,Grayling2024}. SN~2017drh was located near the center of NGC~6384 and is significantly dust reddened, much more than SNe~Ia typically used for cosmological analyses. Thus, we favored using BayeSN because it explicitly models host-galaxy dust reddening and extinction separately from intrinsic SN spectral energy distribution (SED) variations, unlike alternatives like SALT2 \citep{guy_salt2_2007} that aggregate intrinsic and extrinsic SN~Ia color variations. 

Using the BayeSN model trained by \cite{ward_bayesn_2023}, and assuming an $R_V = 3.1$ reddening law, fitting the light curve of SN~2017drh yields a distance modulus to NGC~6384 of $\mu = 31.81 \pm 0.11$ mag, which corresponds to a distance of $23.0 \pm 2.0$ Mpc that we adopt for our analysis of SN~2024pxl. The BayeSN fit to SN~2017drh results in a host-galaxy extinction of $A_V = 2.77\pm 0.04$~mag, confirming the large amount of dust along the line of sight to SN~2017drh. We present the LCO photometry, BayeSN fit, and spectroscopic observations of SN~2017drh in \autoref{sec:appendixA_17drh}.

We also perform a BayeSN fit to SN~2017drh allowing $R_V$ to vary, given that previous observations of heavily dust-extinguished SNe~Ia sometimes indicate low $R_V$ \citep{Krisciunas:2006,Elias-Rosa:2006,Elias-Rosa:2008,Wang:2008,Misra:2008,Amanullah:2015}, possibly due to scattering effects in the surrounding medium or an atypical dust grain size distribution \citep{Wang:2005,Goobar:2008,Kawabata:2014,Gao:2015,Bulla:2018}. In that case we can recover a model fit for SN~2017drh with $R_V = 1.45$ and  $A_V = 1.88 \pm 0.10$ mag that is nearly indistinguishable from the $R_V = 3.1$ model, but with almost 1 mag less total extinction and thus a larger distance modulus, $\mu = 32.70 \pm 0.14$ mag. This would imply a distance of $34.7 \pm 2.3$ Mpc, significantly farther than our adopted estimate. Unfortunately, owing to the lack of NIR data for SN~2017drh, we cannot break the degeneracy between $A_V$ and $R_V$. We thus proceed with our distance estimate using $R_V=3.1$ but caution that the lower-$R_V$ fit and increased distance would imply a factor of $\sim$2 change in the derived luminosity of SN~2024pxl. We note that these distances bracket the distance to SN~2017drh derived by \citet{Hoogendam:2025} of $29 \pm 5$ Mpc using the SNooPy code \citep{Burns:2011}.

The extinction experienced by SN 2024pxl due to the Milky Way in the~direction of NGC~6384 is $A_{V} = 0.338$\,mag, corresponding to $E(B-V) = 0.11$\,mag \citep{2011ApJ...737..103S} and assuming $R_{V}$ = 3.1. Estimating the host-galaxy extinction is more difficult because, unlike the normal SN~Ia 2017drh, we do not have a standard model color for the diverse Type Iax class. However, in our spectroscopic observations of SN~2024pxl, we note the presence of narrow interstellar Na I D absorption lines from both the Milky Way and NGC~6384, with approximately equal strength. Thus, we crudely infer the host-galaxy contribution to the extinction of SN~2024pxl to be similar to the Milky Way extinction and adopt a total reddening and extinction along the line of sight to SN~2024pxl of \( E(B-V) = 0.22 \) mag and \( A_V = 0.68 \) mag. 

A nondetection of SN~2024pxl was reported on 21~July~2024 \citep[JD = 2460512.75;][]{2024TNSTR2553....1R} with a limiting AB magnitude of 19.37 (3$\sigma$ upper limit) in the {\it r} filter. To estimate the explosion epoch of SN~2024pxl, we combined photometric data from the ATLAS and ZTF surveys in four optical bands (ATLAS \textit{c} and \textit{o};  ZTF \textit{g} and \textit{r}). A power-law model of the form

\begin{equation}
    F(t) = A \times (t - t_{\mathrm{exp}})^n,
\end{equation}
\noindent
was used to fit the early-time flux evolution, where $A$ is a band-dependent scaling constant, $t_{\mathrm{exp}}$ is the explosion time, and $n$ is the power-law index. 

\begin{figure}[t]
	\begin{center}
		\includegraphics[width=\columnwidth]{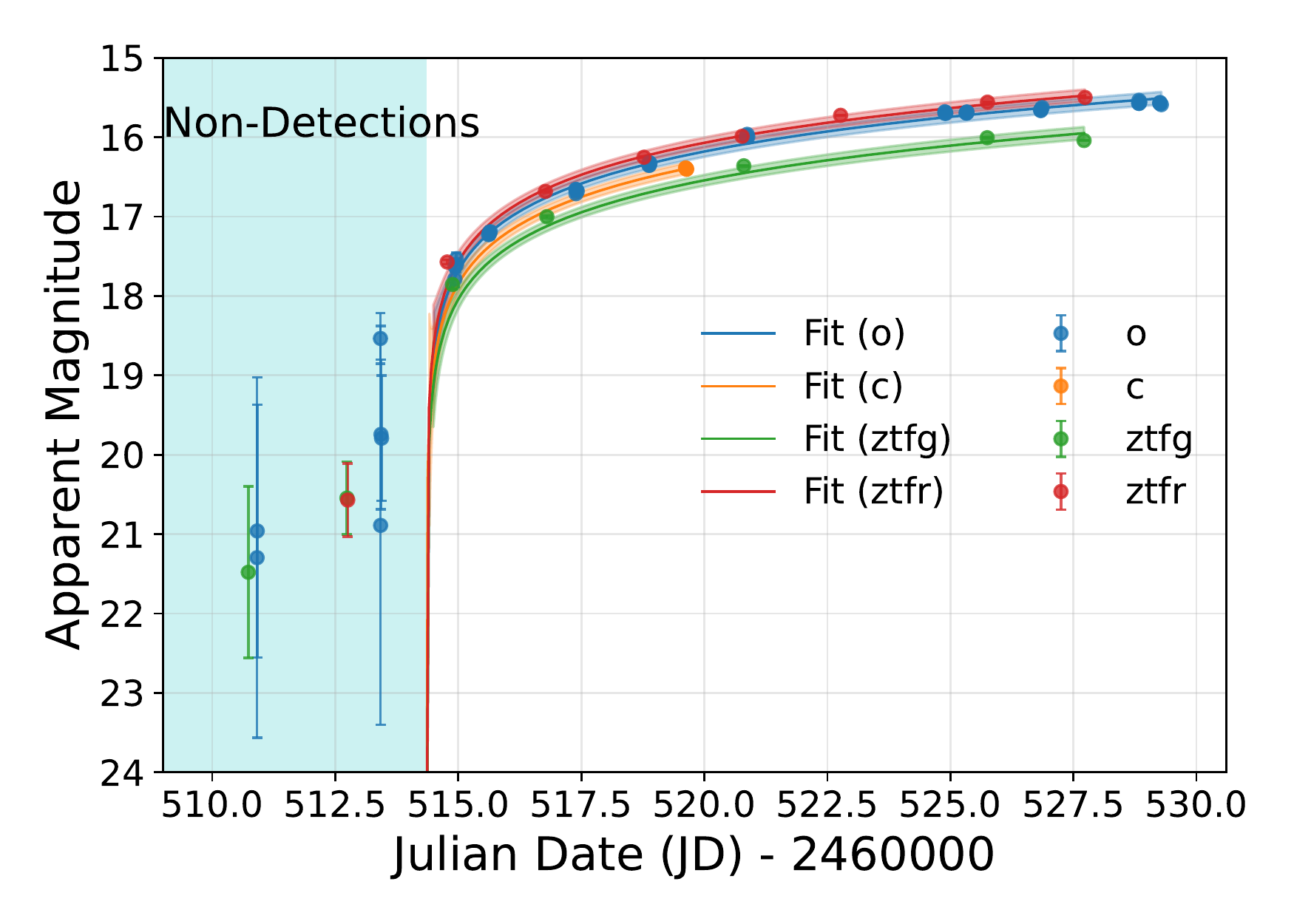}
	\end{center}
	\caption{A power-law fit to the early-time photometry of SN~2024pxl in ATLAS {\it cyan}, {\it orange}, and ZTF {\it g} and {\it r} filters. Nondetections are indicated in the shaded region.}
	\label{fig:SN_2024pxl_power_law_fit}
\end{figure}

The model was simultaneously fit to all the observed fluxes in different bands (\autoref{fig:SN_2024pxl_power_law_fit}). Unlike the commonly adopted fixed $n = 2$ assumption used in early light-curve fits to model a homologously expanding fireball (e.g., \citealt{1982ApJ...253..785A}; \citealt{2011Natur.480..344N}), we allowed $n$ to vary as a free parameter to account for deviations observed in some thermonuclear SN subclasses (e.g., \citealt{2016A&A...589A..89M}). The best-fit result yields a common explosion epoch of \[t_{\mathrm{exp}} = \mathrm{JD}\,2,460,514.36 \pm 0.10,\] with a power-law index of  \[n = 0.634 \pm 0.019,\] indicating a shallower rise than the typical $n = 2$ case. 
We emphasize that the quoted $\pm$0.10 day represents the formal statistical uncertainty from the power-law fit. In practice, the first detection occurs 0.96 days after this fitted zero point, underscoring that the true uncertainty is significantly larger, being influenced by systematic effects such as deviations from the assumed flux rise behavior, and possible undetected flux below the survey limits. Thus, while the formal error quantifies the fit precision, the absolute accuracy of the explosion epoch is less certain. This distinction does not affect the broader discussion in this work.

\begin{figure}[t]
	\begin{center}
		\includegraphics[width=0.9\columnwidth]{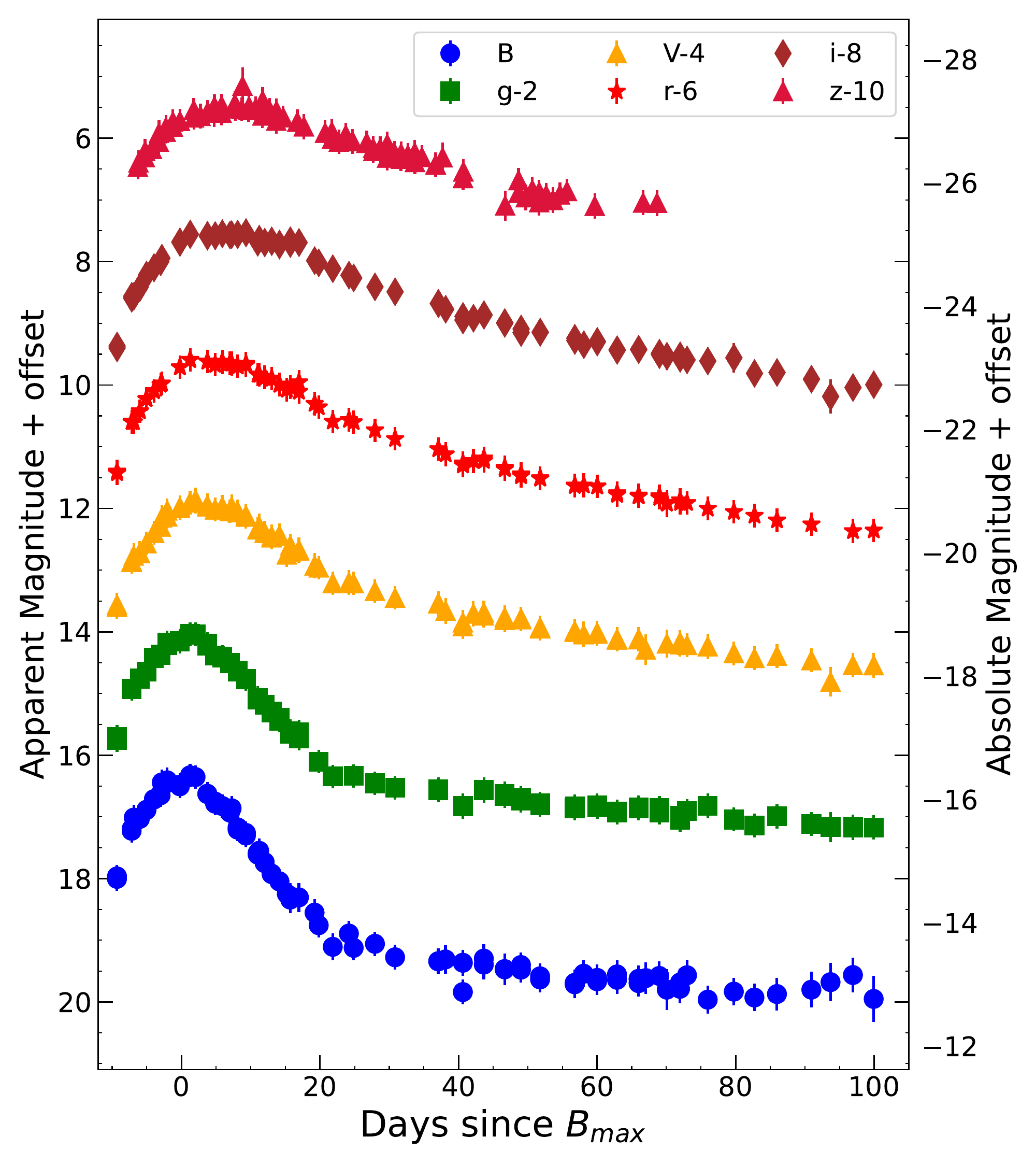}
	\end{center}
	\caption{Light-curve evolution of SN~2024pxl in {\it uBgVrRiIz} and clear bands. Corresponding absolute magnitudes corrected for extinction are also presented on the right-hand ordinate axis.}
	\label{fig:SN_2024pxl_light_curve}
\end{figure}

\begin{figure}[t]
	\begin{center}
		\includegraphics[width=\columnwidth]{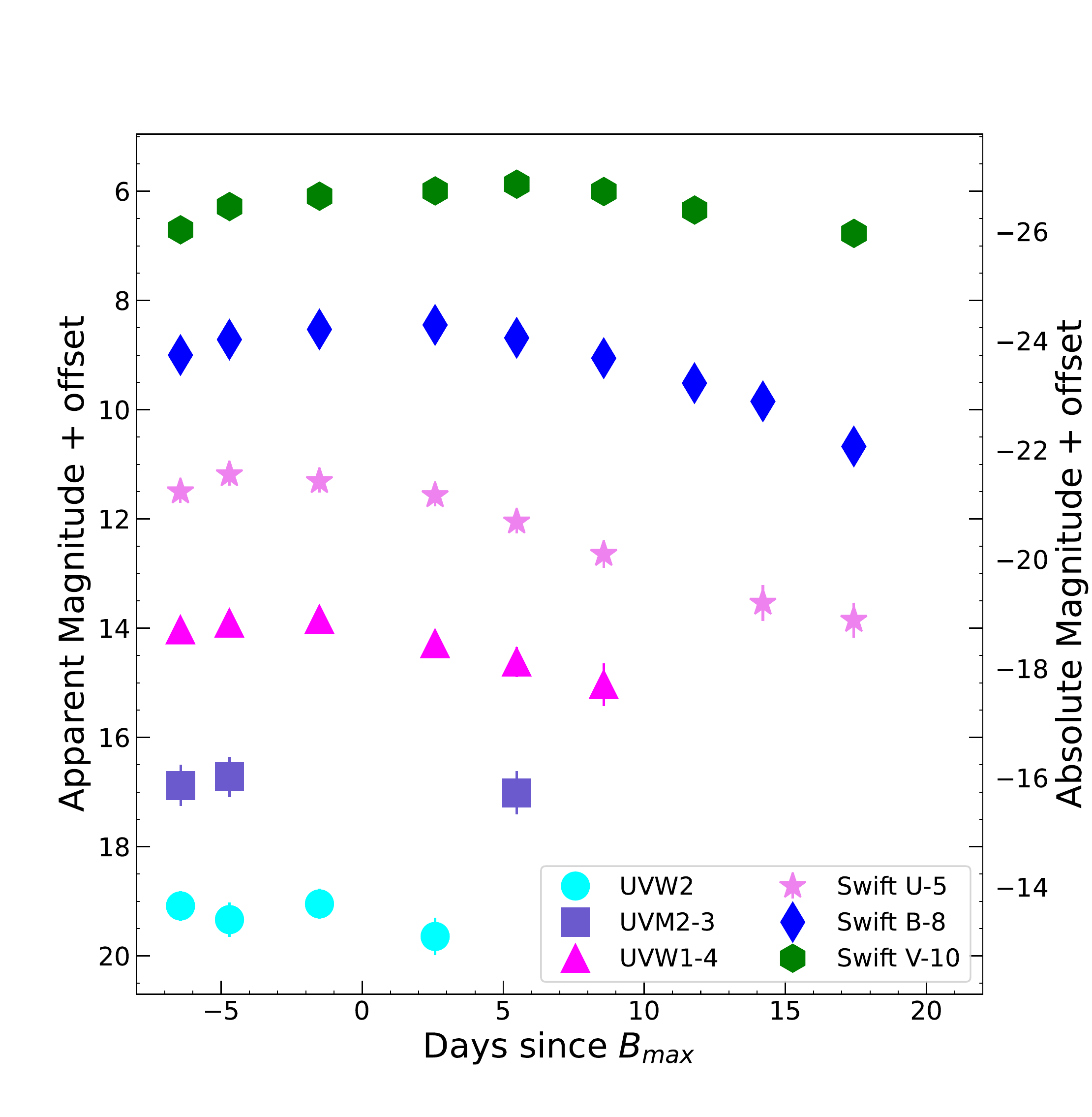}
	\end{center}
	\caption{Light-curve evolution of SN~2024pxl in the {\it Swift} bands.}
	\label{fig:SN_2024pxl_swift_light_curve}
\end{figure}

\begin{figure}
	\begin{center}
		\includegraphics[width=\columnwidth]{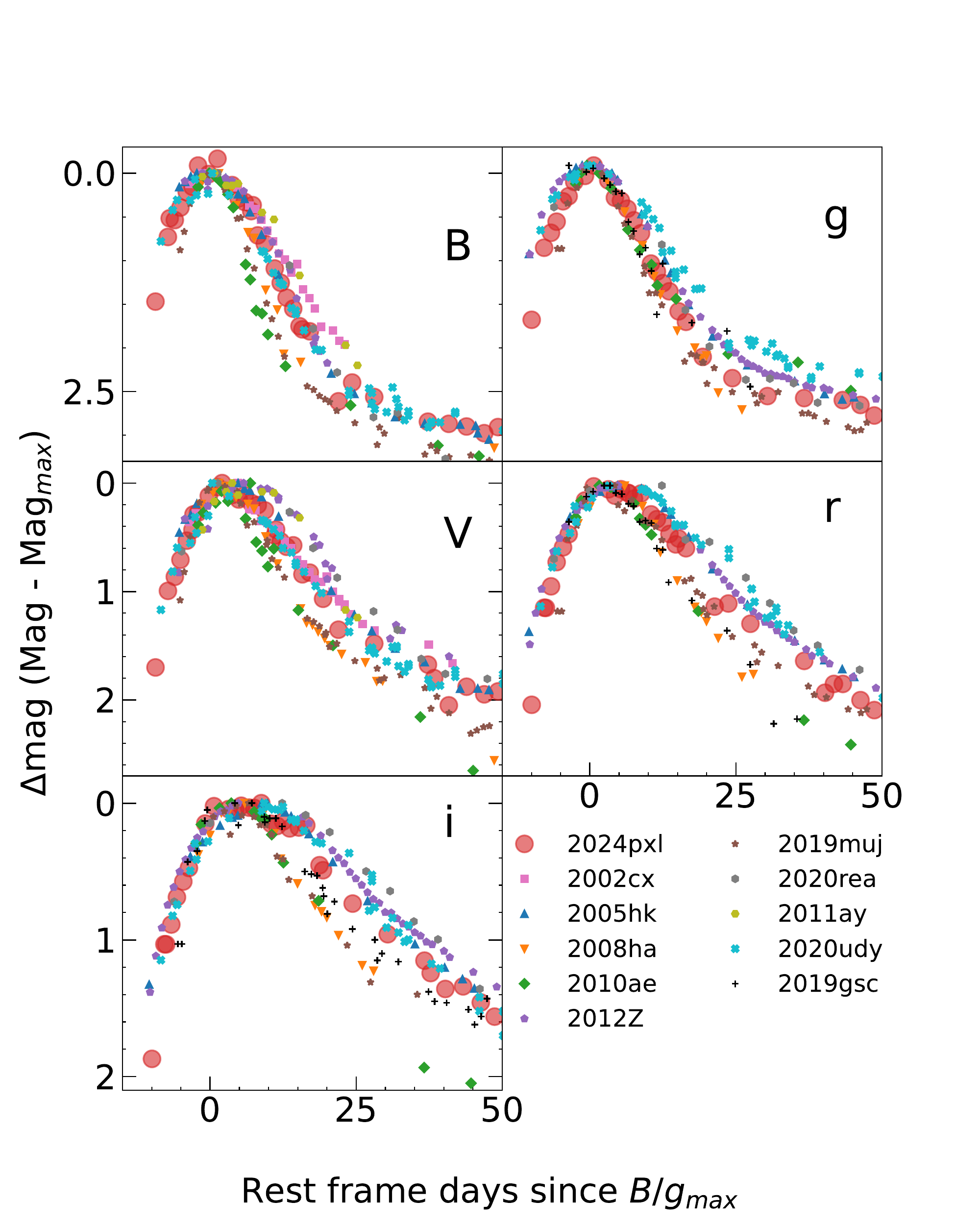}
	\end{center}
	\caption{Comparison of the light-curve evolution of SN~2024pxl in {\it BgVri} bands with other well-studied Type Iax SNe. The {\it B}-band maximum date serves as the reference for comparison plots in both {\it B} and {\it V} bands, while {\it gri}-band comparison plots are referenced to {\it g}-band maximum brightness.}
	\label{fig:SN_2024pxl_comp_light_curve}
\end{figure}

\begin{figure}[t]
	\begin{center}
		\includegraphics[width=\columnwidth]{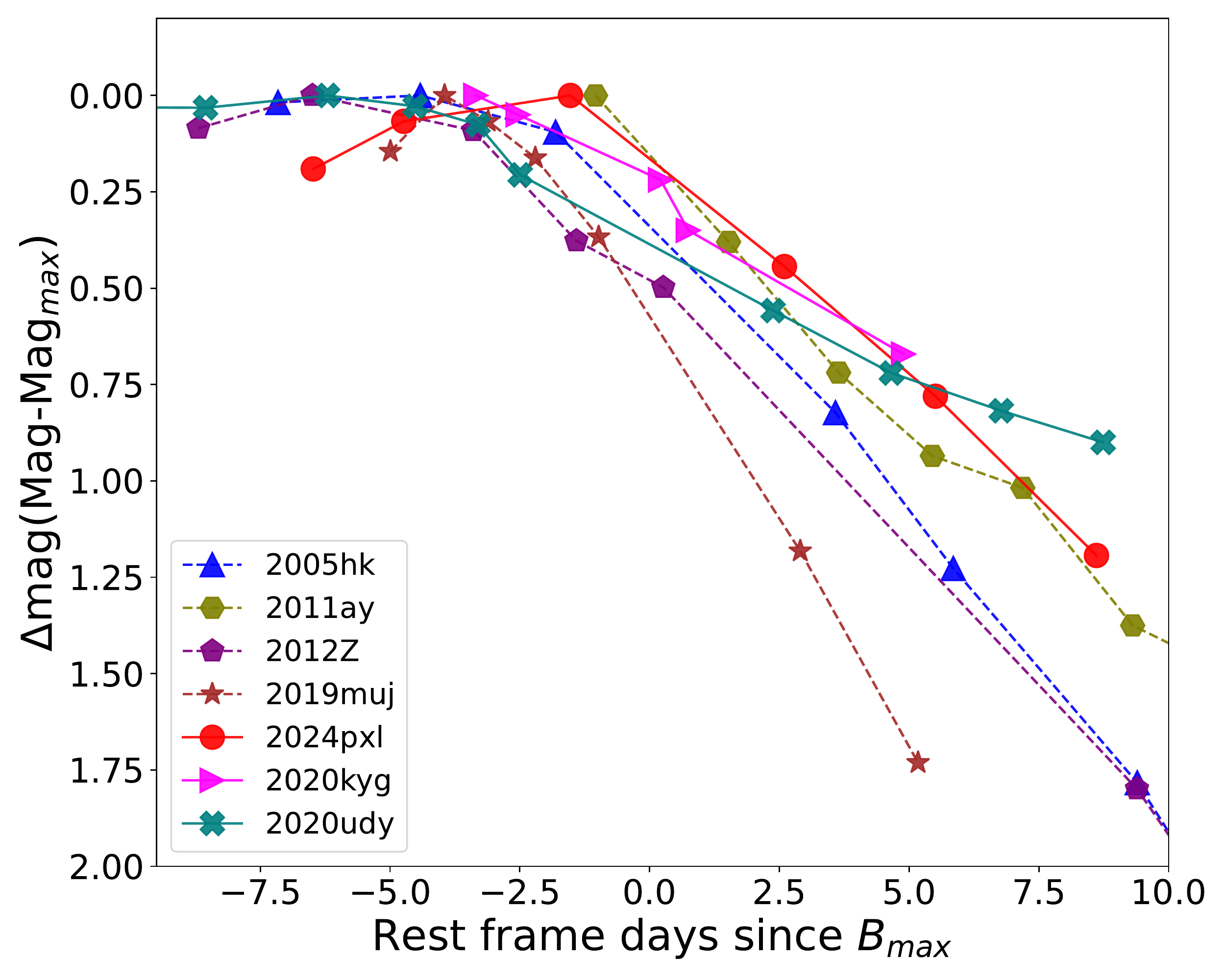}
	\end{center}
	\caption{Comparison of the light-curve evolution of SN~2024pxl in the {\it Swift UVW1} band with other Type Iax SNe.}
	\label{fig:SN_2024pxl_comp_swift_light_curve}
\end{figure}

\section{Light-Curve Properties}
\label{light_curve_properties}

\begin{figure}
	\begin{center}
		\includegraphics[width=\columnwidth]{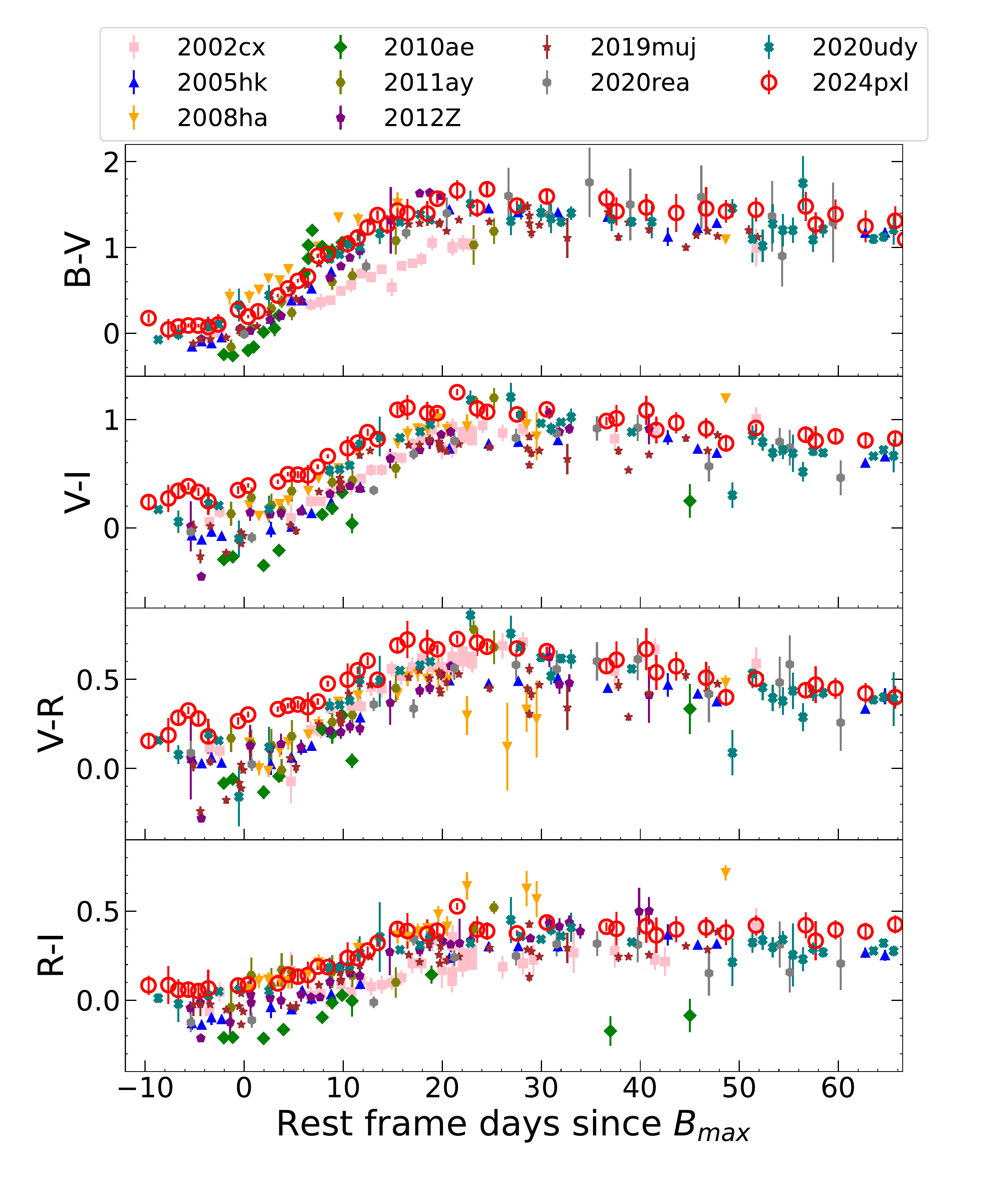}
	\end{center}
	\caption{Comparison of the color (mag) evolution  of SN~2024pxl with other well-studied Type Iax SNe.}
	\label{fig:SN_2024pxl_color_curve}
\end{figure}

\subsection{Light Curves and Color Curves}
\label{light_curves_color_curves}

\autoref{fig:SN_2024pxl_light_curve} presents the evolution of the light curve for SN~2024pxl in the {\it uBgVrRiIz} and clear bands, with dense sampling around maximum brightness in all bands. We use a low-order polynomial fit to estimate the time and magnitude at maximum light in the {\it uBgVrRiIz} bands. Additionally, we estimate the decline in magnitude from the light-curve peak to 15 days after ($\Delta m_{15}$) for these bands. These light-curve parameters for SN~2024pxl are listed in \autoref{tab:lc_param_24pxl}. \autoref{fig:SN_2024pxl_swift_light_curve} displays the evolution of SN~2024pxl in the {\it Swift} UVOT bands. All bands shown in \autoref{fig:SN_2024pxl_swift_light_curve}, except for {\it UVW2} and {\it UVM2}, cover the peak well. We estimate the {\it Swift} UVOT light-curve data parameters in \autoref{tab:lc_param_24pxl_swift}. 

\begin{table*}
\caption{ Light-Curve Parameters of SN 2024pxl  }
\centering
\smallskip
\resizebox{\textwidth}{!}{%
\begin{tabular}{l  c c c c c c c c c}
\hline \hline
SN 2024pxl                          & $u$ band 		  & $B$ band  		 & $g$ band   		& $V$ band  	   & $r$ band 		  & $R$ band         & $i$ band        & $I$ band  		  & $z$ band \\
\hline
JD of maximum light (2,460,000+)    & 523.00$\pm$0.5  &524.6$\pm$0.5     & 525.2$\pm$0.5    & 527.4$\pm$0.5    & 528.14$\pm$0.5   & 529.4$\pm$0.5    & 530.2$\pm$0.5   & 531.1$\pm$0.5    & 531.7$\pm$0.5 \\
Magnitude at maximum (mag)          & 17.02$\pm$0.01  &16.42$\pm$0.02    & 16.10$\pm$0.01   & 15.90$\pm$0.01   & 15.56$\pm$0.01   & 15.44$\pm$0.01   & 15.49$\pm$0.01  & 15.12$\pm$0.02   & 15.47$\pm$0.07  \\
Absolute magnitude at maximum (mag) & $-15.87\pm$0.19 &$-16.30\pm$0.19   & $-16.60\pm$0.19  & $-16.81\pm$0.19  & $-17.13\pm$0.19  & $-17.45\pm$0.19  & $-17.22\pm$0.19 & $-17.77\pm$0.19  & $-17.24\pm$0.21  \\
$\Delta m_{15}$ (mag)               & 2.54$\pm$0.05   &1.80$\pm$0.07     & 1.56$\pm$0.04    & 0.87$\pm$0.06    & 0.49$\pm$0.05    & 0.66$\pm$0.03     & 0.51$\pm$0.02   & 0.40$\pm$0.04    & 0.46$\pm$0.08	\\ 
\hline

\end{tabular}}
\newline
\label{tab:lc_param_24pxl}      
\end{table*}

\begin{table*}
\caption{ Light-Curve Parameters of SN 2024pxl for Swift Data  }
\centering
\smallskip
\begin{tabular}{l  c c c c c c}
\hline \hline
SN 2024pxl                          & UVW2           & UVM2           & UVW1           & Swift $U$   & Swift $B$  & Swift $V$ \\
\hline
JD of maximum light (2,460,000+)      & --   & --   & 521.5$\pm$0.5    & 521.9$\pm$0.5      &525.3$\pm$0.5  & 527.8$\pm$0.5   \\
Magnitude at maximum (mag)          & --   & --   & 17.83$\pm$0.14    & 16.21$\pm$0.08      & 16.46$\pm$0.06 &  15.90$\pm$0.07  \\
Absolute magnitude at maximum (mag) & --   & --   & $-14.91\pm$0.24  & $-16.53\pm$0.20   & $-16.28\pm$0.20     & $-16.84\pm$0.20    \\
$\Delta m_{15}$ (mag)              & --    & --    & 1.46$\pm$0.61     & 2.17$\pm$0.32       & 1.78$\pm$0.15 & 0.95$\pm$0.12 \\ 
\hline

\end{tabular}
\newline
\label{tab:lc_param_24pxl_swift}      
\end{table*}

To further explore the nature of the light-curve evolution of SN~2024pxl, we compare its {\it BgVri} light curves with those of several well-studied Type Iax SNe (\autoref{fig:SN_2024pxl_comp_light_curve}). For this, we choose high-luminosity Type Iax SNe~2002cx \citep{2003PASP..115..453L}, 2005hk \citep{2007PASP..119..360P,2008ApJ...680..580S}, 2011ay \citep{2015MNRAS.453.2103S}, 2012Z \citep{2015A&A...573A...2S}, 2020rea \citep{2022MNRAS.517.5617S}, and 2020udy \citep{2024ApJ...965...73S}; intermediate-luminosity Type Iax SN~2019muj \citep{2021MNRAS.501.1078B}; and low-luminosity Type Iax SNe~2008ha \citep{2009AJ....138..376F}, 2010ae \citep{2014A&A...561A.146S}, and 2019gsc \citep{2020ApJ...892L..24S, 2020MNRAS.496.1132T}. These comparison SNe cover both ends of the luminosity distribution for Type Iax SNe. The rest-frame magnitudes for each SN displayed in \autoref{fig:SN_2024pxl_comp_light_curve} are normalized to the peak in the respective bands. 

In the optical bands, SN~2024pxl consistently exhibits decline rates that are faster than those of the high-luminosity Type Iax SNe and slower than those of the low-luminosity events, placing it between these two groups in terms of photometric evolution. Its behavior is broadly similar to that of the intermediate-luminosity SN~2019muj, although SN~2024pxl evolves somewhat more slowly in several filters. This suggests that SN~2024pxl lies between SN~2019muj and high-luminosity Type Iax SNe in optical decline rates.

In the UV, SN2024pxl displays a slow decline relative to other Type Iax SNe, with a rate comparable to that of a normal Type Ia SN \citep{2010ApJ...721.1627M}. This distinguishes its UV behavior from the luminosity-dependent trend observed in the optical bands. A comparison with {\it Swift} {\it UVW1} light curves of SNe~2005hk, 2011ay, 2012Z, 2019muj, 2020kyg, and 2020udy is presented in \autoref{fig:SN_2024pxl_comp_swift_light_curve}.

\autoref{fig:SN_2024pxl_color_curve} illustrates the evolution of SN~2024pxl’s colors ({\it B-V}, {\it V-I}, {\it V-R}, and {\it R-I}) in comparison with other Type Iax SNe. All colors have been adjusted for total reddening. The color evolution of SN~2024pxl closely resembles that observed in other Type Iax SNe selected for comparison.

\begin{figure}
	\begin{center}
		\includegraphics[width=\columnwidth]{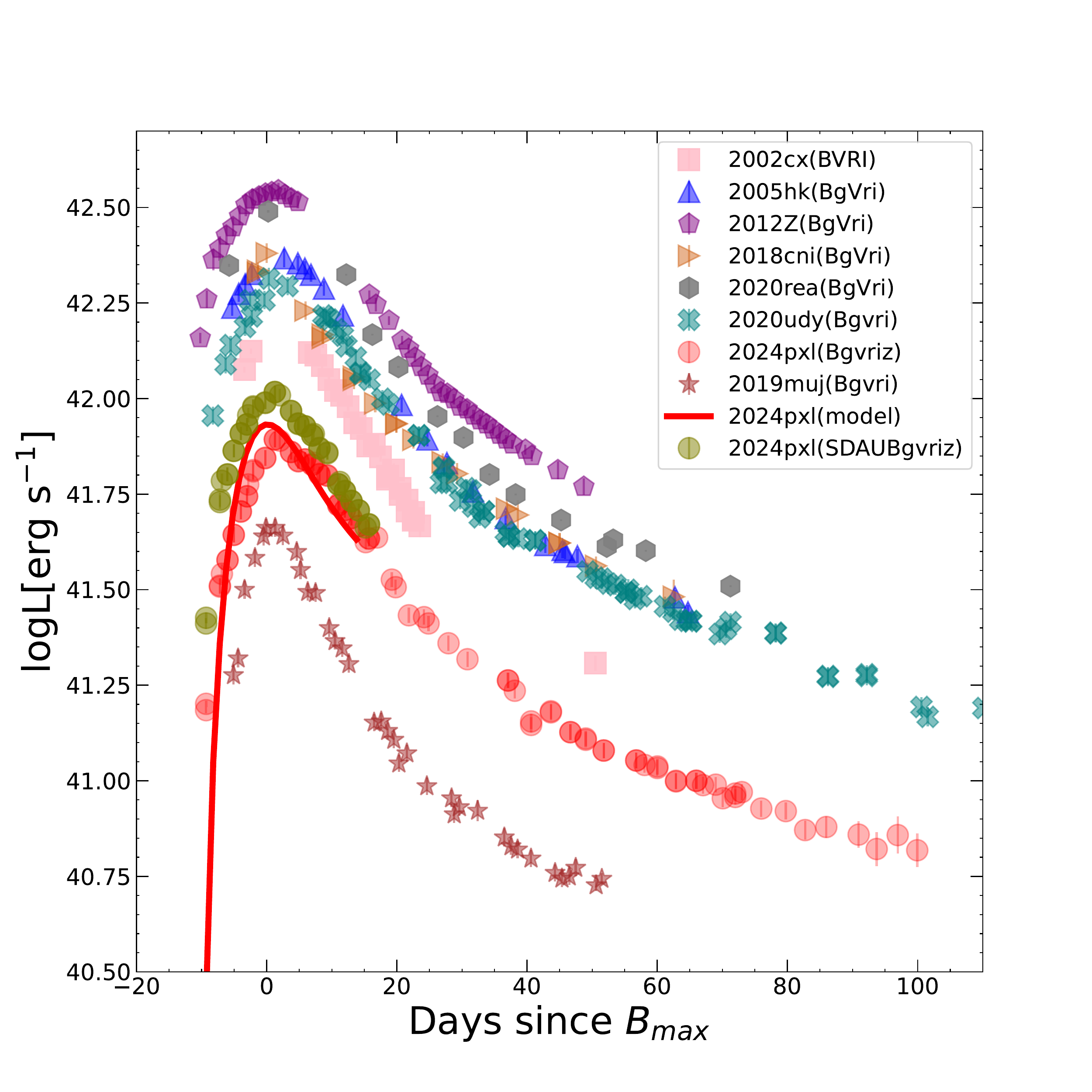}
	\end{center}
	\caption{Comparison of {\it BgVriz} integrated SED light curve of SN~2024pxl (red circles) with other well-studied Type Iax SNe. The solid red line presents the analytical model fit to the integrated SED light curve of SN~2024pxl. We also plot the integrated SED light curve of SN~2024pxl using data in UV bands (olive circles), which lies slightly above the {\it BgVriz} integrated SED light curve.}
	\label{fig:SN_2024pxl_bolometric_light_curve}
\end{figure}

\begin{figure}
	\begin{center}
		\includegraphics[width=\columnwidth]{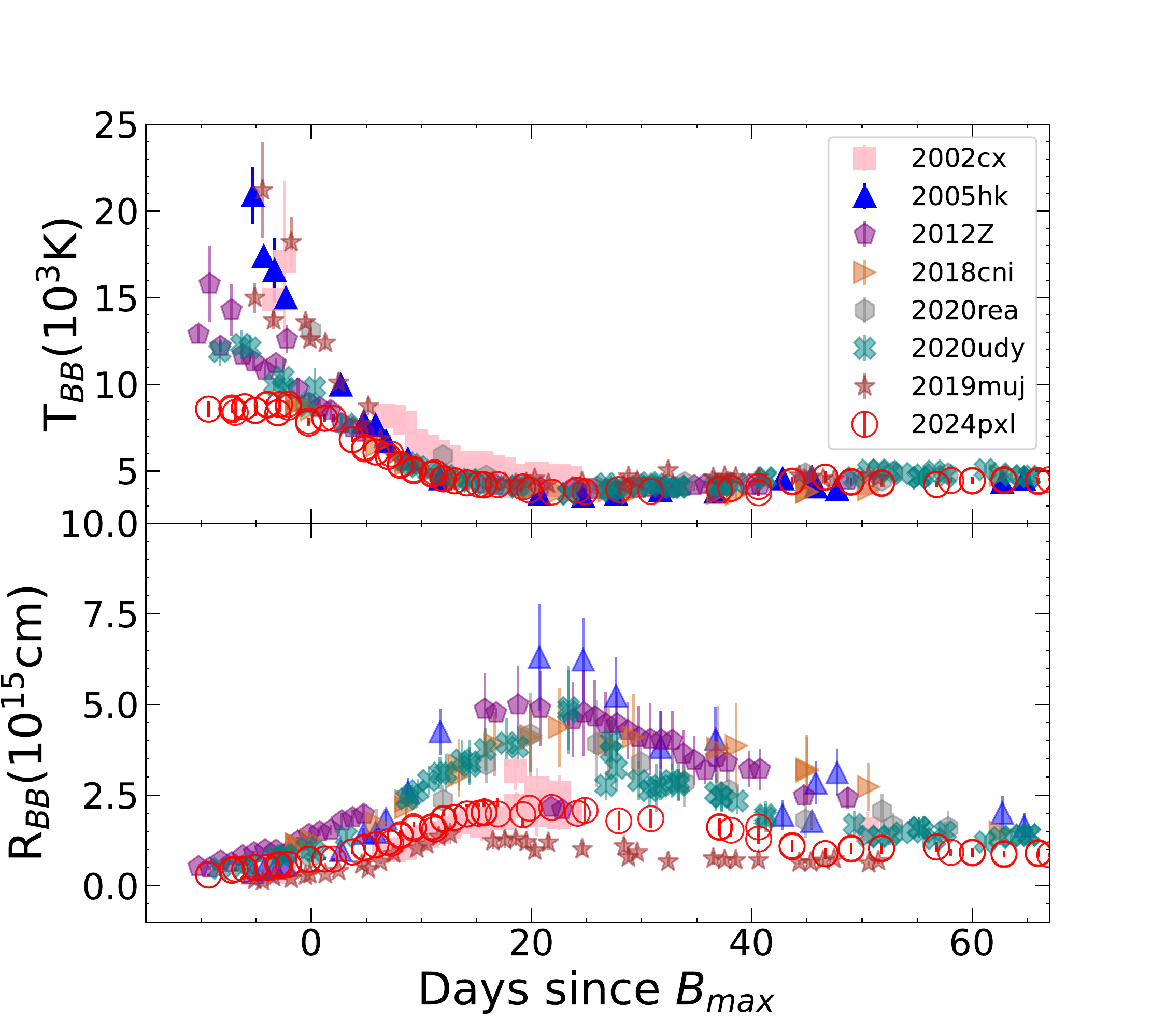}
	\end{center}
	\caption{Evolution of the blackbody temperature and radius of SN~2024pxl (red circles), along with other Type Iax SNe used for comparison.}
	\label{fig:SN_2024pxl_BB_params}
\end{figure}

\subsection{Bolometric Light Curve}
\label{bolometric_light_curve}

The integrated SED light curve of SN~2024pxl is generated from the {\it BgVriz} photometry using \texttt{SuperBol} \citep{2018RNAAS...2..230N}, incorporating the distance and extinction values outlined in \autoref{distance_extinction_explosion_epoch}. \texttt{SuperBol} takes the dereddened fluxes and uncertainties across passbands to construct an SED at each epoch. Luminosities are determined by integrating this spectrum while propagating uncertainties. To account for missing passbands in the UV and NIR, \texttt{SuperBol} fits a blackbody to each SED to derive a correction which is then applied to the integrated SED luminosities.

The integrated SED light curve of the other comparison SNe presented in \autoref{fig:SN_2024pxl_bolometric_light_curve} are constructed using a similar approach. SN 2018cni \citep{2023ApJ...953...93S}, another high-luminosity SN\,Iax, is also included for comparison in Figure \ref{fig:SN_2024pxl_bolometric_light_curve}. The peak integrated SED {\it BgVriz} luminosity of SN~2024pxl is $(7.82 \pm 0.36) \times 10^{41}$\,erg\,s$^{-1}$. SN~2024pxl is less luminous than all the high-luminosity Type Iax SNe, such as SNe~2012Z, 2020rea, and 2020udy, but is brighter than the intermediate-luminosity Type Iax SN~2019muj. SN~2024pxl lies in the gap between the high-luminosity and low-luminosity Type Iax SNe, in agreement with our decline-rate comparison  (Section \ref{light_curves_color_curves}).   

We conduct analytical modeling of the integrated SED light curve for SN~2024pxl using the methods outlined by \cite{1982ApJ...253..785A} and \cite{2008MNRAS.383.1485V}. The basic assumptions of this model consist of a small initial radius, constant optical opacity, spherically symmetric and optically thick ejecta, and the inclusion of $^{56}$Ni in the ejected material. By fitting the integrated SED light curve of SN~2024pxl, we estimate a $^{56}$Ni mass of $0.027 \pm 0.002$\, M$_{\odot}$, an ejecta mass of $0.36 \pm 0.05$\, M$_{\odot}$, and a rise time of 11 days, with uncertainties derived from the covariance matrix. We note that this small formal error represents only the statistical uncertainty from the fit and does not capture systematic effects arising from model assumptions. The true uncertainty is expected to be larger, and the value should be regarded as a model-dependent estimate.

For the fit, we assume a constant opacity of $\kappa_{\rm opt} = 0.1$\,cm$^{2}$\,g$^{-1}$ and a photospheric velocity of 5000\,km\,s$^{-1}$ at maximum brightness. Directly estimating the photospheric expansion velocity from the bottom panel of \autoref{fig:SN_2024pxl_BB_params}, we calculate $dR_{\mathrm{BB}}/dt \approx 10000$\,km\,s$^{-1}$ over the nearly linear expansion period from maximum light to $+$15 days. It is surprising that this value is larger than the typical photospheric absorption-line velocities, perhaps suggesting that the optical depth approaches unity farther out in the ejecta than the lines would predict.

We also compared the evolution of the blackbody temperature and radius of SN~2024pxl with those of other Type Iax SNe (\autoref{fig:SN_2024pxl_BB_params}). The blackbody temperature of SN~2024pxl is lower than that of the different comparison SNe before maximum light. After that, it remains at the lower end of the blackbody temperature distribution. It is possible that this could be the effect of underestimated extinction. The evolution of the blackbody radius of SN~2024pxl is proportional to the luminosity of the SN, which is positioned in an intermediate location in \autoref{fig:SN_2024pxl_BB_params}.  SN~2024pxl's blackbody temperature and radius evolution align well with the characteristic patterns seen in other Type Iax SNe.

For Type Iax SNe, contributions from the unobserved UV and IR passbands remain poorly constrained. However, based on available data for a few Type Iax SNe, the combined UV and NIR contributions to the total bolometric flux have been estimated. \cite{2007PASP..119..360P} demonstrated that the UV flux accounts for approximately 20\% of the UV-through-IR (UVOIR) light curve for an early photospheric phase in the case of SN~2005hk. Similarly, \citet{2015ApJ...806..191Y} estimated that the NIR band makes up about 20\% of the optical+NIR flux around peak light for SN 2012Z. \cite{2016MNRAS.459.1018T} found that the UV+IR flux collectively accounted for $\sim 35$\% of the total emission in SN~2014ck. Moreover, \citet{2020ApJ...892L..24S} and \citet{2022ApJ...925..217D} calculated the ratio of $L_{\rm peak,pseudo}$ (peak integrated SED luminosity) to $L_{\rm peak,bb}$ (peak blackbody luminosity) as 0.69 for SN~2019gsc and 0.62 for SN~2020sck, respectively. In the case of SN~2020kyg, \citet{2022MNRAS.511.2708S} reported that about 60\% of the total bolometric luminosity near maximum light is attributed to the optical emission. 

We have {\it Swift} observations that capture the light-curve evolution of SN~2024pxl around the peak. Using \texttt{SuperBol} for $UVW2$, $UVM2$, $UVW1$, and $U$ filters along with {\it BgVriz} filters, we find that the ratio of peak integrated SED luminosity using {\it BgVriz} and {\it SDAUBgVriz} bands ($S = UVW2$, $D = UVM2$, $A = UVW1$, $Swift U = U$) is $\sim 0.75$, suggesting that the UV may be contributing up to $\sim 25$\% of the bolometric luminosity near peak. Using the same analytical model for the {\it SDAUBgVriz} integrated SED light curve, we find that $\sim 0.03 \pm 0.01$\,M$_{\odot}$ of $^{56}$Ni is synthesized during the explosion of SN~2024pxl. While this result is consistent with our earlier optical-only $^{56}$Ni mass estimate (0.026 $\pm$ 0.002 M$_{\odot}$), the quoted uncertainties represent formal errors from the fitting procedure and do not fully capture systematic effects associated with the simplifying assumptions of the analytic model (e.g., constant opacity, spherical symmetry).

\cite{2014MNRAS.438.1762F} presented numerical calculations for deflagration models of CO WDs with varying explosion energy (parametrized by the number of ignition kernels). These models span a broad range of explosion parameters ($^{56}$Ni mass 0.03--0.38\,M$_{\odot}$, rise time 7.6--14.4 days). The estimated mass of $^{56}$Ni (0.03\,M$_{\odot}$) and rise time (11 days) for SN~2024pxl are consistent with the range provided by these models. However, the peak magnitude in the {\it V} band for SN~2024pxl ($M_{V} = -16.81 \pm 0.19$\,mag) is on the fainter end of the estimation ($-$16.84 to $-$18.96~mag) given by \cite{2014MNRAS.438.1762F}. The mass of $^{56}$Ni in the case of SN~2024pxl matches with the $^{56}$Ni synthesized in the weakest-energy N1def (single ignition point) model presented by \cite{2014MNRAS.438.1762F}. Radiative-transfer modeling based on the N1def model by \cite{2025ApJ...989L..33K} shows good agreement with the spectra of SN~2024pxl from the optical through the MIR. 

For low-luminosity Type Iax SNe, \cite{2015MNRAS.450.3045K} studied the deflagration of a hybrid CONe WD. Our estimated peak magnitude of SN~2024pxl in the {\it B} band ($-16.30 \pm 0.19$~mag) and mass of $^{56}$Ni (0.03\,M$_{\odot}$) exceed the predictions from the hybrid CONe WD model ($B_\mathrm{max} = -13.2$ to $-14.6$\,mag, mass of $^{56}$Ni $= 3.4 \times 10^{-3}$\,M$_{\odot}$) that may apply to lower-luminosity objects. These results also favor the interpretation that SN~2024pxl is an interesting luminosity link, with explosion parameters higher than those predicted for low-luminosity objects but lying at the lower end of the models proposed to explain the high-luminosity members of the class.

\begin{figure*}
	\begin{center}
        \includegraphics[width=0.8\textwidth]{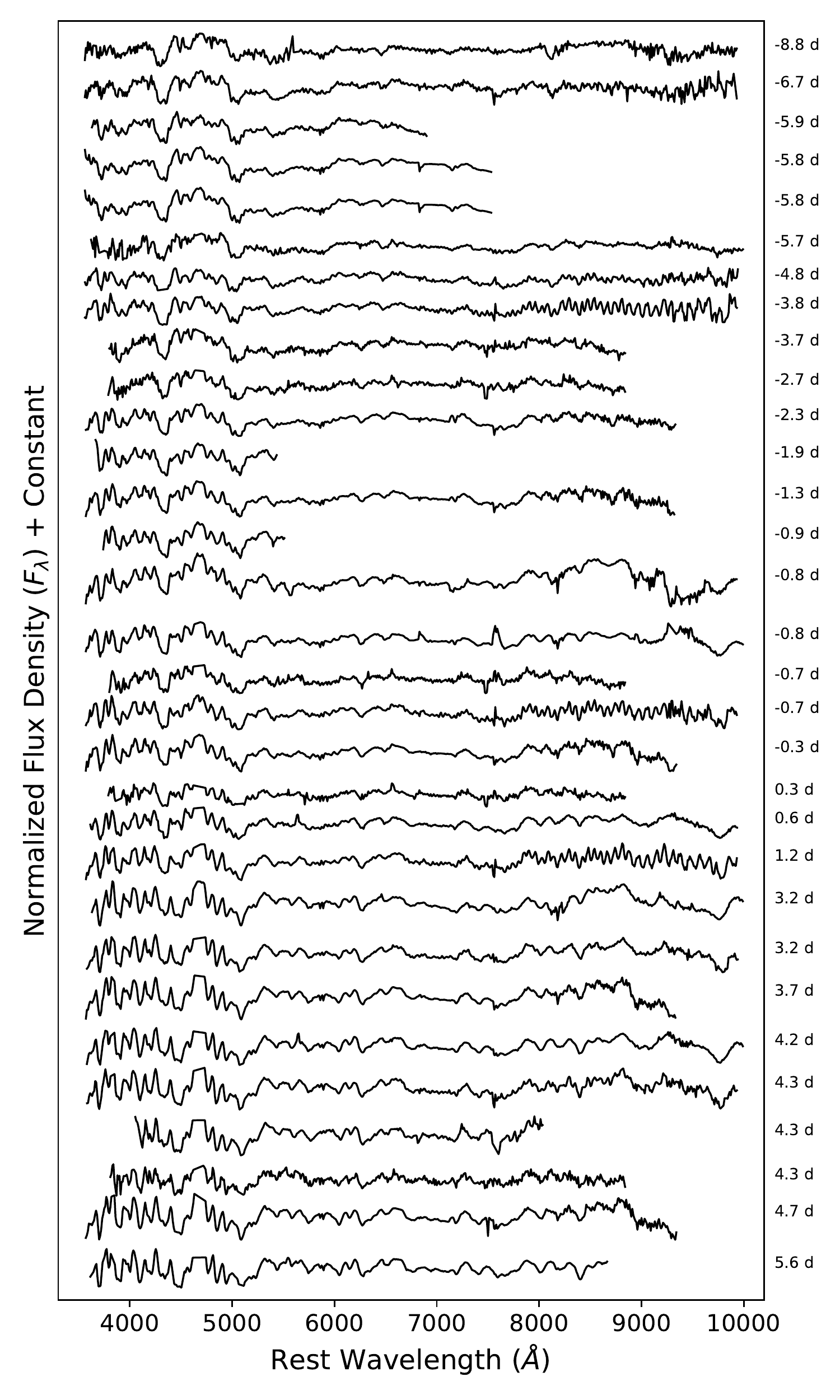}
	\end{center}
    \vspace{-20pt}
	\caption{Spectral evolution of SN~2024pxl spanning from $-8.8$ to 5.6 days after \textit{B}-band maximum brightness.}
	\label{fig:spectral_sequence_1_24pxl}
\end{figure*}

\begin{figure*}
	\begin{center}
        \includegraphics[width=0.8\textwidth]{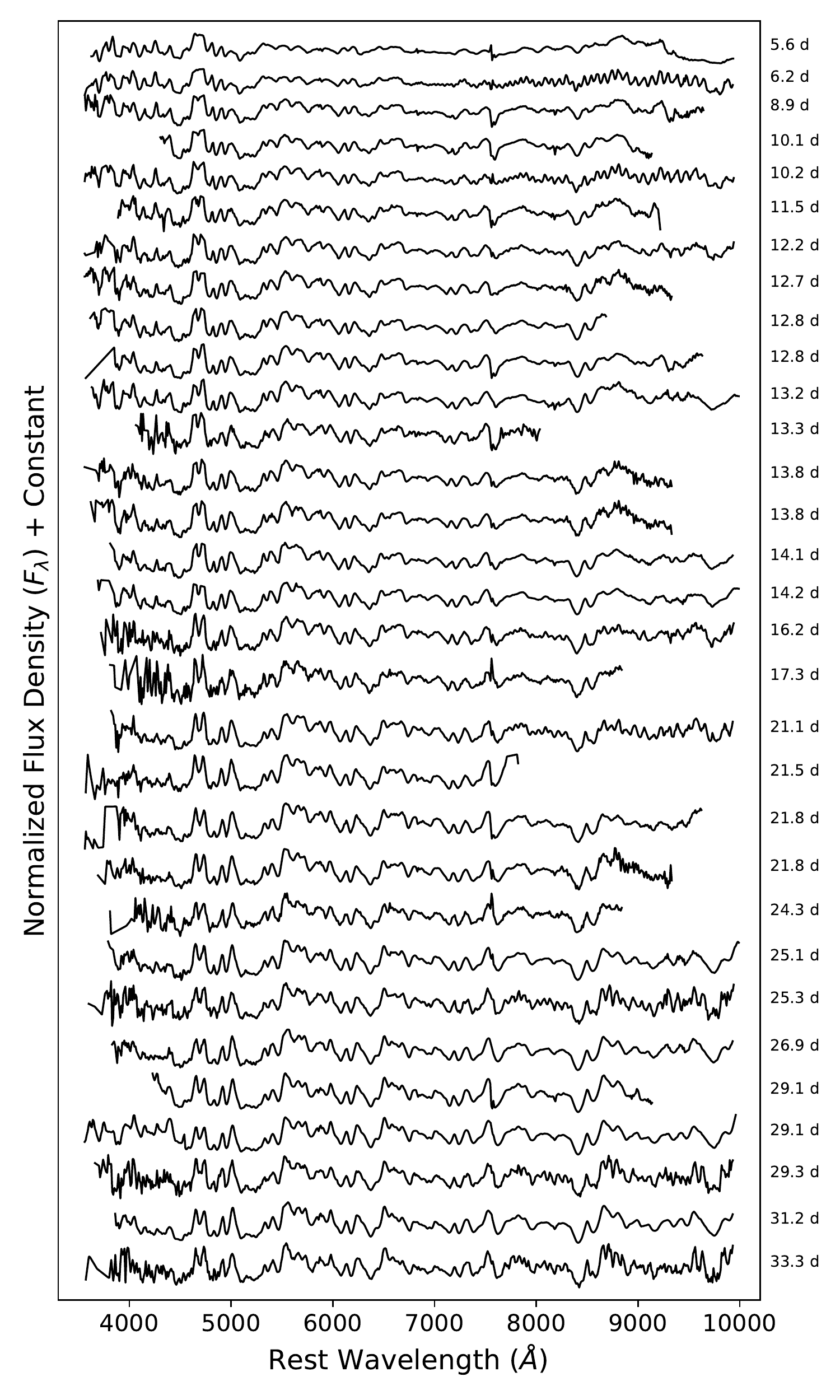}
	\end{center}
    \vspace{-20pt}
	\caption{Spectral evolution of SN~2024pxl spanning from 6.2 to 34.3 days after \textit{B}-band maximum brightness.}
	\label{fig:spectral_sequence_2_24pxl}
\end{figure*}

\begin{figure*}
	\begin{center}
        \includegraphics[width=0.8\textwidth]{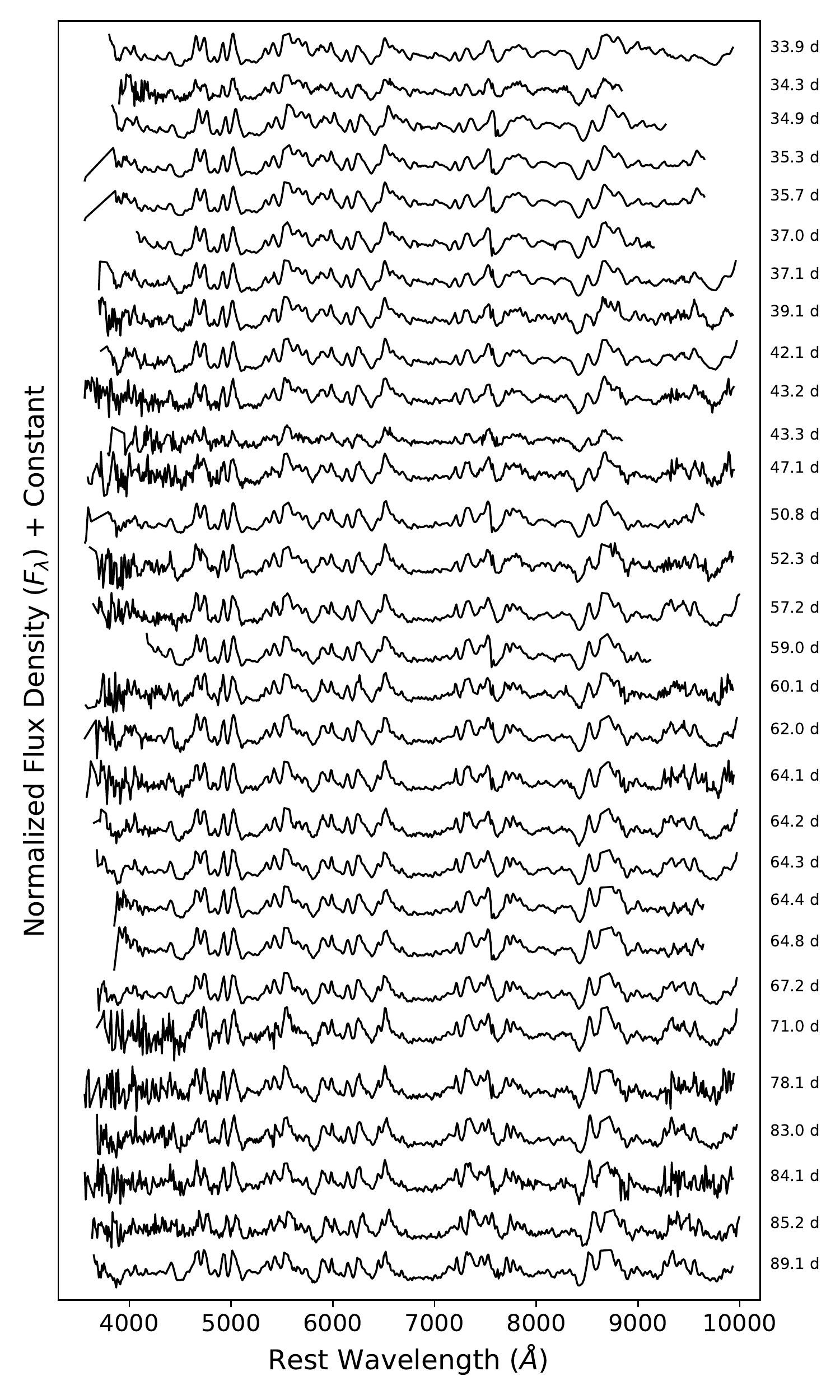}
	\end{center}
    \vspace{-20pt}
	\caption{Spectral evolution of SN~2024pxl spanning from 34.4 to 89.1 days after \textit{B}-band maximum brightness.}
	\label{fig:spectral_sequence_3_24pxl}
\end{figure*}

\section{Spectral Evolution}
\label{spectral_evolution}

\subsection{Spectral Features and Comparisons}
\label{spectral_features_comparison}

Figures \ref{fig:spectral_sequence_1_24pxl}, \ref{fig:spectral_sequence_2_24pxl}, and \ref{fig:spectral_sequence_3_24pxl} present the spectral evolution (\autoref{tab:spectroscopic_observations_24pxl}) of SN~2024pxl from approximately $-$9 to $+$89~days relative to \textit{B}$_\mathrm{max}$. Spectra were obtained nearly daily for almost two months after maximum, producing a rich data set allowing detailed tracking of spectral evolution. We compare the spectral features of SN~2024pxl with those of other well-studied Type Iax SNe at similar pre-maximum, near-maximum, post-maximum, and late phases in Figures \ref{fig:pre_peak_comp_24pxl}, 
\ref{fig:peak_comp_24pxl},
\ref{fig:post_peak_comp_24pxl}, and 
\ref{fig:nebular_comp_24pxl}, respectively.

\subsubsection{Pre-maximum spectra}
\label{pre_maximum_spectra}

\begin{figure}
	\begin{center}
		\includegraphics[width=\columnwidth]{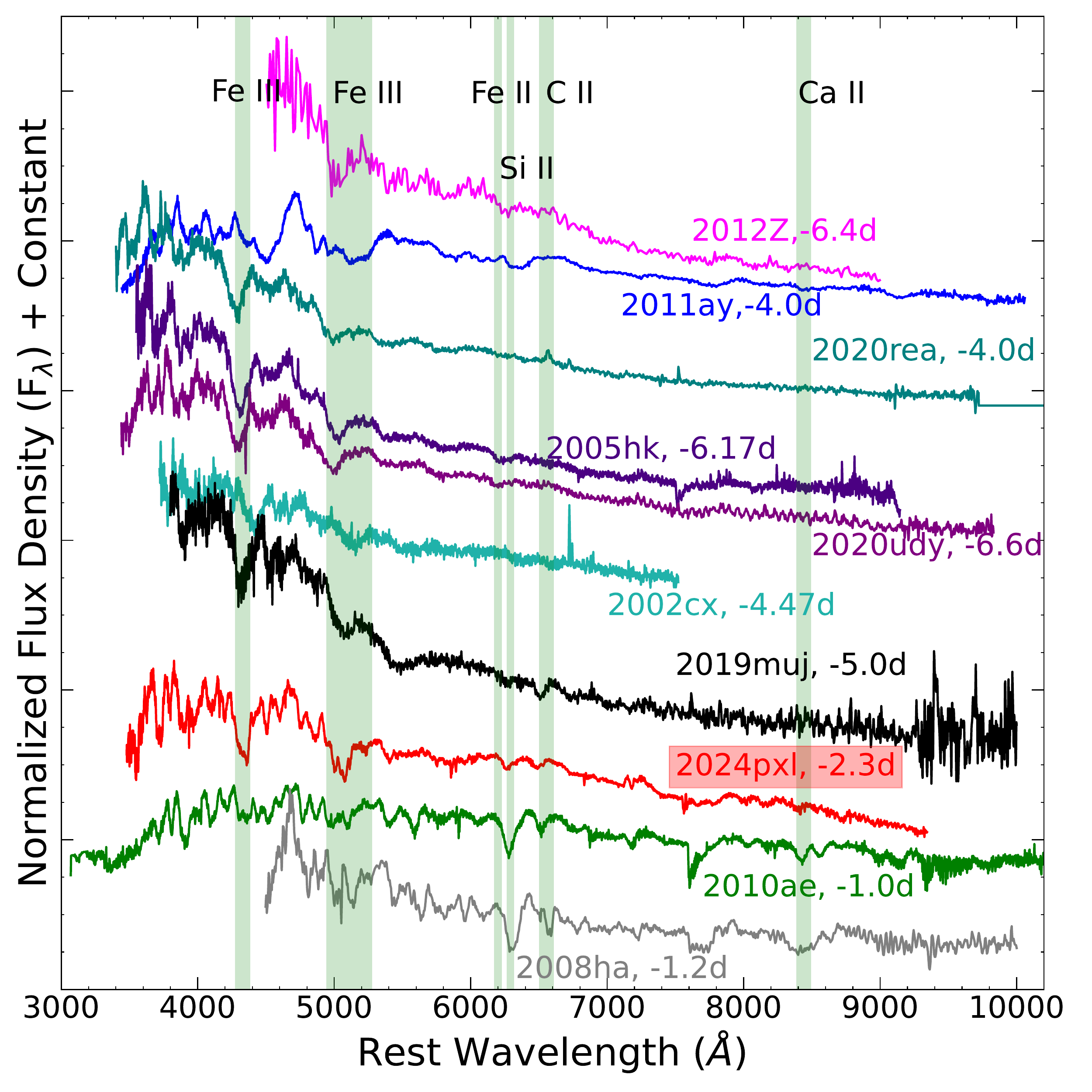}
	\end{center}
	\caption{Pre-peak spectral comparison of SN~2024pxl (red) with other Type Iax SNe at similar epochs.}
	\label{fig:pre_peak_comp_24pxl}
\end{figure}

We observe that at early, prepeak times, SN~2024pxl have a flat continuum consistent with the low temperatures we infer from the photometry (see \autoref{fig:SN_2024pxl_BB_params}). The early-time spectra of SN~2024pxl display features such as \ion{Fe}{3}, \ion{Fe}{2}, and \ion{Si}{2} (\autoref{fig:pre_peak_comp_24pxl}). In agreement with our findings that SN~2024pxl is a photometrically intermediate Type Iax SN, it is spectroscopically intermediate as well, sharing similarities with both high-luminosity Type Iax SNe (e.g., SNe~2012Z, 2005hk, and 2020udy) and low-luminosity objects (e.g., SNe~2010ae and 2008ha). Specifically, SN~2024pxl has weak \ion{Si}{2} features but also exhibits a \ion{C}{2} line at 6580~\AA. In high-luminosity Type Iax SNe, the \ion{Si}{2} line is relatively weak, and \ion{C}{2} is generally not detected. In contrast, in low-luminosity Type Iax SNe, these features are much more prominent. The \ion{Ca}{2}  NIR triplet is present in SN~2024pxl but is weaker than in SNe~2010ae and 2008ha. This combination of weak \ion{Si}{2} and presence of \ion{C}{2} suggests that SN~2024pxl also serves as a spectroscopic transitional link between high-luminosity and low-luminosity Type Iax SNe.

\subsubsection{Near-maximum spectra}
\label{near_maximum_spectra}

\begin{figure}
	\begin{center}
		\includegraphics[width=\columnwidth]{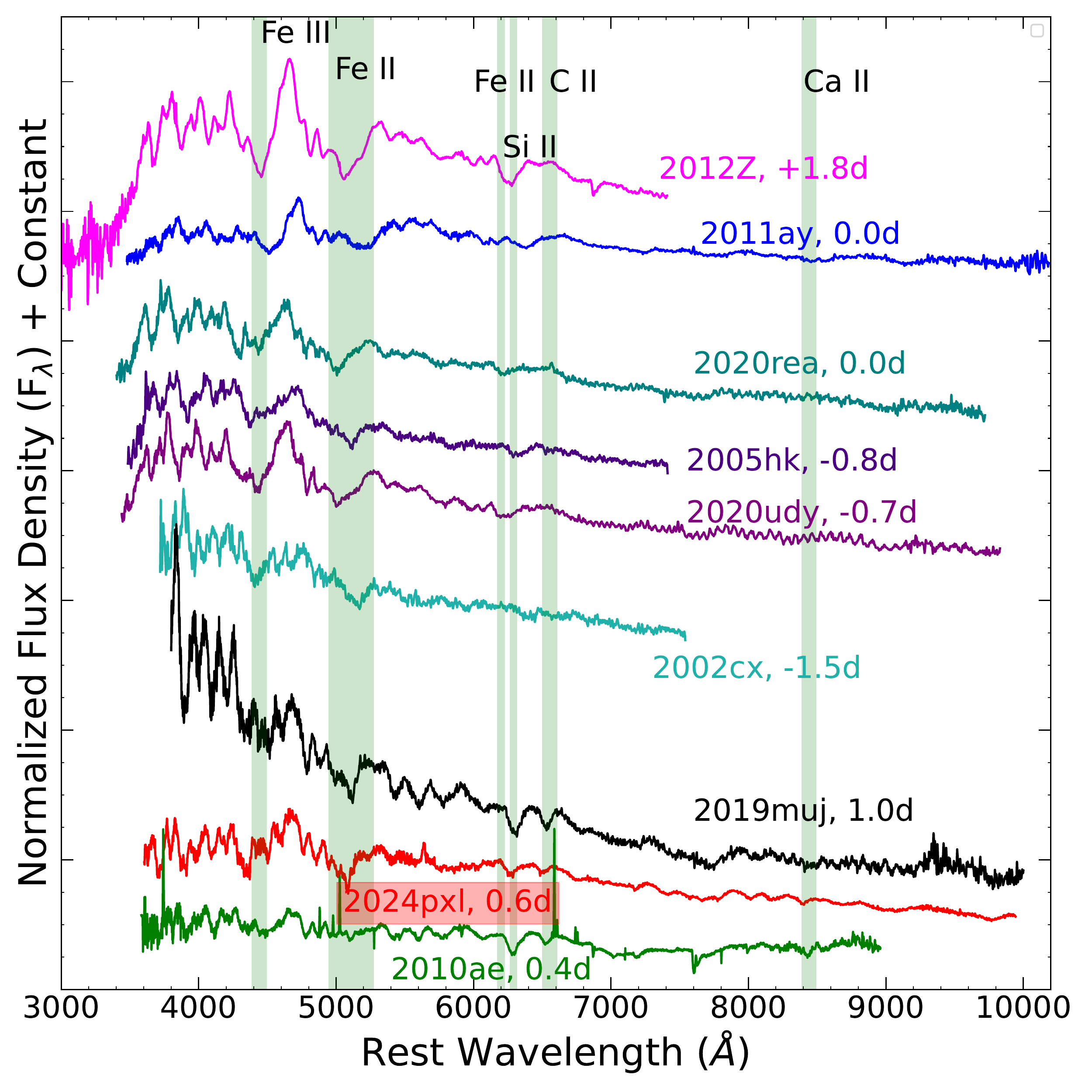}
	\end{center}
	\caption{Near-maximum-light spectral comparison of SN~2024pxl (red) with other Type Iax SNe at similar epochs.}
	\label{fig:peak_comp_24pxl}
\end{figure}

The peak-brightness spectrum of SN~2024pxl (\autoref{fig:peak_comp_24pxl}) shows an increase in the strength of the \ion{Si}{2} line at 6355\,\AA\ and the \ion{Fe}{3} feature near 4400\,\AA. At this epoch, SN~2024pxl resembles SNe~2019muj (intermediate luminosity) and 2020rea (high luminosity). The \ion{Fe}{2} features near 5000\,\AA\ are similar to those in SNe 2002cx, 2005hk (both high luminosity), and 2019muj, while features near 4000\,\AA\ match well with those of SN~2020rea. We detect the \ion{Ca}{2} NIR triplet in SN~2024pxl, which is not detected in high-luminosity objects, but is even more pronounced in the low-luminosity SN~2010ae. The combination of \ion{Ca}{2} NIR features similar to low/intermediate-luminosity objects, and \ion{Fe}{2} characteristics shared with more luminous Type Iax SNe again implies that SN~2024pxl represents a transitional case.

\subsubsection{Post-maximum and late-time spectra}
\label{post_maximum_late_time_spectra}

\begin{figure}
	\begin{center}
		\includegraphics[width=\columnwidth]{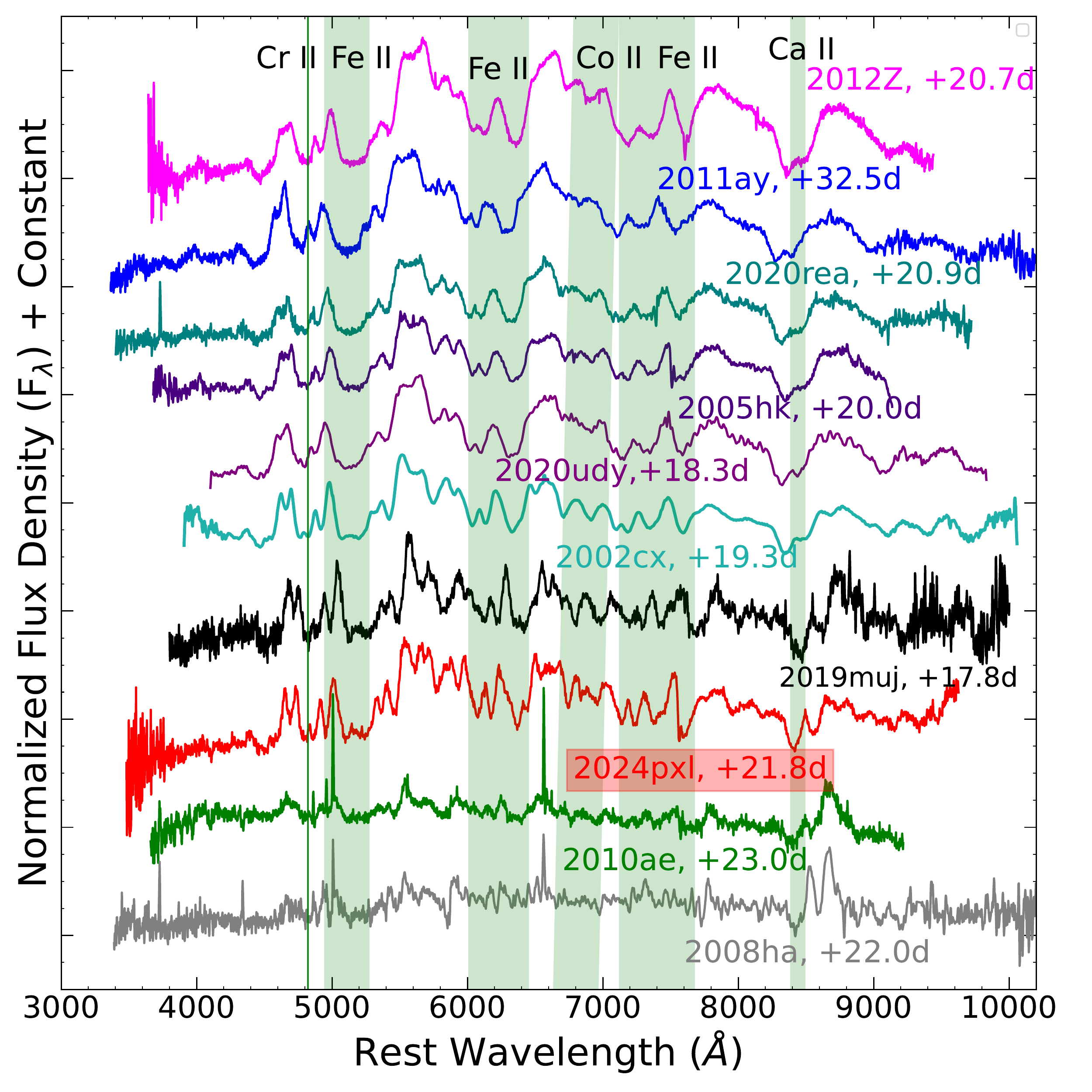}
	\end{center}
	\caption{Post-maximum spectral comparison of SN~2024pxl (red) with other Type Iax SNe at similar epochs.}
	\label{fig:post_peak_comp_24pxl}
\end{figure}

\autoref{fig:post_peak_comp_24pxl} shows the post-maximum spectra of SN~2024pxl in the context of other objects. At these epochs, SN~2024pxl closely resembles SN~2019muj, an intermediate-luminosity Type~Iax SN. During the post-maximum phase, the \ion{Si}{2} and \ion{C}{2} lines vanish, giving way to Fe and Co lines in the 5000-–7000\,\AA\ range. Iron-group elements (IGEs) and a strong \ion{Ca}{2} NIR triplet dominate the post-maximum spectra, similar to other high- and intermediate-luminosity objects. Low-luminosity Type Iax SNe (e.g., SNe 2010ae and 2008ha, shown in \autoref{fig:post_peak_comp_24pxl}) evolve more rapidly and exhibit narrower spectral signatures than SN~2024pxl. The Cr~II feature near 4800~\AA\ and the \ion{Co}{2} lines near 6500~\AA\ in SN~2024pxl share resemblance with high-luminosity SNe 2002cx and 2019muj. Additionally, the \ion{Ca}{2} NIR triplet in SN~2019muj is strikingly similar to that in SN~2024pxl.

\begin{figure}
	\begin{center}
		\includegraphics[width=\columnwidth]{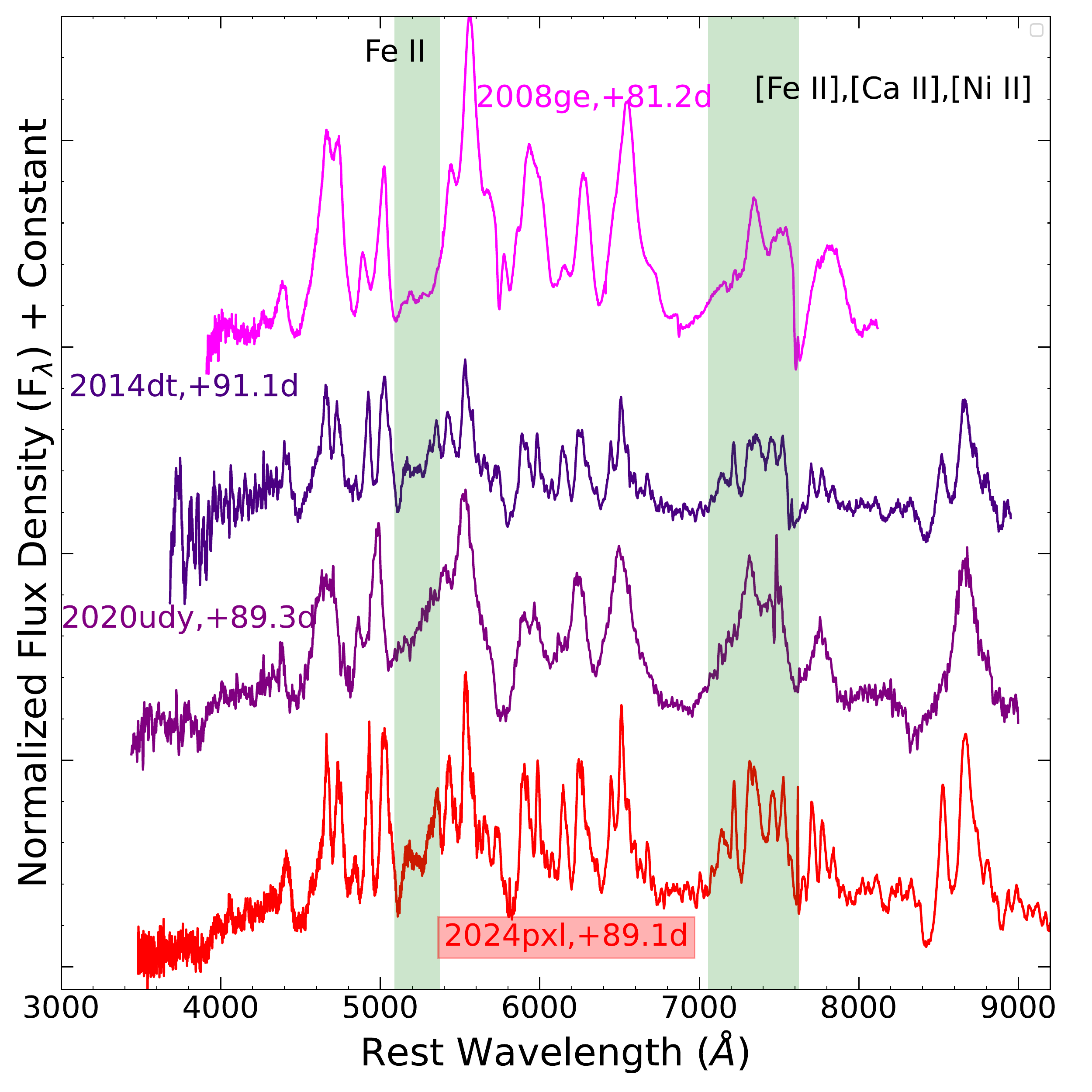}
	\end{center}
	\caption{Late-time spectral comparison of SN~2024pxl (red) with other Type Iax SNe at similar epochs. We highlight a few nebular features in green.}
	\label{fig:nebular_comp_24pxl}
\end{figure}

We compare the late-time spectrum of SN~2024pxl obtained at $\sim+$89~days with spectra of SNe 2008ge \citep{2010AJ....140.1321F}, 2014dt \citep{2018MNRAS.474.2551S}, and 2020udy at similar epochs (\autoref{fig:nebular_comp_24pxl}). At late times, the spectral lines become narrower and several forbidden emission lines such as [\ion{Fe}{2}], [\ion{Ni}{2}], and [\ion{Ca}{2}] emerge along with narrow permitted lines such as \ion{Fe}{2}. The strength of these forbidden lines increases with the cooling of the ejecta \citep{2016MNRAS.461..433F}. Owing to their intrinsically low luminosity, late-time spectra of Type Iax SNe are rare.  The late-time spectra of Type Iax SNe vary widely, with some showing broad emission lines and others dominated by narrow features \citep{2016MNRAS.461..433F}. Of the comparison SNe in \autoref{fig:nebular_comp_24pxl}, SN~2024pxl most closely resembles SN~2014dt, exhibiting narrower lines than the higher-luminosity objects SNe~2020udy and 2008ge. The narrow features seen in the late-phase spectra of Type Iax SNe are thought to originate from the innermost regions \citep{2022ApJ...941...15M,2023ApJ...951...67C}. Additional late-time observations of Type Iax SNe across the luminosity distribution will help to understand the origin and nature of these spectral features more precisely.

Overall, at post-maximum and late times, SN~2024pxl shares similarities with high-luminosity Type Iax SNe and intermediate-luminosity Type Iax SN~2019muj. 

\subsubsection{NIR spectra}
\label{nir_spectra}

\begin{figure}
	\begin{center}
		\includegraphics[width=\columnwidth]{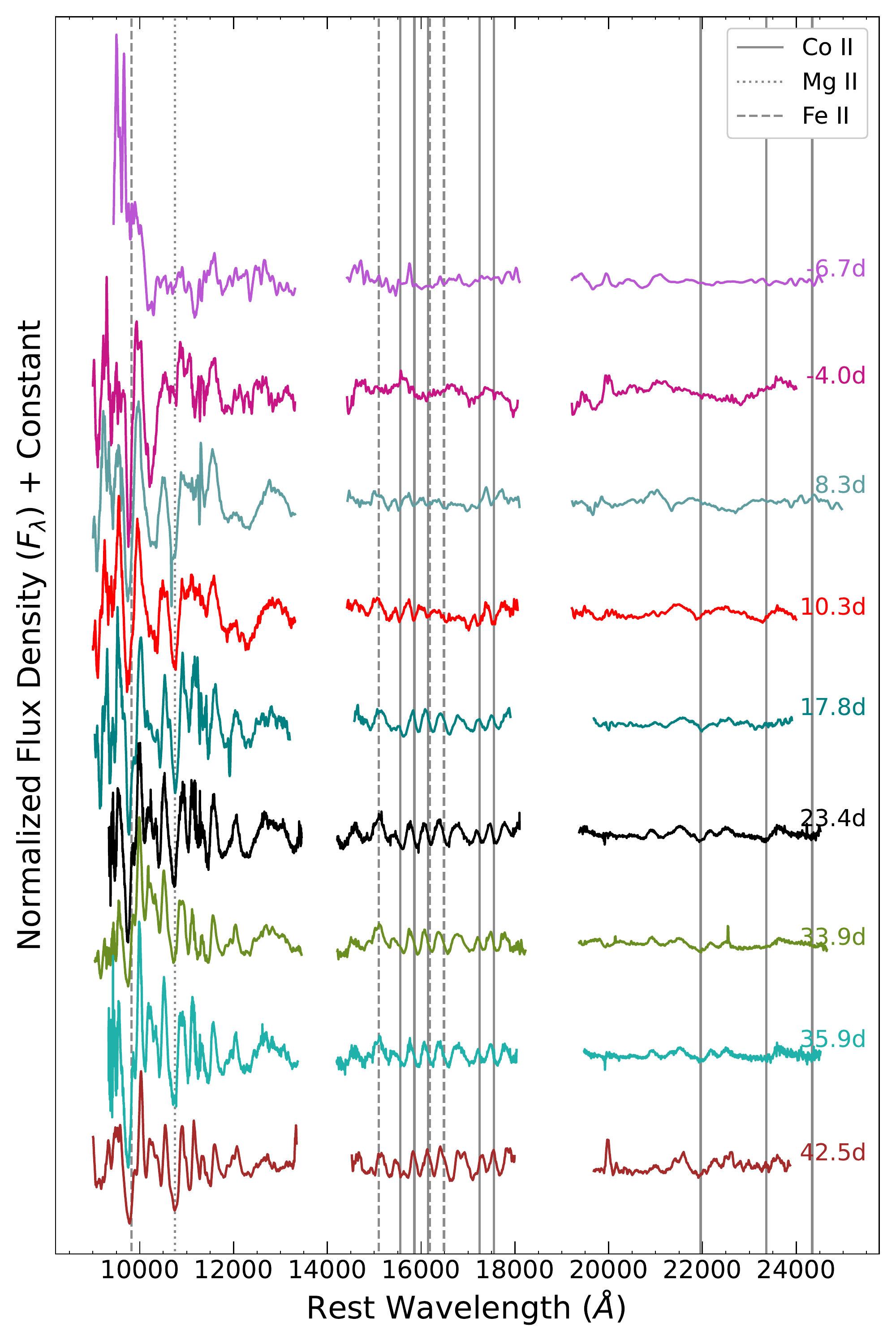}
	\end{center}
	\caption{NIR spectroscopic evolution of SN~2024pxl.}
	\label{fig:nir_spectral_sequence_24pxl}
\end{figure}

\begin{figure}
	\begin{center}
		\includegraphics[width=\columnwidth]{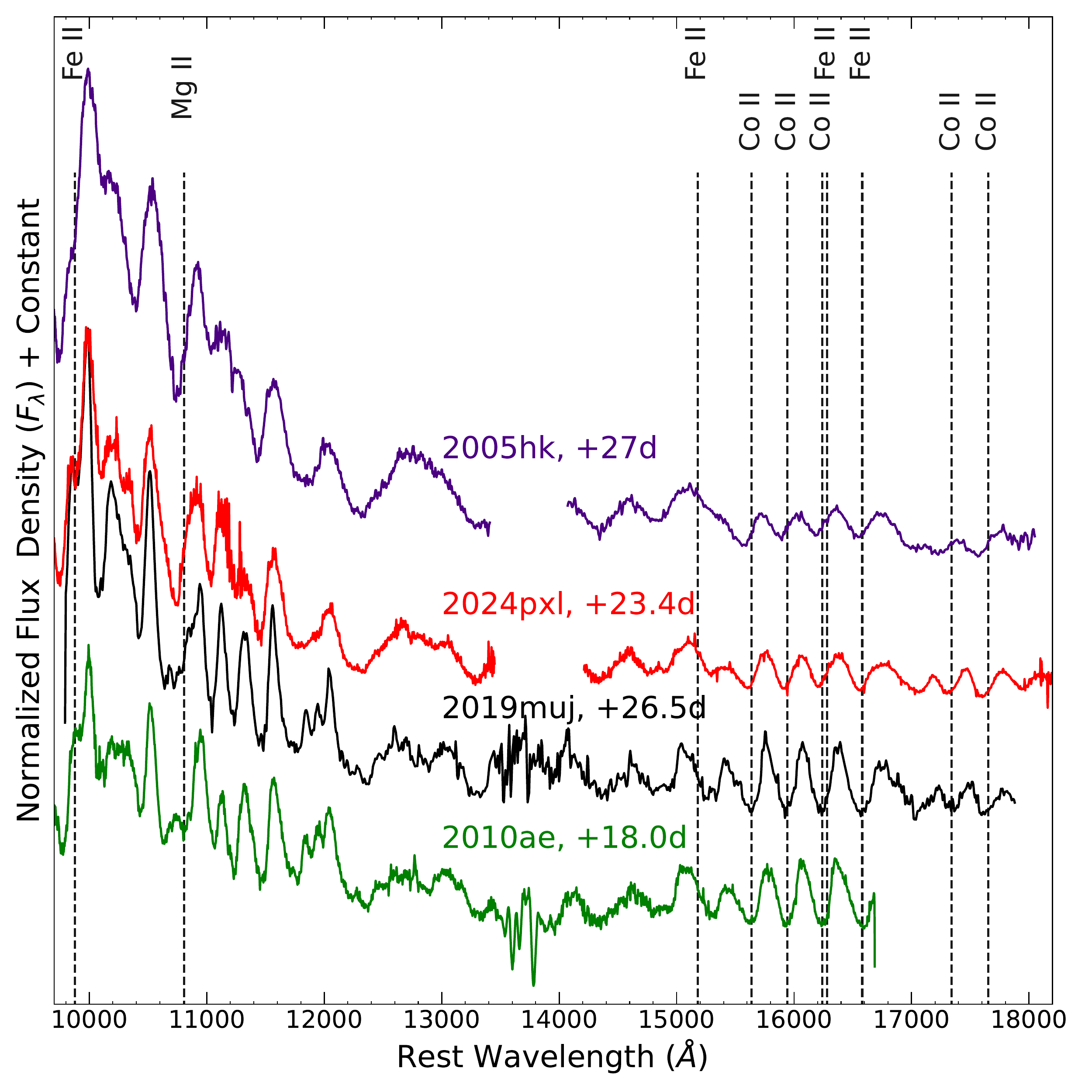}
	\end{center}
	\caption{Comparison of the NIR spectral features of SN~2024pxl with high-luminosity Type Iax SN~2005hk, low-luminosity Type Iax SN~2010ae, and intermediate-luminosity Type Iax SN~2019muj.}
	\label{fig:nir_spectral_comp_24pxl}
\end{figure}

\autoref{fig:nir_spectral_sequence_24pxl} displays the NIR spectral evolution of SN~2024pxl from $-$6.7 to $+$42.5~days post $B$ maximum. The first two NIR spectra of SN~2024pxl show underdeveloped \ion{Fe}{2} features with evidence for the \ion{Mg}{2}~$\lambda$10,952 line. Possible signatures of \ion{C}{1}~$\lambda$10,693 and \ion{C}{1}~$\lambda$11,754 are suggested in the early NIR spectra of SN~2024pxl. As the ejecta cool and expand, the spectrum-formation region recedes to lower velocities and lines become somewhat less blended, with the \ion{Co}{2} lines increasing in prominence over time. Our spectral sequence shows that \ion{Fe}{2} and \ion{Co}{2} lines dominate the NIR. These lines are distinctive features in the NIR spectra of all Type Iax SNe \citep{2014A&A...561A.146S}, and \autoref{fig:nir_spectral_comp_24pxl} displays this similarity in NIR spectra from bright to faint objects. 

The NIR spectrum of SN~2024pxl at $+$23.4~days shows more similarities with the high-luminosity SN~2005hk \citep{2013MNRAS.429.2287K} than with the  intermediate-luminosity SN~2019muj \citep{2021MNRAS.501.1078B} and low-luminosity SN~2010ae \citep{2014A&A...561A.146S}. In particular, the \ion{Mg}{2} absorption, the features at 12,000~\AA\ and 12,700~\AA, and the prominence of the \ion{Co}{2} lines in the 16,000~\AA\ region are differentiating features. Where SN~2024pxl resembles SN~2005hk, SN~2019muj instead resembles SN~2010ae, mainly owing to the lower velocities that allow individual lines to be separated more clearly. This NIR spectral comparison provides additional support that intermediate-luminosity objects of the class (like SNe~2024pxl and 2019muj) are transitional between high- and low-luminosity objects. 

\subsection{Velocity Evolution}
\label{velocity_evolution}

\begin{figure}
	\begin{center}
		\includegraphics[width=\columnwidth]{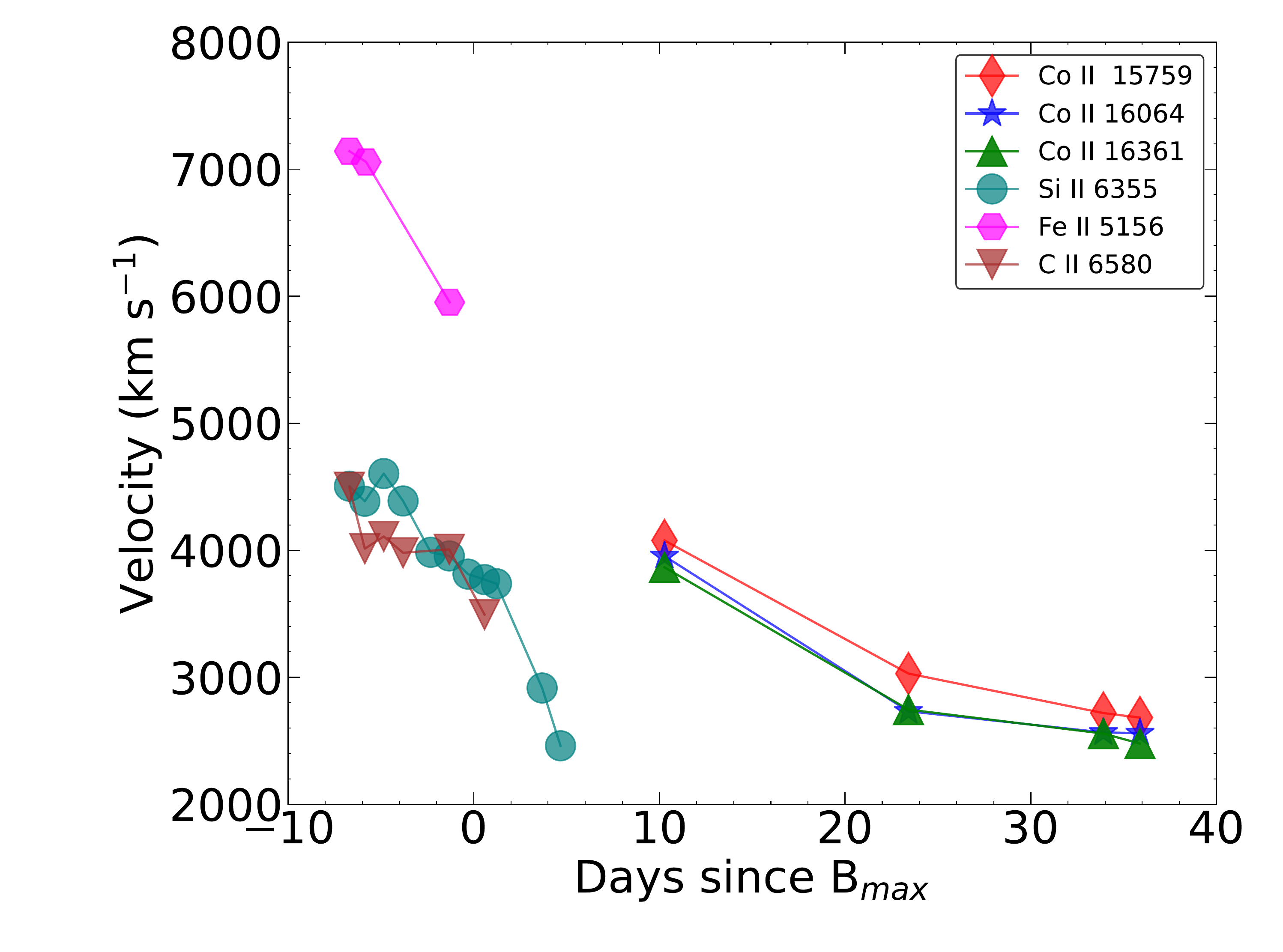}
	\end{center}
	\caption{Evolution of the line velocities of \ion{Si}{2}, \ion{C}{2}, \ion{Fe}{2}, and \ion{Co}{2} lines in SN~2024pxl.}
	\label{fig:SN_2024pxl_velocity_evolution_Si_Fe_Co}
\end{figure}

\begin{figure}
	\begin{center}
		\includegraphics[width=\columnwidth]{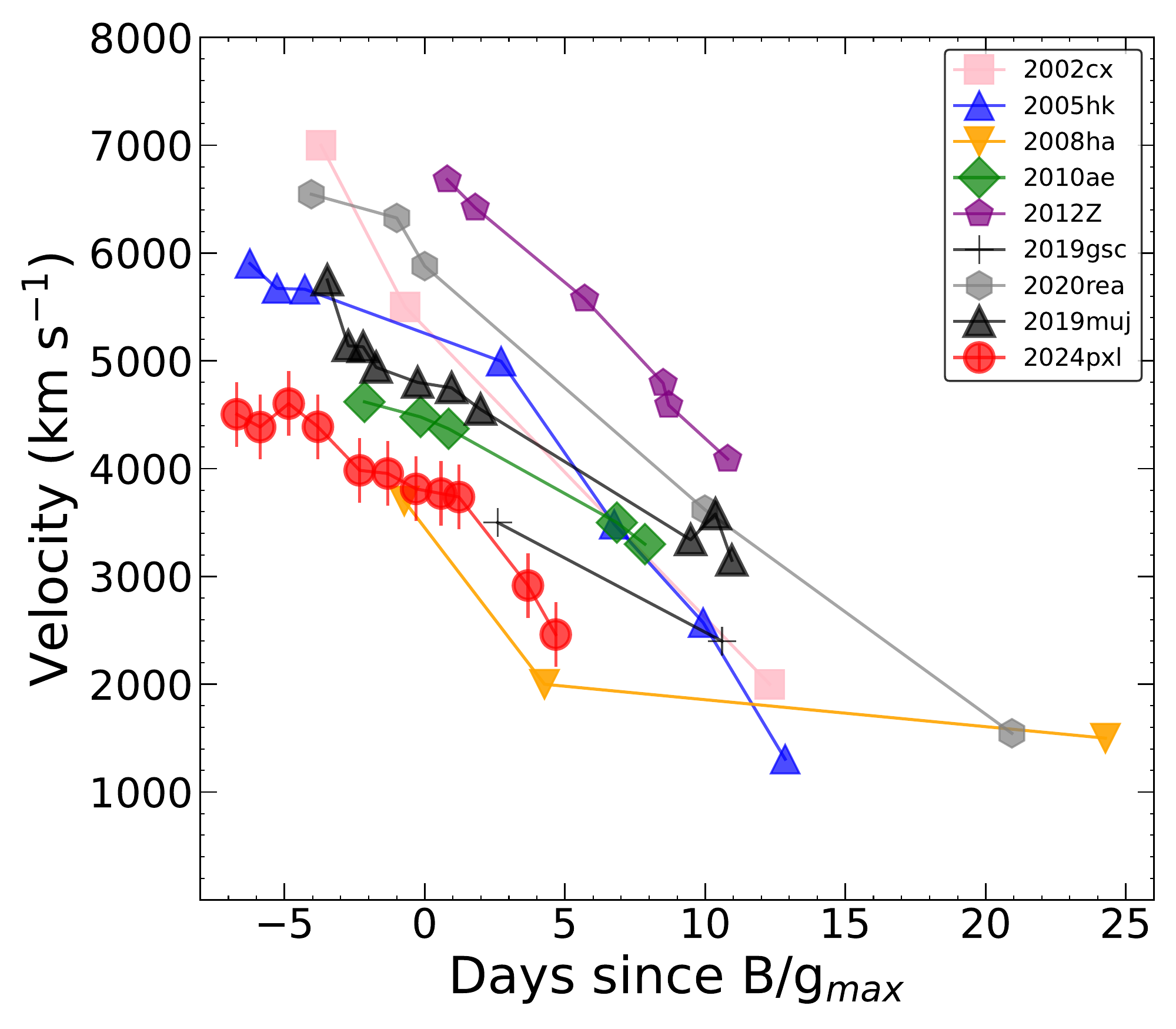}
	\end{center}
	\caption{Evolution of the velocity at maximum absorption of \ion{Si}{2}~$\lambda$6355 for SN~2024pxl and its comparison with other Type Iax SNe.}
	\label{fig:SN_2024pxl_velocity_evolution}
\end{figure}

We probe the evolution of velocities of \ion{C}{2}, \ion{Si}{2}, and \ion{Fe}{2} lines using our optical spectral sequence, and the \ion{Co}{2} lines from our NIR spectra. We measure line velocities by fitting Gaussian profiles to the absorption trough of the P-Cygni profiles of the \ion{C}{2}~$\lambda$6580, \ion{Si}{2}~$\lambda$6355, \ion{Fe}{2}~$\lambda$5156, \ion{Co}{2}~$\lambda$15,759, \ion{Co}{2}~$\lambda$16,064, and \ion{Co}{2}~$\lambda$16,361 lines (\autoref{fig:SN_2024pxl_velocity_evolution_Si_Fe_Co}). This standard method of measuring photospheric line velocities results in \ion{Co}{2} measurements that are consistent with those presented by \cite{2025ApJ...989L..33K}, obtained by fitting blended \ion{Co}{2} lines simultaneously. \autoref{fig:SN_2024pxl_velocity_evolution} shows the \ion{Si}{2} line-velocity evolution for SN~2024pxl alongside several other well-studied Type Iax SNe. The uncertainties presented in \autoref{fig:SN_2024pxl_velocity_evolution} could likely be underestimated, as we only account for measurement uncertainties from our fitting (and not for the effects of line blending, for example). 
    
SN~2024pxl exhibits lower velocities than other SNe shown in \autoref{fig:SN_2024pxl_velocity_evolution} except low-luminosity SN~2008ha. The prominent \ion{Si}{2} line is overlapped by a continuously growing \ion{Fe}{2} feature in the red wing of the \ion{Si}{2} line. As a result, we systematically underestimate the velocities associated with this line over time by considering it as a single absorption feature. In the case of high-luminosity SN~2020udy, \cite{2023MNRAS.525.1210M} showed that the \ion{Si}{2} line is contaminated by the \ion{Fe}{2} line before maximum light, although  \ion{Si}{2} is present in post-maximum spectra. This might be why a significant drop is observed in the \ion{Si}{2} velocities post-maximum in SN~2024pxl. It is likely that more careful modeling of these lines will be required to obtain a more accurate comparison of the \ion{Si}{2} line-velocity evolution across Type Iax SNe.
    
The \ion{Si}{2} and \ion{C}{2} line velocities of SN~2024pxl at peak are 3800~km~s$^{-1}$ and 3500~km~s$^{-1}$, respectively. The measured \ion{Fe}{2} velocities are 7100, 7060, and 5950~km~s$^{-1}$ at epochs $\sim -6.7$, $-$5.8, and $-$1.3~days since {\it B}$_\mathrm{max}$, respectively. At a similar epoch, the estimated velocities using \ion{Fe}{2} lines are higher than the \ion{Si}{2} and \ion{C}{2} line velocities, indicating significant mixing in the burning products \citep{2007PASP..119..360P}. \cite{2025ApJ...989L..33K} also find evidence of mixing in SN~2024pxl from the emission lines of IGEs and intermediate-mass elements (IMEs), which are all centrally peaked and have similar velocity offsets and widths. The measured \ion{Co}{2} line velocities in the NIR region for SN~2024pxl at an epoch of 23.4~days are 3030, 2730, and 2750~km~s$^{-1}$ for \ion{Co}{2}~$\lambda$15,759, \ion{Co}{2}~$\lambda$16,064, and \ion{Co}{2}~$\lambda$16,361, respectively (consistent with one another), and slightly higher than those reported by \cite{2021MNRAS.501.1078B} at an epoch of 26.5 days past {\it B}$_\mathrm{max}$ for intermediate-luminosity SN~2019muj. At a comparable epoch, velocities of \ion{Co}{2} lines in low-luminosity SN~2010ae  \citep{2014A&A...561A.146S} are lower than in SN~2024pxl.

\begin{figure}
	\begin{center}
		\includegraphics[width=\columnwidth]{absmag_delta15_updated.png}
	\end{center}
	\caption{This figure illustrates the distribution of Type~Iax SNe in the absolute magnitude–light-curve decline rate plane in the $r/R$ bands. SN~2024pxl is marked by a red-filled circle.}
	\label{fig:abs_mag_decline_rate_plot}
\end{figure}

\section{Summary}
\label{summary}

SN~2024pxl is a nearby Type Iax SN with an early detection that enabled precise constraints on its rise time. With a peak absolute magnitude of \( M_{V,\text{max}}=-16.81~\pm~0.19 \)~mag, SN~2024pxl is an intermediate-luminosity Type Iax SN similar to SN~2019muj. The color evolution of SN~2024pxl aligns closely with that of other Type Iax SNe. Utilizing \textit{Swift} UV coverage around peak brightness, we estimate the contribution of UV flux to the bolometric luminosity and find that SN~2024pxl's pseudo bolometric luminosity falls between those of SNe~2002cx and 2019muj, placing it in a unique intermediate position in the luminosity range of Type Iax SNe. Analytical modeling of the integrated SED light curve indicates the synthesis of $0.03 \pm 0.01$\,M$_{\odot}$ of $^{56}$Ni and an ejected mass of $0.36 \pm 0.05$\,M$_{\odot}$. The explosion parameters calculated through analytical modeling of the integrated SED light curve are consistent with the N1def model from \cite{2014MNRAS.438.1762F}, a weak deflagration of a near-$M_\mathrm{Ch}$ WD with one ignition point. This is consistent with the results of \cite{2025ApJ...989L..33K} that SN~2024pxl displays spectroscopic similarity to the predictions of the N1def model.

Our spectral comparisons also suggest that SN~2024pxl is a transitional object between high-luminosity and low-luminosity Type Iax SNe. It exhibits many similarities to intermediate-luminosity SN~2019muj and high-luminosity Type Iax SNe. In the early spectral sequence, we also detect \ion{C}{2} that is seen more prominently in less luminous objects. Additionally, the late-time spectra of SN~2024pxl show narrow emission lines around 7300~\AA\, whereas the high-luminosity objects (e.g., SNe~2008ge and 2020udy) display broader features.

In the NIR, SN~2024pxl shows signatures from \ion{Mg}{2}, \ion{Fe}{2}, and \ion{Co}{2} and it resembles the high-luminosity Type Iax SN~2005hk, while intermediate-luminosity Type Iax SN~2019muj more closely resembles the low-luminosity Type Iax SN~2010ae. This is mainly due to more line blending in SN~2024pxl compared with SN~2019muj, caused by wider line widths. Despite the higher velocities indicated by the wider line widths in SN~2024pxl, the evolution of the photospheric velocity of SN~2024pxl shows systematically lower velocities relative to other comparison SNe, except for the very low-luminosity SN~2008ha. \cite{2025ApJ...989L..33K} find potential evidence for bulk motion on the order of a few 100~km~s$^{-1}$ in the ejecta of SN~2024pxl from measurements of MIR forbidden lines, which would impact the true vs. observed photospheric velocity. Given the spectroscopically and photometrically intermediate position of SN~2024pxl between SN~2019muj and brighter Type Iax SNe, we suggest that velocity measurements of the widths of the \ion{Co}{2} lines may be better correlated with luminosity than the photospheric velocity. These spectral properties provide evidence that SN~2024pxl is a transitional Type Iax SN.

We measure the \ion{Fe}{2} velocity to be substantially higher than that of \ion{Si}{2} and \ion{C}{2} at similar epochs. At a slightly later phase, \ion{Co}{2} velocity measurements seem to align with the expected velocity decay from the \ion{Fe}{2} measurements. Higher velocities of IGEs compared to IMEs indicate mixing in the ejecta, consistent with the findings of \cite{2025ApJ...989L..33K}.  

Figure~\ref{fig:abs_mag_decline_rate_plot} shows the distribution of Type~Iax SNe in terms of their peak absolute magnitude and light-curve decline rate in the $r/R$ band. The data are taken from \cite{2023ApJ...953...93S} and references therein, augmented with the recently studied Type~Iax events SN~2022eyw \citep{Das2025}, SN~2022xlp \citep{2025A&A...703A..64B}, SN~2025qe \citep{2025MNRAS.543.3731M}, and SN~2024pxl. \cite{2023ApJ...953...93S} suggested that bright (high-luminosity) Type~Iax SNe have $M_{r} \le -17.1$~mag, while faint (low-luminosity) ones have $M_{r} \ge -14.64$~mag, and further proposed a negative linear correlation for the bright population and a positive correlation for the faint population. SN~2024pxl lies near the faint boundary of the bright Type~Iax group, whereas SN~2022eyw is a luminous member located well within the bright population. In contrast, SNe~2022xlp and 2025qe exhibit intermediate luminosities, bridging the bright and faint subclasses. The newly added events follow the overall trend, with SN~2024pxl positioned close to the lower end of the bright group, underscoring its transitional nature. As additional Type~Iax SNe spanning the full luminosity range become available, these correlations can be tested more robustly.

Overall, SN~2024pxl is a compelling link between high-luminosity and low-luminosity Type Iax SNe. The analysis of UV, optical, and NIR data presented in this work, and the spectral analysis of \textit{JWST} NIR and MIR data by \cite{2025ApJ...989L..33K}, demonstrate that SN~2024pxl is photometrically and spectroscopically consistent with the one-ignition-point weak deflagration model, N1def, from \cite{2014MNRAS.438.1762F}. \cite{2025ApJ...989L..33K} also show that SN~2024pxl shares spectroscopic similarities (aside from the large luminosity difference) with the very faint Type Iax SN~2024vjm. These findings indicate that intermediate-luminosity objects such as SNe~2019muj and 2024pxl are links in a continuous distribution. Future studies of other Type Iax SNe with rich datasets, such as that presented for SN~2024pxl, from both ground- and space-based telescopes will help us better understand the nature of these peculiar SNe. 

Remarkably, the diverse class of Type Iax SNe can span a huge luminosity range. Still, the seemingly continuous properties we observe may yet point to a common (or at least quite similar) progenitor origin and explosion mechanism. This contrasts with normal Type Ia SNe, for which multiple progenitor pathways may be plausible while producing more homogeneous outcomes. Further study of peculiar thermonuclear SNe holds promise to enable a better understanding of which WDs may explode and their explosion mechanisms.

\section{acknowledgments}

We thank the referee for the careful review and valuable suggestions, which have helped refine the manuscript. We also thank St\'ephane Blondin for helpful comments which improved the manuscript.

This work is based in part on observations collected at the European Southern Observatory under ESO programme 114.27JL.001. Partly based on observations made with the Nordic Optical Telescope, owned in collaboration by the University of Turku and Aarhus University. This work also makes use of data gathered with the 6.5~m Magellan telescopes at Las Campanas Observatory, Chile. A FIRE spectrum was obtained through A.P.'s prior support by a Carnegie Fellowship. This work makes use of data from the Las Cumbres Observatory global network of telescopes. The LCO group is supported by National Science Foundation (NSF) grants AST-1911151 and AST-1911225.

The SALT observations were obtained with Rutgers University program 2024-1-MLT-004 (PI L.~A.~Kwok). We are grateful to SALT Astronomer Rosalind Skelton for taking these data. Data reported here were obtained in part at the MMT Observatory, a joint facility of the University of Arizona and the Smithsonian Institution. We sincerely thank the MMT observers and staff for their accommodation of our strict timing requirements to ensure they coincided with {\it JWST} observations. 

Some of the data presented herein were obtained at the W.~M. Keck Observatory, which is operated as a scientific partnership among the California Institute of Technology, the University of California, and the National Aeronautics and Space Administration (NASA). The Observatory was made possible by the generous financial support of the W.~M. Keck Foundation. This work was supported by a NASA Keck PI Data Award (PI D. J. Sand), administered by the NASA Exoplanet Science Institute. The authors wish to recognize and acknowledge the very significant cultural role and reverence that the summit of Maunakea has always had within the Indigenous Hawaiian community. We are most fortunate to have the opportunity to conduct observations from this mountain.

This study employs observations obtained with the Hobby-Eberly Telescope, which is a joint project of the University of Texas at Austin, the Pennsylvania State University, Ludwig-Maximilians-Universit{\"a}t M{\"u}nchen, and Georg-August-Universit{\"a}t G{\"o}ttingen.
The HET is named in honor of its principal benefactors, William P. Hobby and Robert E. Eberly. The Low Resolution Spectrograph 2 (LRS2) was developed and funded by the University of Texas at Austin McDonald Observatory and Department of Astronomy and by Pennsylvania State University. We thank the Leibniz-Institut f{\"u}r Astrophysik Potsdam (AIP) and the Institut f{\"u}r Astrophysik G{\"o}ttingen (IAG) for their contributions to the construction of the integral field units.

KAIT and its ongoing operation were made possible by donations from Sun Microsystems, Inc., the 
Hewlett-Packard Company, AutoScope Corporation, Lick Observatory, the U.S. NSF, the University of California, the Sylvia \& Jim Katzman Foundation, and the TABASGO Foundation.
A major upgrade of the Kast spectrograph on the Shane 3~m telescope at Lick Observatory, led by Brad Holden, was made possible through generous gifts from the Heising-Simons Foundation, William and Marina Kast, and the University of California Observatories. Research at Lick Observatory is partially supported by a generous gift from Google. We would like to express our gratitude to the Lick Observatory staff for their support. Shane 3~m observations were conducted on the stolen land of the Ohlone (Costanoans), Tamyen and Muwekma Ohlone tribes. We gratefully acknowledge the usage of native lands for our science. Observations from coauthor A.J.M. were made under the aegis of the ASTRAL (Astronomy/STEM Alliance with Lick Observatory) consortium, supported by a generous grant from the Gordon and Betty Moore Foundation (PI B. Macintosh).

We thank the staff of IAO, Hanle, CREST, and Hosakote, who made these observations possible. The facilities at IAO and CREST are operated by the Indian Institute of Astrophysics, Bangalore. We thank the Subaru staff for the data taken by the Subaru Telescope (S23A-023). Based in part on observations made with the Gran Telescopio Canarias (GTC), installed at the Spanish Observatorio del Roque de los Muchachos of the Instituto de Astrofísica de Canarias, on the island of La Palma, using OSIRIS and EMIR instruments.

Based (in part) on data acquired at the ANU 2.3~m telescope. The automation of the telescope was made possible through an initial grant provided by the Centre of Gravitational Astrophysics and the Research School of Astronomy and Astrophysics at the Australian National University and through a grant provided by the Australian Research Council through LE230100063. We acknowledge the traditional custodians of the land on which the telescope stands, the Gamilaraay people, and pay our respects to elders past and present.

The HET is a joint project of the University of Texas at Austin, the Pennsylvania State University, Stanford University, Ludwig-Maximilians-Universität München, and Georg-August-Universität Göttingen. The HET is named in honor of its principal benefactors, William P. Hobby and Robert E. Eberly. The Marcario Low-Resolution Spectrograph is named for Mike Marcario of High Lonesome Optics who fabricated several optics for the instrument but died before its completion. The LRS is a joint project of the HET partnership and the Instituto de Astronomía de la Universidad Nacional Autónoma de México.

We acknowledge Weizmann Interactive Supernova data REPository http://wiserep.weizmann.ac.il (WISeREP) \citep{2012PASP..124..668Y}. This research has made use of the CfA Supernova Archive, which is funded in part by the NSF through grant AST-0907903. This research has made use of the NASA/IPAC Extragalactic Database (NED) which is operated by the Jet Propulsion Laboratory, California Institute of Technology, under contract with NASA.

M.S. acknowledges financial support provided under the National Post Doctoral Fellowship (N-PDF; File Number: PDF/2023/002244) by the Science \& Engineering Research Board (SERB), Anusandhan National Research Foundation (ANRF), Government of India. L.A.K. is supported by a CIERA Postdoctoral Fellowship. 

Support for this research at Rutgers University (S.W.J., C.L., M.S.) was provided by NSF award AST-2407567.

A.A.M., C.L., and N.R. are supported by DoE award \#DE-SC0025599. MMT and Keck Observatory access for N.R. and C.L. was supported by Northwestern University and the Center for Interdisciplinary Exploration and Research in Astrophysics (CIERA). A.C.G. and the Fong Group at Northwestern acknowledge support by the NSF under grants AST-1909358, AST-2206494, AST-2308182, and CAREER grant  AST-2047919. 

G.C.A. thanks the Indian National Science Academy for support under the INSA Senior Scientist Programme. A.F. acknowledges support by the European Research Council (ERC) under the European Union’s Horizon 2020 research and innovation program (ERC Advanced Grant KILONOVA \#885281) and the State of Hesse within the Cluster Project ELEMENTS. C.L. acknowledges support from DOE award DE-SC0010008 to Rutgers University. M.R.S. is supported by the STScI Postdoctoral Fellowship.

Time-domain research by the University of Arizona team and D.J.S. is supported by NSF grants 2108032, 2308181, 2407566, and 2432036 and the Heising-Simons Foundation under grant \#2020-1864. Time-domain research by the University of California, Davis team and S.V. is supported by NSF grant AST-2407565. K.A.B. is supported by an LSST-DA Catalyst Fellowship; this publication was thus made possible through the support of grant 62192 from the John Templeton Foundation to LSST-DA. N.F. acknowledges support from the NSF Graduate Research Fellowship Program under grant DGE-2137419. 

J.E.A. is supported by the international Gemini Observatory, a program of NSF's NOIRLab, which is managed by the Association of Universities for Research in Astronomy (AURA) under a cooperative agreement with the NSF, on behalf of the Gemini partnership of Argentina, Brazil, Canada, Chile, the Republic of Korea, and the United States of America. J.A.V. acknowledges the Postgraduate School of the Universidad de Antofagasta for its support and allocated grants. R.D. acknowledges funds by ANID grant FONDECYT Postdoctorado  \#3220449.

A.V.F.’s group at UC Berkeley received financial assistance from the Christopher R. Redlich Fund, as well as donations from Gary and Cynthia Bengier, Clark and Sharon Winslow, Alan Eustace and Kathy Kwan, William Draper, Timothy and Melissa Draper, Briggs and Kathleen Wood, Sanford Robertson (W.Z. is a Bengier-Winslow-Eustace Specialist in Astronomy, T.G.B. is a Draper-Wood-Robertson Specialist in Astronomy, Y.Y. was a Bengier-Winslow-Robertson Fellow in Astronomy), and numerous other donors.

K. Maguire acknowledges funding from Horizon Europe ERC grant 101125877. J.H.T. acknowledges support from EU H2020 ERC grant 758638. T.T. acknowledges support from NSF grant AST-2205314 and the NASA ADAP award 80NSSC23K1130. K. Maede acknowledges support from JSPS KAKENHI grants JP24KK0070, JP24H01810, and JP20H00174, and from JSPS Bilateral Joint Research Project (JPJSBP120229923). K. Misra acknowledges support from the BRICS grant DST/ICD/BRICS/Call-5/CoNMuTraMO/2023 (G) funded by the DST, India. J.V. is supported by NKFIH-OTKA grant K142534. G.C.A. thanks the Indian National Science Academy for support under the INSA Senior Scientist Programme. M.R.S. is supported by the STScI Postdoctoral Fellowship. B.B. received support from the Hungarian National Research, Development and Innovation Office grants OTKA PD-147091. L.G. acknowledges financial support from AGAUR, CSIC, MCIN and AEI 10.13039/501100011033 under projects PID2023-151307NB-I00, PIE 20215AT016, CEX2020-001058-M, ILINK23001, COOPB2304, and 2021-SGR-01270. H.K. was funded by the Research Council of Finland projects 324504, 328898, and 353019. D.A.H., G.H., and C.M. were supported by NSF grants AST-1313484 and AST-1911225.

%






\bibliography{ms}{}
\bibliographystyle{aasjournal}

\begin{appendix}

\begin{longtable*}
{>{\centering\arraybackslash}p{0.10\textwidth}
>{\centering\arraybackslash}p{0.06\textwidth}
>{\raggedright\arraybackslash}p{0.06\textwidth}
>{\centering\arraybackslash}p{0.10\textwidth}
>{\centering\arraybackslash}p{0.10\textwidth}
>{\centering\arraybackslash}p{0.10\textwidth}
>{\centering\arraybackslash}p{0.10\textwidth}
>{\centering\arraybackslash}p{0.10\textwidth}
>{\centering\arraybackslash}p{0.10\textwidth}}
\caption{Optical photometric data of SN~2024pxl obtained with the LCO and Nickel telescopes. Photometry from the other facilities is provided in the online supplementary material.}
\\

\hline
Date & JD$^\dagger$ & Phase$^\ddagger$  & $B$ & $g$ & $V$ & $r$ & $i$ & Telescope \\
 & & (Days) & (mag) & (mag)  & (mag) & (mag) & (mag) &    \\
\hline  
\endfirsthead

\multicolumn{9}{c}%
    {{\bfseries \tablename\ \thetable{} -- Continued from previous page}} \\
    \hline
    Date    &   JD$^\dagger$   &   Phase$^\ddagger$ 	&   B       &      g        &    V     &  r   	    &   i      &     Telescope             \\
    \hline
    \endhead

\hline 
\multicolumn{9}{l}{\textsuperscript{} $^\dagger$ JD 2,460,000+ ,
$^\ddagger$Phase  calculated with respect to $B_{\rm max} = 2,460,524.61$.
} \\
\multicolumn{9}{r}{\textit{Continued on next page}} \\ 
\hline
    \endfoot
    
    \hline
    \endlastfoot

2024-07-23 & 515.32 &   $-$9.3 & 18.01 ± 0.03 & 17.76 ± 0.03 & 17.60 ± 0.03 & 17.43 ± 0.02 & 17.37 ± 0.02 &       LCO \\
2024-07-23 & 515.32 &   $-$9.3 & 17.97 ± 0.03 & 17.70 ± 0.02 & 17.56 ± 0.03 & 17.40 ± 0.02 & 17.40 ± 0.03 &       LCO \\
2024-07-25 & 517.44 &   $-$7.2 & 17.19 ± 0.02 & 16.93 ± 0.01 & 16.87 ± 0.02 & 16.59 ± 0.01 & 16.59 ± 0.02 &       LCO \\
2024-07-25 & 517.44 &   $-$7.2 & 17.23 ± 0.02 & 16.93 ± 0.01 & 16.86 ± 0.02 & 16.60 ± 0.01 & 16.56 ± 0.02 &       LCO \\
2024-07-26 & 517.74 &   $-$6.9 & 17.01 ± 0.09 &          --- & 16.77 ± 0.08 & 16.60 ± 0.06 & 16.56 ± 0.07 &    Nickel \\
2024-07-27 & 518.61 &   $-$6.0 & 17.03 ± 0.01 & 16.76 ± 0.01 & 16.72 ± 0.01 & 16.42 ± 0.02 & 16.42 ± 0.03 &       LCO \\
2024-07-27 & 518.61 &   $-$6.0 & 17.03 ± 0.01 & 16.76 ± 0.01 & 16.73 ± 0.01 & 16.42 ± 0.02 & 16.42 ± 0.03 &       LCO \\
2024-07-28 & 519.56 &   $-$5.1 & 16.89 ± 0.01 & 16.65 ± 0.01 & 16.56 ± 0.01 &          --- &          --- &       LCO \\
2024-07-28 & 519.56 &   $-$5.1 & 16.89 ± 0.01 &          --- & 16.57 ± 0.01 &          --- &          --- &       LCO \\
2024-07-28 & 519.57 &   $-$5.0 &          --- & 16.64 ± 0.01 &          --- & 16.23 ± 0.02 & 16.22 ± 0.01 &       LCO \\
2024-07-28 & 519.57 &   $-$5.0 &          --- &          --- &          --- & 16.22 ± 0.01 & 16.22 ± 0.02 &       LCO \\
2024-07-29 & 520.65 &   $-$4.0 & 16.72 ± 0.01 & 16.43 ± 0.01 & 16.40 ± 0.01 & 16.09 ± 0.02 & 16.10 ± 0.02 &       LCO \\
2024-07-29 & 520.65 &   $-$4.0 & 16.71 ± 0.01 & 16.42 ± 0.01 & 16.39 ± 0.01 & 16.10 ± 0.02 & 16.10 ± 0.02 &       LCO \\
2024-07-30 & 521.61 &   $-$3.0 & 16.65 ± 0.01 & 16.37 ± 0.01 & 16.30 ± 0.01 & 15.99 ± 0.02 & 15.99 ± 0.03 &       LCO \\
2024-07-30 & 521.61 &   $-$3.0 & 16.65 ± 0.01 & 16.37 ± 0.01 & 16.29 ± 0.01 & 15.99 ± 0.03 & 16.00 ± 0.03 &       LCO \\
2024-07-30 & 521.79 &   $-$2.8 & 16.44 ± 0.08 &          --- & 16.15 ± 0.08 & 15.97 ± 0.04 & 15.94 ± 0.06 &    Nickel \\
2024-07-31 & 522.56 &   $-$2.1 & 16.45 ± 0.07 &          --- & 16.03 ± 0.03 &          --- &          --- &       LCO \\
2024-07-31 & 522.56 &   $-$2.1 & 16.41 ± 0.08 &          --- & 16.13 ± 0.04 &          --- &          --- &       LCO \\
2024-07-31 & 522.56 &   $-$2.0 &          --- & 16.18 ± 0.03 &          --- &          --- &          --- &       LCO \\
2024-07-31 & 522.56 &   $-$2.0 &          --- & 16.21 ± 0.06 &          --- &          --- &          --- &       LCO \\
2024-08-01 & 524.39 &   $-$0.2 & 16.48 ± 0.01 & 16.16 ± 0.01 & 16.00 ± 0.01 & 15.71 ± 0.01 & 15.69 ± 0.02 &       LCO \\
2024-08-01 & 524.40 &   $-$0.2 & 16.51 ± 0.01 & 16.15 ± 0.01 & 15.98 ± 0.01 & 15.71 ± 0.01 & 15.68 ± 0.02 &       LCO \\
2024-08-03 & 525.86 &    1.2 & 16.33 ± 0.01 &          --- &          --- &          --- &          --- &       LCO \\
2024-08-03 & 525.86 &    1.2 & 16.33 ± 0.01 &          --- &          --- &          --- &          --- &       LCO \\
2024-08-03 & 525.87 &    1.3 &          --- & 16.05 ± 0.01 & 15.91 ± 0.01 & 15.59 ± 0.01 & 15.56 ± 0.01 &       LCO \\
2024-08-03 & 525.88 &    1.3 &          --- & 16.04 ± 0.01 & 15.90 ± 0.01 & 15.60 ± 0.01 & 15.55 ± 0.01 &       LCO \\
2024-08-04 & 526.65 &    2.0 & 16.35 ± 0.01 & 16.04 ± 0.05 & 15.89 ± 0.02 &          --- &          --- &       LCO \\
2024-08-04 & 526.65 &    2.0 &          --- &          --- & 15.86 ± 0.06 &          --- &          --- &       LCO \\
2024-08-05 & 528.36 &    3.7 & 16.63 ± 0.01 & 16.20 ± 0.01 & 15.95 ± 0.01 & 15.62 ± 0.01 &          --- &       LCO \\
2024-08-05 & 528.36 &    3.7 & 16.63 ± 0.01 & 16.22 ± 0.01 & 15.97 ± 0.01 &          --- &          --- &       LCO \\
2024-08-05 & 528.36 &    3.8 &          --- &          --- &          --- & 15.62 ± 0.01 & 15.59 ± 0.01 &       LCO \\
2024-08-05 & 528.36 &    3.8 &          --- &          --- &          --- &          --- & 15.57 ± 0.01 &       LCO \\
2024-08-06 & 529.48 &    4.9 & 16.79 ± 0.01 & 16.38 ± 0.01 & 16.02 ± 0.01 & 15.67 ± 0.01 & 15.58 ± 0.01 &       LCO \\
2024-08-06 & 529.49 &    4.9 & 16.75 ± 0.01 & 16.38 ± 0.01 & 16.01 ± 0.01 & 15.67 ± 0.01 & 15.58 ± 0.01 &       LCO \\
2024-08-08 & 530.51 &    5.9 & 16.82 ± 0.01 & 16.42 ± 0.01 & 15.99 ± 0.01 & 15.63 ± 0.02 & 15.55 ± 0.02 &       LCO \\
2024-08-08 & 530.51 &    5.9 & 16.82 ± 0.01 & 16.41 ± 0.01 & 15.97 ± 0.01 & 15.62 ± 0.02 & 15.55 ± 0.02 &       LCO \\
2024-08-09 & 531.62 &    7.0 & 16.93 ± 0.01 & 16.50 ± 0.01 & 16.04 ± 0.01 & 15.64 ± 0.01 & 15.57 ± 0.01 &       LCO \\
2024-08-09 & 531.62 &    7.0 & 16.93 ± 0.01 & 16.51 ± 0.01 & 16.04 ± 0.01 & 15.65 ± 0.01 &          --- &       LCO \\
2024-08-09 & 531.86 &    7.2 & 16.86 ± 0.07 &          --- & 15.97 ± 0.07 &          --- &          --- &    Nickel \\
2024-08-09 & 531.86 &    7.3 &          --- &          --- &          --- & 15.65 ± 0.05 & 15.56 ± 0.05 &    Nickel \\
2024-08-10 & 532.72 &    8.1 & 17.21 ± 0.01 & 16.63 ± 0.01 & 16.04 ± 0.01 & 15.69 ± 0.01 & 15.56 ± 0.01 &       LCO \\
2024-08-10 & 532.73 &    8.1 & 17.17 ± 0.02 & 16.65 ± 0.01 & 16.06 ± 0.01 & 15.70 ± 0.01 & 15.55 ± 0.01 &       LCO \\
2024-08-11 & 533.93 &    9.3 & 17.30 ± 0.02 & 16.77 ± 0.01 & 16.14 ± 0.01 & 15.68 ± 0.01 & 15.53 ± 0.01 &       LCO \\
2024-08-11 & 533.93 &    9.3 & 17.26 ± 0.02 & 16.77 ± 0.01 & 16.11 ± 0.01 & 15.65 ± 0.01 & 15.53 ± 0.01 &       LCO \\
2024-08-13 & 535.62 &   11.0 & 17.59 ± 0.02 & 17.10 ± 0.01 & 16.33 ± 0.01 & 15.83 ± 0.01 & 15.68 ± 0.01 &       LCO \\
2024-08-13 & 535.62 &   11.0 & 17.61 ± 0.03 & 17.07 ± 0.01 & 16.34 ± 0.01 & 15.83 ± 0.01 & 15.69 ± 0.01 &       LCO \\
2024-08-13 & 535.85 &   11.2 & 17.55 ± 0.06 &          --- & 16.29 ± 0.08 & 15.83 ± 0.05 & 15.63 ± 0.06 &    Nickel \\
2024-08-14 & 536.63 &   12.0 & 17.75 ± 0.02 & 17.18 ± 0.01 & 16.39 ± 0.01 & 15.88 ± 0.01 & 15.70 ± 0.01 &       LCO \\
2024-08-14 & 536.64 &   12.0 & 17.73 ± 0.03 & 17.18 ± 0.01 & 16.40 ± 0.01 & 15.87 ± 0.01 & 15.68 ± 0.01 &       LCO \\
2024-08-15 & 537.64 &   13.0 & 17.92 ± 0.04 & 17.31 ± 0.02 & 16.47 ± 0.02 & 15.90 ± 0.01 & 15.66 ± 0.01 &       LCO \\
2024-08-15 & 537.64 &   13.0 & 17.93 ± 0.04 & 17.29 ± 0.02 & 16.45 ± 0.02 & 15.89 ± 0.01 & 15.68 ± 0.01 &       LCO \\
2024-08-16 & 538.76 &   14.1 & 18.05 ± 0.04 & 17.40 ± 0.02 & 16.44 ± 0.02 &          --- &          --- &       LCO \\
2024-08-16 & 538.76 &   14.1 & 18.05 ± 0.04 & 17.44 ± 0.02 & 16.43 ± 0.01 &          --- &          --- &       LCO \\
2024-08-16 & 538.76 &   14.2 &          --- &          --- &          --- & 15.99 ± 0.01 & 15.71 ± 0.01 &       LCO \\
2024-08-16 & 538.76 &   14.2 &          --- &          --- &          --- & 15.99 ± 0.01 & 15.73 ± 0.01 &       LCO \\
2024-08-17 & 539.83 &   15.2 & 18.25 ± 0.09 &          --- & 16.75 ± 0.07 & 16.07 ± 0.05 &          --- &    Nickel \\
2024-08-17 & 540.33 &   15.7 & 18.28 ± 0.10 & 17.61 ± 0.03 & 16.60 ± 0.03 & 16.04 ± 0.02 & 15.71 ± 0.02 &       LCO \\
2024-08-17 & 540.33 &   15.7 & 18.34 ± 0.11 & 17.65 ± 0.06 & 16.70 ± 0.03 & 16.03 ± 0.01 & 15.66 ± 0.01 &       LCO \\
2024-08-19 & 541.58 &   17.0 & 18.31 ± 0.14 & 17.73 ± 0.06 & 16.67 ± 0.07 & 15.95 ± 0.07 & 15.69 ± 0.03 &       LCO \\
2024-08-19 & 541.58 &   17.0 &          --- & 17.63 ± 0.08 & 16.69 ± 0.05 & 16.11 ± 0.05 & 15.69 ± 0.05 &       LCO \\
2024-08-21 & 543.84 &   19.2 & 18.55 ± 0.12 &          --- & 16.93 ± 0.08 & 16.30 ± 0.06 & 15.98 ± 0.05 &    Nickel \\
2024-08-21 & 544.43 &   19.8 & 18.76 ± 0.05 & 18.11 ± 0.02 & 16.96 ± 0.02 & 16.36 ± 0.02 & 16.02 ± 0.03 &       LCO \\
2024-08-23 & 546.46 &   21.8 & 19.11 ± 0.12 & 18.34 ± 0.03 & 17.21 ± 0.02 & 16.59 ± 0.01 &          --- &       LCO \\
2024-08-23 & 546.46 &   21.9 &          --- &          --- &          --- &          --- & 16.11 ± 0.01 &       LCO \\
2024-08-26 & 548.80 &   24.2 & 18.89 ± 0.09 &          --- & 17.20 ± 0.07 & 16.56 ± 0.04 & 16.22 ± 0.05 &    Nickel \\
2024-08-26 & 549.48 &   24.9 & 19.12 ± 0.08 & 18.34 ± 0.02 & 17.22 ± 0.02 & 16.60 ± 0.03 & 16.26 ± 0.04 &       LCO \\
2024-08-30 & 552.55 &   27.9 & 19.06 ± 0.06 & 18.46 ± 0.02 & 17.34 ± 0.02 & 16.73 ± 0.01 & 16.41 ± 0.01 &       LCO \\
2024-09-01 & 555.45 &   30.8 & 19.28 ± 0.07 & 18.53 ± 0.02 & 17.45 ± 0.02 & 16.87 ± 0.02 & 16.49 ± 0.02 &       LCO \\
2024-09-08 & 561.70 &   37.1 & 19.34 ± 0.09 & 18.57 ± 0.03 & 17.54 ± 0.02 & 17.04 ± 0.01 & 16.68 ± 0.01 &       LCO \\
2024-09-08 & 561.70 &   37.1 & 19.34 ± 0.07 & 18.55 ± 0.02 & 17.54 ± 0.03 & 17.04 ± 0.01 & 16.68 ± 0.01 &       LCO \\
2024-09-09 & 562.76 &   38.1 & 19.31 ± 0.14 &          --- & 17.66 ± 0.09 & 17.12 ± 0.06 & 16.77 ± 0.07 &    Nickel \\
2024-09-11 & 565.25 &   40.6 & 19.36 ± 0.09 & 18.83 ± 0.09 & 17.91 ± 0.08 & 17.30 ± 0.04 & 16.95 ± 0.05 &       LCO \\
2024-09-11 & 565.25 &   40.6 & 19.84 ± 0.07 &          --- & 17.85 ± 0.08 & 17.27 ± 0.04 & 16.89 ± 0.04 &       LCO \\
2024-09-13 & 566.76 &   42.1 &          --- &          --- & 17.71 ± 0.09 &          --- &          --- &    Nickel \\
2024-09-13 & 566.77 &   42.2 &          --- &          --- &          --- & 17.23 ± 0.06 & 16.92 ± 0.07 &    Nickel \\
2024-09-14 & 568.26 &   43.6 & 19.40 ± 0.15 &          --- &          --- &          --- &          --- &       LCO \\
2024-09-14 & 568.26 &   43.6 & 19.30 ± 0.14 &          --- &          --- &          --- &          --- &       LCO \\
2024-09-14 & 568.28 &   43.7 &          --- & 18.57 ± 0.07 & 17.74 ± 0.05 & 17.23 ± 0.03 & 16.86 ± 0.03 &       LCO \\
2024-09-14 & 568.28 &   43.7 &          --- & 18.56 ± 0.06 & 17.69 ± 0.05 & 17.19 ± 0.03 & 16.87 ± 0.04 &       LCO \\
2024-09-17 & 571.29 &   46.7 & 19.47 ± 0.17 & 18.63 ± 0.05 & 17.81 ± 0.05 & 17.36 ± 0.03 & 17.00 ± 0.02 &       LCO \\
2024-09-17 & 571.29 &   46.7 & 19.47 ± 0.17 & 18.66 ± 0.05 & 17.77 ± 0.06 & 17.34 ± 0.04 & 16.99 ± 0.03 &       LCO \\
2024-09-20 & 573.64 &   49.0 & 19.40 ± 0.08 & 18.74 ± 0.03 & 17.78 ± 0.03 & 17.45 ± 0.02 & 17.15 ± 0.05 &       LCO \\
2024-09-20 & 573.64 &   49.0 & 19.48 ± 0.09 & 18.70 ± 0.05 & 17.80 ± 0.03 & 17.48 ± 0.02 & 17.09 ± 0.03 &       LCO \\
2024-09-22 & 576.40 &   51.8 & 19.64 ± 0.10 & 18.78 ± 0.03 & 17.95 ± 0.03 & 17.51 ± 0.02 & 17.14 ± 0.03 &       LCO \\
2024-09-22 & 576.40 &   51.8 & 19.59 ± 0.09 & 18.81 ± 0.04 & 17.93 ± 0.03 & 17.52 ± 0.02 & 17.15 ± 0.03 &       LCO \\
2024-09-27 & 581.40 &   56.8 & 19.71 ± 0.13 & 18.86 ± 0.05 & 17.99 ± 0.03 & 17.64 ± 0.02 & 17.24 ± 0.03 &       LCO \\
2024-09-27 & 581.40 &   56.8 & 19.69 ± 0.11 & 18.83 ± 0.05 & 18.00 ± 0.03 & 17.63 ± 0.02 & 17.28 ± 0.04 &       LCO \\
2024-09-29 & 582.70 &   58.1 & 19.54 ± 0.11 &          --- & 18.04 ± 0.09 & 17.63 ± 0.06 & 17.35 ± 0.08 &    Nickel \\
2024-10-01 & 584.64 &   60.0 & 19.67 ± 0.12 & 18.81 ± 0.04 & 18.03 ± 0.03 & 17.64 ± 0.02 & 17.30 ± 0.03 &       LCO \\
2024-10-01 & 584.64 &   60.0 & 19.61 ± 0.12 & 18.84 ± 0.03 & 18.01 ± 0.04 & 17.64 ± 0.02 & 17.29 ± 0.02 &       LCO \\
2024-10-04 & 587.51 &   62.9 & 19.55 ± 0.12 & 18.93 ± 0.04 & 18.13 ± 0.04 & 17.78 ± 0.03 & 17.43 ± 0.02 &       LCO \\
2024-10-04 & 587.51 &   62.9 & 19.64 ± 0.13 & 18.91 ± 0.04 & 18.11 ± 0.04 & 17.75 ± 0.02 & 17.44 ± 0.02 &       LCO \\
2024-10-07 & 590.62 &   66.0 & 19.63 ± 0.11 & 18.86 ± 0.03 & 18.13 ± 0.04 & 17.79 ± 0.02 & 17.43 ± 0.02 &       LCO \\
2024-10-07 & 590.62 &   66.0 & 19.70 ± 0.11 & 18.85 ± 0.04 & 18.11 ± 0.04 & 17.80 ± 0.02 & 17.42 ± 0.03 &       LCO \\
2024-10-08 & 591.65 &   67.0 & 19.61 ± 0.17 &          --- & 18.29 ± 0.16 &          --- &          --- &    Nickel \\
2024-10-10 & 593.61 &   69.0 & 19.58 ± 0.14 & 18.86 ± 0.06 &          --- & 17.83 ± 0.03 & 17.48 ± 0.02 &       LCO \\
2024-10-10 & 593.61 &   69.0 &          --- & 18.93 ± 0.04 &          --- & 17.80 ± 0.03 & 17.51 ± 0.03 &       LCO \\
2024-10-11 & 594.70 &   70.1 & 19.80 ± 0.28 &          --- & 18.19 ± 0.12 & 17.91 ± 0.08 & 17.53 ± 0.12 &    Nickel \\
2024-10-11 & 594.69 &   70.1 &          --- &          --- &          --- & 17.92 ± 0.11 &          --- &    Nickel \\
2024-10-13 & 596.60 &   72.0 & 19.79 ± 0.14 & 18.98 ± 0.05 & 18.17 ± 0.05 & 17.91 ± 0.03 & 17.53 ± 0.03 &       LCO \\
2024-10-13 & 596.60 &   72.0 & 19.68 ± 0.13 & 19.05 ± 0.06 & 18.20 ± 0.05 & 17.87 ± 0.03 & 17.57 ± 0.02 &       LCO \\
2024-10-14 & 597.60 &   73.0 & 19.57 ± 0.17 & 18.91 ± 0.04 & 18.22 ± 0.06 & 17.91 ± 0.03 & 17.59 ± 0.03 &       LCO \\
2024-10-17 & 600.59 &   76.0 & 19.96 ± 0.13 & 18.82 ± 0.07 & 18.24 ± 0.08 & 18.00 ± 0.04 & 17.61 ± 0.04 &       LCO \\
2024-10-20 & 604.36 &   79.7 & 19.83 ± 0.11 & 19.04 ± 0.04 & 18.35 ± 0.04 & 18.06 ± 0.03 &          --- &       LCO \\
2024-10-20 & 604.36 &   79.8 &          --- &          --- &          --- &          --- & 17.56 ± 0.15 &       LCO \\
2024-10-23 & 607.34 &   82.7 & 19.93 ± 0.12 & 19.14 ± 0.05 & 18.43 ± 0.05 & 18.12 ± 0.03 & 17.81 ± 0.04 &       LCO \\
2024-10-27 & 610.59 &   86.0 & 19.87 ± 0.19 & 18.99 ± 0.05 & 18.39 ± 0.04 & 18.19 ± 0.03 & 17.80 ± 0.02 &       LCO \\
2024-11-01 & 615.56 &   90.9 & 19.80 ± 0.22 & 19.12 ± 0.04 & 18.46 ± 0.05 & 18.26 ± 0.03 & 17.90 ± 0.04 &       LCO \\
2024-11-03 & 618.32 &   93.7 & 19.68 ± 0.25 & 19.17 ± 0.15 & 18.81 ± 0.15 &          --- & 18.19 ± 0.20 &       LCO \\
2024-11-07 & 621.55 &   96.9 & 19.56 ± 0.21 & 19.17 ± 0.08 & 18.54 ± 0.06 & 18.37 ± 0.06 & 18.04 ± 0.04 &       LCO \\
2024-11-10 & 624.55 &   99.9 & 19.95 ± 0.32 & 19.17 ± 0.06 & 18.54 ± 0.07 & 18.36 ± 0.03 & 17.99 ± 0.03 &       LCO \\


\hline    
\multicolumn{9}{l}{\textsuperscript{} $^\dagger$ JD 2,460,000+ .
$^\ddagger$Phase  calculated with respect to $B_{\rm max} = 2,460,524.61$ .
} \\

\label{tab:photometric_observational_log_2024pxl}                                                        

\end{longtable*}

\begin{longtable}{lccc l}
\caption{z-band Photometry of SN~2024pxl} \label{tab:z_band_photometry_24pxl} \\
\hline\hline
Date & JD$^\dagger$ & Phase$^\ddagger$ (days) & $z$ (mag) & Telescope \\
\hline
\endfirsthead

\caption[]{z-band Photometry of SN~2024pxl (cont.)} \\
\hline\hline
Date & JD$^\dagger$ & Phase$^\ddagger$ (days) & $z$ (mag) & Telescope \\
\hline
\endhead

\hline
\endfoot

\hline
\endlastfoot

2024-07-26 & 517.83 & $-$6.3 & 16.47 $\pm$ 0.05 & Konkoly \\
2024-07-26 & 517.92 & $-$6.2 & 16.39 $\pm$ 0.06 & Baja \\
2024-07-27 & 518.83 & $-$5.3 & 16.31 $\pm$ 0.05 & Konkoly \\
2024-07-27 & 518.86 & $-$5.2 & 16.21 $\pm$ 0.06 & Baja \\
2024-07-28 & 519.84 & $-$4.3 & 16.18 $\pm$ 0.06 & Konkoly \\
2024-07-29 & 520.83 & $-$3.3 & 16.06 $\pm$ 0.05 & Konkoly \\
2024-07-29 & 520.88 & $-$3.2 & 15.91 $\pm$ 0.09 & Baja \\
2024-07-30 & 521.83 & $-$2.3 & 15.90 $\pm$ 0.04 & Konkoly \\
2024-07-30 & 521.87 & $-$2.2 & 15.82 $\pm$ 0.06 & Baja \\
2024-07-31 & 522.87 & $-$1.2 & 15.82 $\pm$ 0.03 & Konkoly \\
2024-07-31 & 522.88 & $-$1.2 & 15.73 $\pm$ 0.07 & Baja \\
2024-08-01 & 523.89 & $-$0.2 & 15.72 $\pm$ 0.07 & Baja \\
2024-08-03 & 525.89 &    1.8 & 15.54 $\pm$ 0.06 & Baja \\
2024-08-03 & 525.91 &    1.8 & 15.66 $\pm$ 0.05 & Konkoly \\
2024-08-04 & 526.86 &    2.8 & 15.64 $\pm$ 0.07 & Konkoly \\
2024-08-05 & 527.91 &    3.8 & 15.57 $\pm$ 0.05 & Konkoly \\
2024-08-06 & 528.86 &    4.8 & 15.59 $\pm$ 0.13 & Konkoly \\
2024-08-06 & 528.87 &    4.8 & 15.48 $\pm$ 0.06 & Baja \\
2024-08-07 & 529.89 &    5.8 & 15.59 $\pm$ 0.07 & Baja \\
2024-08-07 & 529.91 &    5.8 & 15.47 $\pm$ 0.04 & Konkoly \\
2024-08-09 & 531.83 &    7.7 & 15.44 $\pm$ 0.07 & Baja \\
2024-08-09 & 531.88 &    7.8 & 15.52 $\pm$ 0.04 & Konkoly \\
2024-08-10 & 532.81 &    8.7 & 15.53 $\pm$ 0.04 & Konkoly \\
2024-08-10 & 532.93 &    8.8 & 15.15 $\pm$ 0.24 & Baja \\
2024-08-11 & 533.81 &    9.7 & 15.54 $\pm$ 0.03 & Konkoly \\
2024-08-11 & 533.87 &    9.8 & 15.49 $\pm$ 0.04 & Baja \\
2024-08-12 & 534.81 &   10.7 & 15.52 $\pm$ 0.05 & Baja \\
2024-08-13 & 535.81 &   11.7 & 15.63 $\pm$ 0.05 & Konkoly \\
2024-08-13 & 535.83 &   11.7 & 15.39 $\pm$ 0.12 & Baja \\
2024-08-14 & 536.80 &   12.7 & 15.57 $\pm$ 0.06 & Konkoly \\
2024-08-14 & 536.83 &   12.7 & 15.55 $\pm$ 0.07 & Baja \\
2024-08-15 & 537.80 &   13.7 & 15.72 $\pm$ 0.08 & Konkoly \\
2024-08-15 & 537.83 &   13.7 & 15.55 $\pm$ 0.07 & Baja \\
2024-08-16 & 538.80 &   14.7 & 15.66 $\pm$ 0.04 & Konkoly \\
2024-08-18 & 540.84 &   16.7 & 15.73 $\pm$ 0.08 & Konkoly \\
2024-08-19 & 541.79 &   17.7 & 15.81 $\pm$ 0.08 & Konkoly \\
2024-08-22 & 544.85 &   20.8 & 15.91 $\pm$ 0.08 & Konkoly \\
2024-08-23 & 545.82 &   21.7 & 15.89 $\pm$ 0.07 & Baja \\
2024-08-23 & 545.84 &   21.7 & 16.01 $\pm$ 0.07 & Konkoly \\
2024-08-24 & 546.83 &   22.7 & 16.06 $\pm$ 0.06 & Konkoly \\
2024-08-25 & 547.83 &   23.7 & 15.94 $\pm$ 0.05 & Konkoly \\
2024-08-26 & 548.83 &   24.7 & 16.05 $\pm$ 0.07 & Konkoly \\
2024-08-28 & 550.83 &   26.7 & 16.08 $\pm$ 0.05 & Konkoly \\
2024-08-28 & 550.86 &   26.8 & 16.08 $\pm$ 0.07 & Baja \\
2024-08-29 & 551.78 &   27.7 & 16.16 $\pm$ 0.07 & Baja \\
2024-08-29 & 551.82 &   27.7 & 16.21 $\pm$ 0.05 & Konkoly \\
2024-08-30 & 552.80 &   28.7 & 16.17 $\pm$ 0.05 & Baja \\
2024-08-30 & 552.82 &   28.7 & 16.19 $\pm$ 0.06 & Konkoly \\
2024-08-31 & 553.80 &   29.7 & 16.32 $\pm$ 0.08 & Baja \\
2024-08-31 & 553.82 &   29.7 & 16.09 $\pm$ 0.08 & Konkoly \\
2024-09-01 & 554.83 &   30.7 & 16.28 $\pm$ 0.08 & Baja \\
2024-09-01 & 554.85 &   30.8 & 16.24 $\pm$ 0.07 & Konkoly \\
2024-09-02 & 555.80 &   31.7 & 16.33 $\pm$ 0.06 & Baja \\
2024-09-02 & 555.82 &   31.7 & 16.25 $\pm$ 0.06 & Konkoly \\
2024-09-03 & 556.79 &   32.7 & 16.27 $\pm$ 0.06 & Baja \\
2024-09-04 & 557.79 &   33.7 & 16.23 $\pm$ 0.08 & Baja \\
2024-09-04 & 557.82 &   33.7 & 16.39 $\pm$ 0.06 & Konkoly \\
2024-09-05 & 558.82 &   34.7 & 16.31 $\pm$ 0.05 & Konkoly \\
2024-09-07 & 560.81 &   36.7 & 16.42 $\pm$ 0.06 & Konkoly \\
2024-09-07 & 560.83 &   36.7 & 16.43 $\pm$ 0.07 & Baja \\
2024-09-08 & 561.81 &   37.7 & 16.31 $\pm$ 0.15 & Konkoly \\
2024-09-11 & 564.76 &   40.7 & 16.65 $\pm$ 0.06 & Baja \\
2024-09-11 & 564.81 &   40.7 & 16.54 $\pm$ 0.08 & Konkoly \\
2024-09-17 & 570.84 &   46.7 & 17.09 $\pm$ 0.15 & Baja \\
2024-09-19 & 572.78 &   48.7 & 16.68 $\pm$ 0.07 & Baja \\
2024-09-19 & 572.78 &   48.7 & 16.89 $\pm$ 0.09 & Konkoly \\
2024-09-20 & 573.82 &   49.7 & 16.96 $\pm$ 0.09 & Konkoly \\
2024-09-21 & 574.76 &   50.7 & 16.84 $\pm$ 0.08 & Konkoly \\
2024-09-21 & 574.76 &   50.7 & 16.88 $\pm$ 0.11 & Baja \\
2024-09-22 & 575.75 &   51.6 & 16.87 $\pm$ 0.06 & Baja \\
2024-09-22 & 575.75 &   51.7 & 17.04 $\pm$ 0.09 & Konkoly \\
2024-09-23 & 576.76 &   52.7 & 16.92 $\pm$ 0.07 & Konkoly \\
2024-09-24 & 577.74 &   53.6 & 17.00 $\pm$ 0.09 & Konkoly \\
2024-09-25 & 578.75 &   54.6 & 16.91 $\pm$ 0.05 & Baja \\
2024-09-26 & 579.77 &   55.7 & 16.86 $\pm$ 0.07 & Baja \\
2024-09-30 & 583.78 &   59.7 & 17.10 $\pm$ 0.07 & Baja \\
2024-10-07 & 590.75 &   66.7 & 17.05 $\pm$ 0.07 & Baja \\
2024-10-09 & 592.76 &   68.7 & 17.05 $\pm$ 0.10 & Baja \\

\end{longtable}

\vspace{1ex}
\noindent\footnotesize{
$^\dagger$JD = JD $-$ 2,460,000 \\
$^\ddagger$Phase  calculated with respect to $B_{\rm max} = 2,460,524.61$.
}

\begin{table}[ht]
\centering
\caption{UV and Optical Observations of SN~2024pxl}
\begin{tabular}{lcccccccc}
\hline
Date & JD$^\dagger$ & Phase$^\ddagger$ & $UVW2$ & $UVM2$ & $UVW1$ & $Swift~U$ & $Swift~B$ & $Swift~V$ \\
 &  & (days) & (mag) & (mag) & (mag) & (mag) & (mag) & (mag) \\
\hline
2024-07-26 & 518.17 & $-$6.44 & 19.09 ± 0.20 & 19.88 ± 0.32 & 18.03 ± 0.14 & 16.50 ± 0.08 & 17.00 ± 0.07 & 16.71 ± 0.10 \\
2024-07-28 & 519.90 & $-$4.71 & --- & --- & 17.90 ± 0.14 & 16.18 ± 0.08 & 16.72 ± 0.07 & --- \\
2024-07-28 & 519.91 & $-$4.70 & 19.34 ± 0.25 & 19.72 ± 0.32 & --- & --- & --- & 16.28 ± 0.09 \\
2024-07-31 & 523.09 & $-$1.52 & --- & --- & 17.84 ± 0.13 & 16.31 ± 0.08 & 16.53 ± 0.06 & --- \\
2024-07-31 & 523.10 & $-$1.51 & 19.05 ± 0.20 & --- & --- & --- & --- & 16.09 ± 0.08 \\
2024-08-04 & 527.19 & 2.58 & --- & --- & 18.28 ± 0.16 & 16.57 ± 0.08 & 16.45 ± 0.06 & --- \\
2024-08-04 & 527.20 & 2.59 & 19.64 ± 0.28 & --- & --- & --- & --- & 15.99 ± 0.07 \\
2024-08-07 & 530.09 & 5.48 & --- & 20.01 ± 0.35 & 18.62 ± 0.20 & 17.05 ± 0.10 & 16.68 ± 0.07 & 15.87 ± 0.07 \\
2024-08-10 & 533.17 & 8.56 & --- & --- & 19.03 ± 0.34 & 17.64 ± 0.17 & 17.06 ± 0.09 & --- \\
2024-08-10 & 533.18 & 8.57 & --- & --- & --- & --- & --- & 16.00 ± 0.09 \\
2024-08-13 & 536.39 & 11.78 & --- & --- & --- & --- & 17.51 ± 0.10 & 16.34 ± 0.09 \\
2024-08-16 & 538.81 & 14.20 & --- & --- & --- & 18.54 ± 0.27 & --- & --- \\
2024-08-16 & 538.82 & 14.21 & --- & --- & --- & --- & 17.85 ± 0.12 & --- \\
2024-08-19 & 542.04 & 17.43 & --- & --- & --- & 18.85 ± 0.26 & 18.67 ± 0.16 & 16.77 ± 0.09 \\
\hline \hline
\end{tabular}
\newline
$^\dagger$JD 2,460,000+ .
$^\ddagger$Phase  calculated with respect to $B_{\rm max} = 2,460,524.61$ .
\label{tab:swift_photometry_24pxl} 
\end{table}

\newpage
\begin{longtable*}{c c c l }
\caption{Log of optical spectroscopy of SN 2024pxl} \label{tab:spectroscopic_observations_24pxl} \\
\hline \hline
Date & JD$^\dagger$ & Phase$^\ddagger$ (days) & Telescope/Instrument \\
\hline
\endfirsthead
\caption{Log of optical spectroscopy of SN 2024pxl (cont.)} \\
\hline \hline
Date & JD$^\dagger$ & Phase$^\ddagger$ (days) & Telescope/Instrument \\
\hline
\endhead
\hline
\endfoot
\hline
\endlastfoot
2024-07-24 & 515.84 & $-$8.77 & P200/DBSP \\
2024-07-26 & 517.95 & $-$6.66 & FTN/FLOYDS \\
2024-07-27 & 518.74 & $-$5.87 & HET/LRS \\
2024-07-27 & 518.83 & $-$5.78 & Gemini-N/GMOS-N \\
2024-07-27 & 518.9  & $-$5.71 & Lick/Kast \\
2024-07-28 & 519.76 & $-$4.85 & FTN/FLOYDS \\
2024-07-29 & 520.80 & $-$3.81 & FTN/FLOYDS \\
2024-07-29 & 520.90 & $-$3.71 & ANU/WiFeS \\
2024-07-30 & 521.87 & $-$2.74 & ANU/WiFeS \\
2024-07-30 & 522.29 & $-$2.32 & SALT/RSS \\
2024-07-31 & 522.75 & $-$1.86 & Lick/Kast \\
2024-07-31 & 523.29 & $-$1.32 & SALT/RSS \\
2024-08-01 & 523.73 & $-$0.88 & Lick/Kast \\
2024-08-01 & 523.78 & $-$0.83 & P200/DBSP \\
2024-08-01 & 523.79 & $-$0.82 & Keck/LRIS \\
2024-08-01 & 523.87 & $-$0.74 & ANU/WiFeS \\
2024-08-01 & 523.89 & $-$0.72 & FTN/FLOYDS \\
2024-08-01 & 524.30 & $-$0.31 & SALT/RSS \\
2024-08-02 & 524.92 & 0.31 & ANU/WiFeS \\
2024-08-02 & 525.19 & 0.58 & Lick/Kast \\
2024-08-03 & 525.83 & 1.22 & FTN/FLOYDS \\
2024-08-05 & 527.79 & 3.18 & FTN/FLOYDS \\
2024-08-05 & 527.79 & 3.18 & HET/LRS \\
2024-08-05 & 528.29 & 3.68 & SALT/RSS \\
2024-08-06 & 528.77 & 4.16 & Lick/Kast \\
2024-08-06 & 528.88 & 4.27 & FTN/FLOYDS \\
2024-08-06 & 528.92 & 4.31 & LT/SPRAT \\
2024-08-06 & 528.93 & 4.32 & ANU/WiFeS \\
2024-08-06 & 529.28 & 4.67 & SALT/RSS \\
2024-08-07 & 530.20 & 5.59 & Lick/Kast \\
2024-08-07 & 530.20 & 5.59 & GTC/OSIRIS \\
2024-08-08 & 530.80 & 6.19 & FTN/FLOYDS \\
2024-08-10 & 533.47 & 8.86 & NOT/ALFOSC \\
2024-08-12 & 534.75 & 10.14 & MMT/Binospec \\
2024-08-12 & 534.79 & 10.18 & FTN/FLOYDS \\
2024-08-13 & 536.13 & 11.52 & HCT/HFOSC \\
2024-08-14 & 536.81 & 12.20 & FTN/FLOYDS \\
2024-08-14 & 537.27 & 12.66 & SALT/RSS \\
2024-08-14 & 537.36 & 12.75 & Lick/Kast \\
2024-08-14 & 537.41 & 12.80 & NOT/ALFOSC \\
2024-08-15 & 537.79 & 13.18 & HET/LRS \\
2024-08-15 & 537.90 & 13.29 & LT/SPRAT \\
2024-08-15 & 538.36 & 13.75 & SALT/RSS \\
2024-08-15 & 538.36 & 13.75 & SALT/RSS \\
2024-08-16 & 538.74 & 14.13 & P200/DBSP \\
2024-08-16 & 538.80 & 14.19 & Lick/Kast \\
2024-08-18 & 540.82 & 16.21 & FTN/FLOYDS \\
2024-08-19 & 541.88 & 17.27 & ANU/WiFeS \\
2024-08-23 & 545.76 & 21.15 & FTN/FLOYDS \\
2024-08-23 & 546.10 & 21.49 & HCT/HFOSC \\
2024-08-23 & 546.43 & 21.82 & NOT/ALFOSC \\
2024-08-23 & 546.43 & 21.82 & SALT/RSS \\
2024-08-26 & 548.88 & 24.27 & ANU/WiFeS \\
2024-08-27 & 549.75 & 25.14 & Lick/Kast \\
2024-08-27 & 549.94 & 25.33 & FTN/FLOYDS \\
2024-08-29 & 551.52 & 26.91 & Lick/Kast \\
2024-08-31 & 553.66 & 29.05 & MMT/Binospec \\
2024-08-31 & 553.73 & 29.12 & Keck/LRIS \\
2024-08-31 & 553.93 & 29.32 & FTN/FLOYDS \\
2024-09-02 & 555.83 & 31.22 & Keck/LRIS \\
2024-09-04 & 557.88 & 33.27 & FTN/FLOYDS \\
2024-09-05 & 558.52 & 33.91 & GTC/OSIRIS \\
2024-09-05 & 558.92 & 34.31 & ANU/WiFeS \\
2024-09-05 & 559.00 & 34.39 & Lick/Kast \\
2024-09-05 & 559.49 & 34.88 & VLT/XShooter \\
2024-09-06 & 559.92&  35.31 & NOT/ALFOSC \\
2024-09-06 & 560.35&  35.74 & NOT/ALFOSC \\
2024-09-08 & 561.65 & 37.04 & MMT/Binospec \\
2024-09-08 & 561.71 & 37.10 & Lick/Kast \\ 
2024-09-10 & 563.76 & 39.15 & FTN/FLOYDS \\
2024-09-11 & 565.00 & 40.39 & Lick/Kast \\
2024-09-13 & 566.67 & 42.06 & Lick/Kast \\  
2024-09-14 & 567.79 & 43.18 & FTN/FLOYDS \\
2024-09-14 & 567.91 & 43.30 & ANU/WiFeS \\
2024-09-18 & 571.72 & 47.11 & FTN/FLOYDS \\
2024-09-21 & 575.36 & 50.75 & NOT/ALFOSC \\
2024-09-23 & 576.87 & 52.26 & FTN/FLOYDS \\
2024-09-28 & 581.79 & 57.18 & Lick/Kast \\
2024-09-30 & 583.58 & 58.97 & MMT/Binospec \\
2024-10-01 & 584.74 & 60.13 & FTN/FLOYDS \\
2024-10-03 & 586.65 & 62.04 & Lick/Kast \\ 
2024-10-04 & 587.92 & 63.31 & Lick/Kast \\
2024-10-05 & 588.74 & 64.13 & FTN/FLOYDS \\
2024-10-05 & 588.79 & 64.18 & Keck/LRIS \\
2024-10-05 & 588.92 & 64.31 & Lick/Kast \\
2024-10-05 & 589.00 & 64.39 & NOT/ALFOSC \\
2024-10-05 & 589.37 & 64.76 & NOT/ALFOSC \\
2024-10-08 & 591.79 & 67.18 & Keck/LRIS \\
2024-10-10 & 593.91 & 69.31 & Lick/Kast \\
2024-10-12 & 595.66 & 71.05 & Lick/Kast \\
2024-10-19 & 602.72 & 78.11 & FTN/FLOYDS \\
2024-10-24 & 607.63 & 83.02 & Lick/Kast \\  
2024-10-25 & 608.71 & 84.10 & FTN/FLOYDS \\
2024-10-26 & 609.79 & 85.18 & Lick/Kast \\
2024-10-30 & 613.70 & 89.09 & Keck/LRIS \\
\end{longtable*}

{\footnotesize
$^\dagger$JD 2,460,000+ . \\
$^\ddagger$Phase calculated with respect to $B_{\rm max} = 2,460,524.61$ .
}

\begin{table*}
\caption{Log of NIR Spectroscopy of SN 2024pxl}
\centering
\smallskip
\begin{tabular}{c c c c }
\hline \hline
Date     & JD$^\dagger$     & Phase$^\ddagger$              & Telescope/Instrument       \\
          &                  &(Days)                                    &                 \\
\hline
2024-07-26   & 517.95  &    $-$6.66                    & Keck/NIRES  \\
2024-07-28   & 520.63  &    $-$3.98                    & Magellan/FIRE  \\
2024-08-10   & 532.95  &    8.34                       & Magellan/FIRE  \\
2024-08-12   & 534.88  &    10.27                      & IRTF/SpeX  \\
2024-08-19   & 542.43  &    17.82                      & GTC/EMIR   \\
2024-08-26   & 548.02  &    23.41                      & Soar/TripleSpec  \\
2024-09-05   & 559.49  &    34.88                      & VLT/XShooter$^\ast$ \\
2024-09-06   & 560.49  &    35.88              	       & Soar/TripleSpec  \\
2024-09-13   & 567.07  &    42.46         	           & GTC/EMIR   \\

\hline                                   
\end{tabular}
\newline
$^\dagger$JD 2,460,000+ .
$^\ddagger$Phase  calculated with respect to $B_{\rm max} = 2,460,524.61$ .
$^\ast$This is the same spectrum as referred to in Table \ref{tab:spectroscopic_observations_24pxl}. 
\label{tab:nir_spectroscopic_observations_24pxl}     
\end{table*}

\section{Optical photometry and spectroscopy of SN 2017drh}
\label{sec:appendixA_17drh}

\begin{figure}[ht]
	\begin{center}
		\includegraphics[width=0.6\columnwidth]{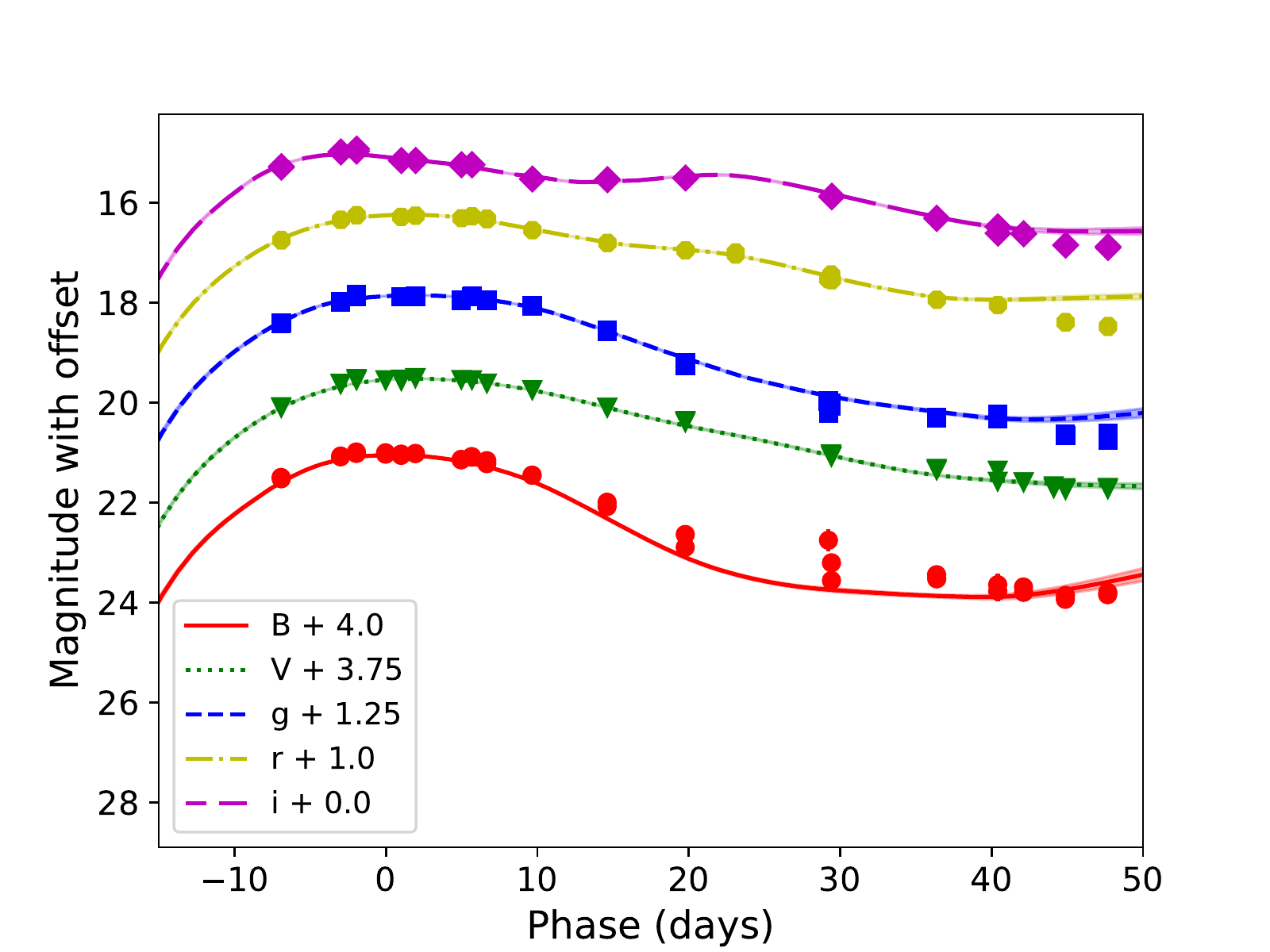}
	\end{center}
	\caption{LCO light curves of 2017drh with offsets. The overplotted solid curve corresponds to the BayeSN model where $R_V = 3.1$, while the dashed curve represents the BayeSN model with $R_V = 1.45$. Unfortunately, without NIR data, we cannot distinguish well between the two models. The phases are given in the rest frame of SN~2017drh.}
	\label{fig:SN_2017drh_light_curve}
\end{figure}

We present the optical Las Cumbres Observatory photometry and FLOYDS spectroscopy for SN 2017drh, which exploded in the same host galaxy, NGC 6384, as SN~2024pxl. The spectra were reduced using the FLOYDS reduction pipeline \citep{valenti_floyds_2014}. The photometric observations were taken as part of the Global Supernova Project (GSP) collaboration. These data were reduced with \textsc{lcogtsnpipe} \citep{valenti_lco_2016}, a PyRAF-based image-reduction pipeline. In Figure~\ref{fig:SN_2017drh_light_curve}, the LCO data are shown with the BayeSN light curve fits that also included data from \citet{Stahl2019}.

\begin{figure}[ht]
	\begin{center}
		\includegraphics[width=0.7\columnwidth]{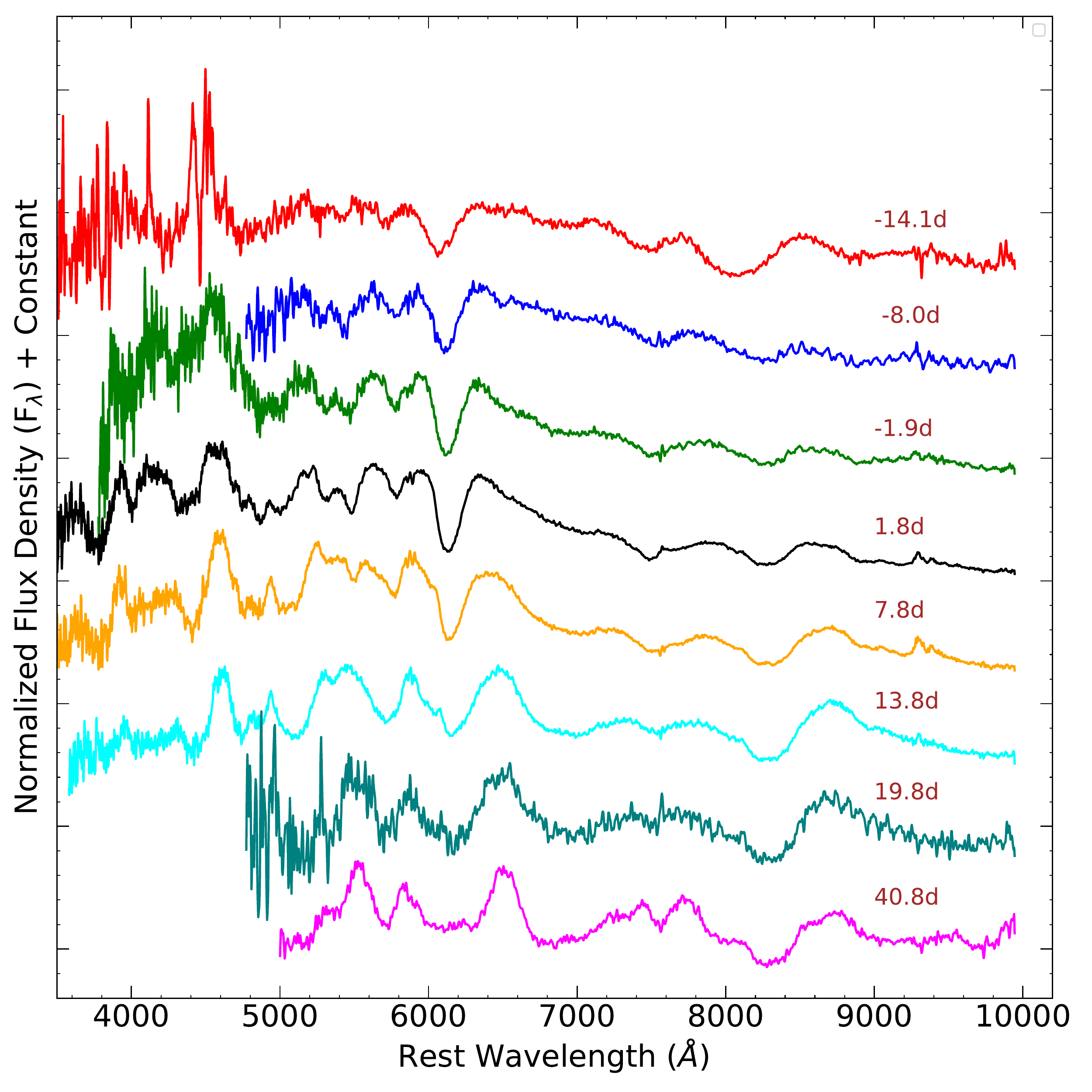}
	\end{center}
	\caption{FTN/FLOYDS spectra of 2017drh from a phase of $-$14.1 days to 40.8 days. A full list of spectroscopic observations is presented in ~\autoref{tab:spectroscopic_observations_17drh}.}
	\label{fig:SN_2017drh_spectral_evolution}
\end{figure}

\begin{longtable*}{>{\centering\arraybackslash}p{0.10\textwidth}
>{\centering\arraybackslash}p{0.06\textwidth}
>{\raggedright\arraybackslash}p{0.06\textwidth}
>{\centering\arraybackslash}p{0.10\textwidth}
>{\centering\arraybackslash}p{0.10\textwidth}
>{\centering\arraybackslash}p{0.10\textwidth}
>{\centering\arraybackslash}p{0.10\textwidth}
>{\centering\arraybackslash}p{0.10\textwidth}
>{\centering\arraybackslash}p{0.10\textwidth}}

\caption{Photometric Observations of SN 2017drh}
\\
\hline
Date & JD$^\dagger$ & Phase$^\ddagger$  & $U$ & $B$ & $g$ & $V$ & $r$ & $i$ \\
 & & (Days) & (mag) & (mag)  & (mag) & (mag) & (mag) & (mag)   \\
\hline  
\endfirsthead

\multicolumn{9}{c}%
    {{\bfseries \tablename\ \thetable{} -- Continued from previous page}} \\
    \hline
    Date    &   JD$^\dagger$   &   Phase$^\ddagger$ 	&   $U$       &      $B$        &    $g$     &  $V$   	    &   $r$      &     $i$             \\
    \hline
    \endhead

\hline 
\multicolumn{9}{l}{\textsuperscript{} $^\dagger$JD 2,457,000+ ,
$^\ddagger$Phase  calculated with respect to $B_{\rm max} = 2,457,891.2$.
} \\
\multicolumn{9}{r}{\textit{Cont. next page}} \\ 
\hline
    \endfoot
    
    \hline
    \endlastfoot

2017-05-10 & 883.54 & $-$7.66 & 18.51$\pm$0.26 & 17.46$\pm$0.03 & --- & --- & --- & --- \\
2017-05-10 & 883.55 & $-$7.65 & --- & 17.52$\pm$0.05 & 17.16$\pm$0.02 & 16.35$\pm$0.02 & --- & --- \\
2017-05-10 & 883.56 & $-$7.64 & --- & --- & 17.15$\pm$0.02 & --- & 15.74$\pm$0.01 & 15.27$\pm$0.02 \\
2017-05-13 & 887.48 & $-$3.72 & 18.37$\pm$0.08 & --- & --- & --- & --- & --- \\
2017-05-13 & 887.49 & $-$3.71 & 18.41$\pm$0.08 & 17.09$\pm$0.03 & --- & --- & --- & --- \\
2017-05-14 & 887.50 & $-$3.70 & --- & 17.07$\pm$0.04 & 16.73$\pm$0.02 & 15.87$\pm$0.02 & --- & --- \\
2017-05-14 & 887.51 & $-$3.69 & --- & --- & --- & --- & 15.35$\pm$0.01 & 14.99$\pm$0.01 \\
2017-05-15 & 888.53 & $-$2.67 & 18.01$\pm$0.21 & --- & --- & --- & --- & --- \\
2017-05-15 & 888.54 & $-$2.66 & 17.96$\pm$0.19 & 16.99$\pm$0.04 & --- & --- & --- & --- \\
2017-05-15 & 888.55 & $-$2.65 & --- & --- & 16.59$\pm$0.04 & 15.98$\pm$0.02 & --- & --- \\
2017-05-15 & 888.56 & $-$2.64 & --- & --- & --- & --- & 15.25$\pm$0.02 & 14.97$\pm$0.03 \\
2017-05-16 & 889.56 & $-$1.64 & 18.69$\pm$0.09 & --- & --- & --- & --- & --- \\
2017-05-16 & 890.47 & $-$0.73 & 18.32$\pm$0.07 & --- & --- & --- & --- & --- \\
2017-05-16 & 890.48 & $-$0.72 & --- & 16.95$\pm$0.03 & --- & 15.75$\pm$0.02 & --- & --- \\
2017-05-18 & 891.50 & 0.30 & 18.41$\pm$0.06 & --- & --- & --- & --- & --- \\
2017-05-18 & 891.51 & 0.31 & 18.40$\pm$0.06 & 17.06$\pm$0.02 & --- & 15.86$\pm$0.02 & --- & --- \\
2017-05-18 & 891.52 & 0.32 & --- & --- & 16.63$\pm$0.01 & 15.81$\pm$0.01 & 15.28$\pm$0.01 & --- \\
2017-05-18 & 891.53 & 0.33 & --- & --- & --- & --- & 15.28$\pm$0.01 & 15.16$\pm$0.01 \\
2017-05-18 & 892.45 & 1.25 & 18.11$\pm$0.19 & --- & --- & --- & --- & --- \\
2017-05-18 & 892.46 & 1.26 & --- & 16.98$\pm$0.02 & --- & 15.76$\pm$0.01 & --- & --- \\
2017-05-18 & 892.47 & 1.27 & --- & --- & 16.61$\pm$0.01 & --- & 15.26$\pm$0.01 & --- \\
2017-05-18 & 892.48 & 1.28 & --- & --- & --- & --- & --- & 15.15$\pm$0.01 \\
2017-05-21 & 895.49 & 4.29 & 18.21$\pm$0.18 & --- & --- & --- & --- & --- \\
2017-05-22 & 895.50 & 4.30 & 18.23$\pm$0.18 & 17.21$\pm$0.03 & --- & --- & --- & --- \\
2017-05-22 & 895.51 & 4.31 & --- & --- & 16.69$\pm$0.01 & 15.80$\pm$0.02 & --- & --- \\
2017-05-22 & 895.52 & 4.32 & --- & --- & --- & --- & 15.31$\pm$0.01 & 15.24$\pm$0.01 \\
2017-05-22 & 896.19 & 4.99 & 18.76$\pm$0.07 & --- & --- & --- & --- & --- \\
2017-05-22 & 896.20 & 5.00 & 18.70$\pm$0.08 & 17.08$\pm$0.02 & --- & --- & --- & --- \\
2017-05-22 & 896.21 & 5.01 & --- & 17.23$\pm$0.02 & 16.63$\pm$0.01 & 15.84$\pm$0.02 & --- & --- \\
2017-05-22 & 896.22 & 5.02 & --- & --- & 16.62$\pm$0.01 & --- & 15.27$\pm$0.01 & 15.23$\pm$0.01 \\
2017-05-23 & 897.19 & 5.99 & 18.55$\pm$0.21 & --- & --- & --- & --- & --- \\
2017-05-23 & 897.20 & 6.00 & --- & 17.45$\pm$0.03 & --- & 16.02$\pm$0.02 & --- & --- \\
2017-05-23 & 897.21 & 6.01 & --- & --- & 16.71$\pm$0.01 & 15.86$\pm$0.01 & 15.32$\pm$0.05 & --- \\
2017-05-23 & 897.22 & 6.02 & --- & --- & --- & --- & 15.33$\pm$0.01 & --- \\
2017-05-26 & 900.20 & 9.00 & 19.21$\pm$0.08 & --- & --- & --- & --- & --- \\
2017-05-26 & 900.21 & 9.01 & 18.98$\pm$0.32 & 17.45$\pm$0.03 & --- & --- & --- & --- \\
2017-05-26 & 900.22 & 9.02 & --- & --- & 16.82$\pm$0.01 & 16.07$\pm$0.02 & --- & --- \\
2017-05-26 & 900.23 & 9.03 & --- & --- & --- & --- & 15.55$\pm$0.01 & 15.53$\pm$0.01 \\
2017-05-31 & 905.18 & 13.98 & 19.08$\pm$0.34 & --- & --- & --- & --- & --- \\
2017-05-31 & 905.19 & 13.99 & --- & 18.38$\pm$0.04 & --- & --- & --- & --- \\
2017-05-31 & 905.20 & 14.00 & --- & --- & 17.33$\pm$0.02 & 16.42$\pm$0.02 & --- & --- \\
2017-05-31 & 905.21 & 14.01 & --- & --- & --- & --- & 15.81$\pm$0.01 & 15.54$\pm$0.01 \\
2017-06-05 & 910.37 & 19.17 & 20.21$\pm$0.35 & --- & --- & --- & --- & --- \\
2017-06-05 & 910.38 & 19.18 & 20.37$\pm$0.30 & 18.64$\pm$0.12 & --- & --- & --- & --- \\
2017-06-05 & 910.39 & 19.19 & --- & --- & 18.00$\pm$0.04 & 16.61$\pm$0.03 & --- & --- \\
2017-06-05 & 910.40 & 19.20 & --- & --- & --- & --- & 15.96$\pm$0.02 & 15.51$\pm$0.02 \\
2017-06-09 & 913.73 & 22.53 & --- & --- & --- & --- & 16.04$\pm$0.03 & --- \\
2017-06-09 & 913.74 & 22.54 & --- & --- & --- & 16.91$\pm$0.04 & --- & --- \\
2017-06-09 & 913.75 & 22.55 & --- & 18.97$\pm$0.22 & --- & 16.80$\pm$0.04 & --- & --- \\
2017-06-09 & 913.76 & 22.56 & --- & 19.21$\pm$0.25 & --- & --- & --- & --- \\
2017-06-15 & 919.88 & 28.68 & --- & --- & 18.72$\pm$0.07 & --- & 16.53$\pm$0.03 & --- \\
2017-06-15 & 919.89 & 28.69 & --- & 18.76$\pm$0.23 & 18.96$\pm$0.15 & --- & --- & --- \\
2017-06-15 & 920.06 & 28.86 & --- & --- & 18.83$\pm$0.06 & 17.49$\pm$0.04 & --- & --- \\
2017-06-15 & 920.07 & 28.87 & --- & --- & --- & --- & 16.43$\pm$0.04 & --- \\
2017-06-15 & 920.08 & 28.88 & --- & --- & 18.75$\pm$0.09 & --- & --- & --- \\
2017-06-15 & 920.09 & 28.89 & --- & 19.56$\pm$0.18 & --- & 17.51$\pm$0.04 & --- & --- \\
2017-06-15 & 920.10 & 28.90 & 20.63$\pm$0.35 & --- & --- & --- & --- & --- \\
2017-06-15 & 920.11 & 28.91 & --- & --- & --- & --- & 16.55$\pm$0.02 & 15.87$\pm$0.02 \\
2017-06-22 & 927.07 & 35.87 & --- & 19.71$\pm$0.07 & --- & --- & --- & --- \\
2017-06-22 & 927.08 & 35.88 & --- & 19.82$\pm$0.07 & --- & 17.91$\pm$0.05 & --- & 16.31$\pm$0.03 \\
2017-06-22 & 927.09 & 35.89 & --- & --- & 19.04$\pm$0.06 & --- & 16.94$\pm$0.06 & --- \\
2017-06-26 & 931.11 & 39.91 & 20.98$\pm$0.48 & --- & --- & --- & --- & --- \\
2017-06-26 & 931.12 & 39.92 & 20.25$\pm$0.43 & --- & --- & --- & --- & --- \\
2017-06-26 & 931.13 & 39.93 & --- & 19.64$\pm$0.23 & 19.07$\pm$0.11 & 17.83$\pm$0.09 & --- & --- \\
2017-06-26 & 931.14 & 39.94 & --- & --- & 18.98$\pm$0.07 & --- & 17.05$\pm$0.03 & 16.48$\pm$0.06 \\
2017-06-26 & 931.15 & 39.95 & --- & --- & --- & --- & --- & 16.61$\pm$0.12 \\
2017-06-28 & 932.83 & 41.63 & 19.72$\pm$0.43 & --- & --- & --- & --- & --- \\
2017-06-28 & 932.84 & 41.64 & 19.93$\pm$0.54 & 19.69$\pm$0.18 & --- & --- & --- & --- \\
2017-06-28 & 932.85 & 41.65 & --- & 19.80$\pm$0.17 & --- & 17.84$\pm$0.05 & --- & 16.62$\pm$0.03 \\
2017-06-28 & 932.86 & 41.66 & --- & --- & --- & --- & --- & 16.63$\pm$0.02 \\
2017-06-30 & 934.85 & 43.65 & --- & --- & --- & 18.00$\pm$0.03 & --- & --- \\
2017-07-01 & 935.61 & 44.41 & 21.06$\pm$0.26 & --- & --- & --- & --- & --- \\
2017-07-01 & 935.62 & 44.42 & 21.01$\pm$0.21 & 19.94$\pm$0.13 & --- & --- & --- & --- \\
2017-07-01 & 935.63 & 44.43 & --- & 19.86$\pm$0.16 & 19.38$\pm$0.07 & 18.00$\pm$0.05 & --- & --- \\
2017-07-01 & 935.64 & 44.44 & --- & --- & 19.41$\pm$0.07 & --- & 17.40$\pm$0.03 & 16.85$\pm$0.03 \\
2017-07-01 & 935.65 & 44.45 & --- & --- & --- & --- & --- & 16.85$\pm$0.03 \\
2017-07-03 & 938.43 & 47.23 & --- & 19.85$\pm$0.19 & --- & --- & --- & --- \\
2017-07-03 & 938.44 & 47.24 & --- & 20.25$\pm$0.17 & --- & 17.98$\pm$0.05 & --- & --- \\
2017-07-03 & 938.45 & 47.25 & --- & --- & 19.50$\pm$0.10 & --- & 17.49$\pm$0.03 & --- \\
2017-07-03 & 938.46 & 47.26 & --- & --- & --- & --- & 17.47$\pm$0.03 & 16.90$\pm$0.03 \\
2017-07-09 & 943.67 & 52.47 & --- & 20.39$\pm$0.35 & --- & --- & --- & --- \\
2017-07-09 & 943.68 & 52.48 & --- & --- & 19.12$\pm$0.26 & 18.23$\pm$0.08 & --- & --- \\
2017-07-09 & 943.69 & 52.49 & --- & --- & 18.65$\pm$0.22 & --- & 17.42$\pm$0.14 & --- \\
2017-07-09 & 943.70 & 52.50 & --- & --- & --- & --- & --- & 16.93$\pm$0.12 \\
2017-07-15 & 950.05 & 58.85 & --- & 19.80$\pm$0.19 & --- & --- & --- & --- \\
2017-07-15 & 950.06 & 58.86 & --- & --- & 19.59$\pm$0.12 & 18.24$\pm$0.06 & --- & --- \\
2017-07-15 & 950.07 & 58.87 & --- & --- & 19.51$\pm$0.11 & --- & 17.78$\pm$0.03 & --- \\
2017-07-15 & 950.08 & 58.88 & --- & --- & --- & --- & --- & 17.32$\pm$0.04 \\
2017-07-21 & 956.29 & 65.09 & --- & 20.26$\pm$0.08 & --- & --- & --- & --- \\
2017-07-21 & 956.30 & 65.10 & --- & 20.29$\pm$0.08 & --- & 18.50$\pm$0.03 & --- & --- \\
2017-07-21 & 956.31 & 65.11 & --- & --- & 19.78$\pm$0.11 & --- & 18.12$\pm$0.05 & --- \\
2017-07-21 & 956.32 & 65.12 & --- & --- & --- & --- & 18.14$\pm$0.05 & 17.59$\pm$0.05 \\
2017-07-27 & 962.34 & 71.14 & --- & 20.39$\pm$0.10 & --- & --- & --- & --- \\
2017-07-27 & 962.35 & 71.15 & --- & --- & 19.74$\pm$0.09 & 18.65$\pm$0.04 & --- & --- \\
2017-07-27 & 962.36 & 71.16 & --- & --- & 19.70$\pm$0.10 & --- & 18.31$\pm$0.05 & --- \\
2017-07-27 & 962.37 & 71.17 & --- & --- & --- & --- & --- & 17.82$\pm$0.06 \\
2017-08-02 & 968.32 & 77.12 & --- & 20.65$\pm$0.38 & --- & --- & --- & --- \\
2017-08-02 & 968.33 & 77.13 & --- & 20.88$\pm$0.48 & --- & 18.65$\pm$0.11 & --- & --- \\
2017-08-02 & 968.34 & 77.14 & --- & --- & 20.09$\pm$0.18 & --- & 18.47$\pm$0.09 & --- \\
2017-08-02 & 968.35 & 77.15 & --- & --- & --- & --- & 18.59$\pm$0.10 & --- \\
2017-08-08 & 973.98 & 82.78 & --- & 19.86$\pm$0.21 & --- & --- & --- & --- \\
2017-08-08 & 973.99 & 82.79 & --- & --- & 19.53$\pm$0.13 & 19.36$\pm$0.18 & --- & --- \\
2017-08-08 & 974.00 & 82.80 & --- & --- & 20.04$\pm$0.14 & --- & 18.50$\pm$0.11 & --- \\
2017-08-08 & 974.01 & 82.81 & --- & --- & --- & --- & --- & 18.16$\pm$0.11 \\
2017-08-14 & 979.98 & 88.78 & --- & 20.15$\pm$0.21 & --- & --- & --- & --- \\
2017-08-14 & 979.99 & 88.79 & --- & 20.58$\pm$0.14 & --- & 18.96$\pm$0.10 & --- & --- \\
2017-08-14 & 980.00 & 88.80 & --- & --- & 19.18$\pm$0.26 & --- & --- & --- \\
2017-08-19 & 984.55 & 93.35 & --- & 20.03$\pm$0.14 & --- & --- & --- & --- \\
2017-08-19 & 984.56 & 93.36 & --- & 20.66$\pm$0.18 & --- & 19.27$\pm$0.10 & --- & --- \\
2017-08-19 & 984.57 & 93.37 & --- & --- & 19.98$\pm$0.09 & --- & 19.24$\pm$0.10 & --- \\
2017-08-19 & 984.58 & 93.38 & --- & --- & --- & --- & 19.31$\pm$0.11 & 18.60$\pm$0.10 \\
2017-08-25 & 990.56 & 99.36 & --- & 20.26$\pm$0.18 & --- & 19.38$\pm$0.08 & --- & --- \\
2017-08-25 & 990.57 & 99.37 & --- & --- & 20.02$\pm$0.11 & 19.52$\pm$0.07 & --- & --- \\
2017-08-25 & 990.58 & 99.38 & --- & --- & --- & --- & 19.26$\pm$0.10 & 18.79$\pm$0.10 \\
2017-08-25 & 990.59 & 99.39 & --- & --- & --- & --- & --- & 18.88$\pm$0.09 \\
2017-08-31 & 996.94 & 105.74 & --- & 19.53$\pm$0.09 & --- & 19.32$\pm$0.09 & --- & --- \\
2017-08-31 & 996.95 & 105.75 & --- & --- & 19.86$\pm$0.10 & 20.44$\pm$0.50 & --- & --- \\
2017-08-31 & 996.96 & 105.76 & --- & --- & --- & --- & 19.35$\pm$0.09 & 19.09$\pm$0.11 \\
2017-08-31 & 996.97 & 105.77 & --- & --- & --- & --- & --- & 19.93$\pm$0.22 \\
2017-09-06 & 1002.92 & 111.72 & --- & 21.22$\pm$0.32 & --- & --- & --- & --- \\
2017-09-06 & 1002.93 & 111.73 & --- & --- & 20.37$\pm$0.13 & 19.72$\pm$0.25 & --- & --- \\
2017-09-06 & 1002.94 & 111.74 & --- & --- & 20.61$\pm$0.19 & --- & 20.00$\pm$0.16 & --- \\
2017-09-06 & 1002.95 & 111.75 & --- & --- & --- & --- & --- & 19.10$\pm$0.16 \\
2017-09-12 & 1008.63 & 117.43 & --- & 20.83$\pm$0.15 & --- & --- & --- & --- \\
2017-09-12 & 1008.64 & 117.44 & --- & --- & 20.07$\pm$0.13 & 19.87$\pm$0.18 & --- & --- \\
2017-09-12 & 1008.65 & 117.45 & --- & --- & 20.50$\pm$0.17 & --- & 19.87$\pm$0.13 & --- \\
2017-09-12 & 1008.66 & 117.46 & --- & --- & --- & --- & --- & 19.24$\pm$0.13 \\
2017-09-16 & 1012.88 & 121.68 & --- & 20.35$\pm$0.21 & --- & --- & --- & --- \\
2017-09-16 & 1012.89 & 121.69 & --- & --- & 21.08$\pm$0.14 & 19.82$\pm$0.14 & --- & --- \\
2017-09-16 & 1012.90 & 121.70 & --- & --- & 20.07$\pm$0.11 & --- & --- & --- \\
2017-09-16 & 1012.91 & 121.71 & --- & --- & --- & --- & 20.18$\pm$0.20 & 19.53$\pm$0.17 \\
2017-09-16 & 1012.92 & 121.72 & --- & --- & --- & --- & --- & 19.46$\pm$0.17 \\
2017-10-09 & 1035.59 & 144.39 & --- & 21.34$\pm$0.17 & --- & --- & --- & --- \\
2017-10-09 & 1035.60 & 144.40 & --- & --- & --- & 20.29$\pm$0.10 & --- & --- \\
2017-10-09 & 1035.61 & 144.41 & --- & --- & 20.78$\pm$0.12 & --- & --- & --- \\
2017-10-09 & 1035.62 & 144.42 & --- & --- & --- & --- & 20.55$\pm$0.19 & 20.10$\pm$0.25 \\
2017-10-09 & 1035.63 & 144.43 & --- & --- & --- & --- & --- & 21.80$\pm$0.48 \\
2017-10-17 & 1043.56 & 152.36 & --- & 21.72$\pm$0.40 & --- & --- & --- & --- \\
2017-10-17 & 1043.57 & 152.37 & --- & 21.47$\pm$0.25 & --- & 20.61$\pm$0.15 & --- & --- \\
2017-10-17 & 1043.58 & 152.38 & --- & --- & --- & 20.19$\pm$0.12 & --- & --- \\
2017-10-17 & 1043.59 & 152.39 & --- & --- & 21.67$\pm$0.14 & --- & 20.37$\pm$0.21 & --- \\
2017-10-17 & 1043.60 & 152.40 & --- & --- & --- & --- & 20.61$\pm$0.19 & 19.87$\pm$0.22 \\
2018-03-07 & 1184.92 & 293.72 & --- & 21.15$\pm$0.44 & --- & --- & --- & --- \\
2018-03-07 & 1184.93 & 293.73 & --- & 20.96$\pm$0.42 & --- & --- & --- & --- \\
2018-03-07 & 1184.94 & 293.74 & --- & --- & --- & 21.46$\pm$0.45 & --- & --- \\
2018-03-07 & 1184.95 & 293.75 & --- & --- & 22.27$\pm$0.30 & 21.19$\pm$0.41 & --- & --- \\
2018-03-07 & 1184.96 & 293.76 & --- & --- & 23.12$\pm$0.31 & --- & --- & --- \\
2018-03-07 & 1184.97 & 293.77 & --- & --- & --- & --- & 22.14$\pm$0.16 & --- \\
2018-03-07 & 1184.98 & 293.78 & --- & --- & --- & --- & --- & 21.41$\pm$0.23 \\
2018-03-07 & 1184.99 & 293.79 & --- & --- & --- & --- & --- & 21.70$\pm$0.24 \\

\hline    
\multicolumn{9}{l}{\textsuperscript{} $^\dagger$ JD 2,457,000+ .
$^\ddagger$Phase calculated with respect to $B_{\rm max} = 2,457,891.2$ .
} \\

\label{tab:photometric_table_2017drh}
                                                                            
\end{longtable*} 

\begin{table*}[ht]
\caption{Log of Spectroscopy of SN 2017drh}
\centering
\smallskip
\begin{tabular}{c c c c }
\hline \hline
Date     & JD$^\dagger$     & Phase$^\ddagger$                  & Telescope/Instrument       \\
          &                  &(days)                                    &                 \\
\hline
2017-05-03  & 877.07     & $-$14.12         & FTN/FLOYDS  \\
2017-05-09  & 883.19     & $-$8.00          & FTN/FLOYDS \\
2017-05-15  & 889.21     & $-$1.98          & FTN/FLOYDS  \\
2017-05-19  & 893.05     & 1.86             & FTN/FLOYDS  \\
2017-05-25  & 899.06     & 7.86             & FTN/FLOYDS  \\
2017-05-31  & 905.04     & 13.85            & FTN/FLOYDS \\
2017-06-06  & 911.06     & 19.86            & FTN/FLOYDS  \\
2017-06-27  & 932.01     & 40.81            & FTN/FLOYDS  \\

\hline                                   
\end{tabular}
\newline
$^\dagger$JD 2,457,000+ .
$^\ddagger$ Phase  calculated with respect to $B_{\rm max} = 2,457,891.20$ .
\label{tab:spectroscopic_observations_17drh}     
\end{table*}




\end{appendix}

\end{document}